\documentclass[english
]{aa}
\usepackage{epsf,amsfonts,amssymb,graphicx,fancyheadings,caption}
\usepackage{natbib}
\usepackage{babel}
\usepackage{txfonts}
\usepackage{float}
\bibpunct{(}{)}{;}{a}{}{,}

\begin{document}

\title{New Magellanic Cloud R Coronae Borealis and DY Per type stars from the EROS-2 database: the connection between RCBs, DYPers and ordinary carbon stars
\thanks{Based on observations made with the CNRS/INSU MARLY telescope at the European
Southern Observatory, La Silla, Chile.}
}

\author{
P.~Tisserand\inst{1,2},
P.R.~Wood\inst{1},
J.B.~Marquette\inst{3},
C.~Afonso\inst{2}\thanks{Now at Max-Planck-Institut f\"ur Astronomie, Koenigstuhl 17, D-69117
 Heidelberg, Germany},
J.N.~Albert\inst{4},
J.~Andersen\inst{5},
R.~Ansari\inst{4},
\'{E}.~Aubourg\inst{2},
P.~Bareyre\inst{2},
J.P.~Beaulieu\inst{3},
X.~Charlot\inst{2},
C.~Coutures\inst{2,3},
R.~Ferlet\inst{3},
P.~Fouqu\'{e}\inst{6},
J.F.~Glicenstein\inst{2},
B.~Goldman\inst{2}$^{\star\star}$,
A.~Gould\inst{8},
M.~Gros\inst{2},
J.~de Kat\inst{2},
\'{E}.~Lesquoy\inst{2,3},
C.~Loup\inst{3}\thanks{Now at Observatoire Astronomique de Strasbourg, UMR 7550, 11 rue de
 l'Universit\'e, 67000 Strasbourg, France},
C.~Magneville\inst{2},
\'{E}.~Maurice\inst{9},
A.~Maury\inst{7}\thanks{Now at San Pedro de Atacama Celestial Exploration,
Casilla 21, San Pedro de Atacama, Chile},
A.~Milsztajn\inst{2}\thanks{deceased},
M.~Moniez\inst{4},
N.~Palanque-Delabrouille\inst{2},
O.~Perdereau\inst{4},
J.~Rich\inst{2},
P.~Schwemling\inst{10},
M.~Spiro\inst{2},
and A.~Vidal-Madjar\inst{3}
}

\institute{
Research School of Astronomy and Astrophysics, Australian National University, Cotter Rd, Weston Creek
 ACT 2611, Australia \and
CEA, DSM, DAPNIA, Centre d'\'{E}tudes de Saclay, 91191 Gif-sur-Yvette Cedex, France \and
Institut d'Astrophysique de Paris, UMR7095 CNRS, Universite Pierre \& Marie Curie, 98 bis boulevard
 Arago, 75014 Paris, France \and
Laboratoire de l'Acc\'{e}l\'{e}rateur Lin\'{e}aire, IN2P3 CNRS, Universit\'{e} de Paris-Sud, 91405 Orsay
 Cedex, France \and
The Niels Bohr Institute, Astronomy Group, Juliane Maries Vej 30, 2100 Copenhagen, Denmark \and
Observatoire Midi-Pyr\'en\'ees, UMR 5572, 14 avenue Edouard Belin, 31400 Toulouse, France \and
European Southern Observatory, Casilla 19001, Santiago 19, Chile \and
Department of Astronomy, Ohio State University, Columbus, OH 43210, U.S.A. \and
Observatoire de Marseille, 2 place Le Verrier, 13248 Marseille Cedex 04, France \and
LPNHE, IN2P3 CNRS and Universit\'{e}s Paris 6 \& Paris 7, 4 place
 Jussieu, 75252 Paris Cedex 05, France
}

\offprints{P. Tisserand; \email{tisseran@mso.anu.edu.au}}

\date{Received ; Accepted}

 %% Real number of objects analysed: LMC: 62244518 ; SMC: 7906448
\abstract {R Coronae Borealis stars (RCB) are a rare type of evolved carbon-rich supergiant stars that are increasingly thought to result from the merger of two white dwarfs, called the Double degenerate scenario. This scenario is also studied as a source, at higher mass, of type Ia Supernovae (SnIa) explosions. Therefore a better understanding of RCBs composition would help to constrain simulations of such events.}
{We searched for and studied RCB stars in the EROS Magellanic Clouds database. We also extended our research to DY Per type stars (DYPers) that are expected to be cooler RCBs (T$\sim$3500 K) and much more numerous than their hotter counterparts. With the aim of studying possible evolutionary connections between RCBs and DYPers, and also ordinary carbon stars, we compared their publically available broad band photometry in the optical, near, and mid-infrared.} 
{The light curves of $\sim$70 millions stars, monitored for 6.7 years (from July 1996 to February 2003), have been analysed to search for the main signature of RCBs and DYPers: a large (up to 9 mags) drop in luminosity. Carbon stars with fading episodes were also found by inspecting numerous light curves of objects that presented an infrared excess in the 2MASS and Spitzer- SAGE and S$^3$MC databases. Follow-up optical spectroscopy was used to confirm each photometric candidate found.} 
{We have discovered and confirmed 6 new Magellanic Cloud RCB stars and 7 new DYPers, but also listed new candidates: 3 RCBs and 14 DYPers. Optical and infrared colour magnitude diagrams that give new insights into these two sets of stars are discussed. We estimated a range of Magellanic RCB shell temperatures between 360 and 600 K.} 
{We confirm the wide range of absolute luminosity known for RCB stars, $M_V\sim$-5.2 to -2.6. Our study further shows that mid-infrared surveys are ideal to search for RCB stars, since they have thinner and cooler circumstellar shells than classical post-AGB stars.
In addition, by increasing the number of known DYPers by $\sim$400\%, we have been able to shed light on the similarities in the spectral energy distribution between DYPers and ordinary carbon stars. We also observed that DYPer circumstellar shells are fainter and hotter than those of RCBs. This suggests that DYPers may simply be ordinary carbon stars with ejection events, but more abundance analysis is necessary to give a status on a possible evolutionnary connexion between RCBs and DYPers.}


\authorrunning{P. Tisserand et al. (EROS coll.)}
\titlerunning{new Magellanic R Coronae Borealis stars from EROS-2}

\maketitle

\section{Introduction \label{sec_intro}}

R Coronae Borealis (RCB) stars are hydrogen-deficient and carbon-rich supergiant stars. They are known to undergo unpredictable fast declines in brightness, up to 9 magnitudes in the visible, as their cores are shadowed by newly formed dust clouds along the line of sight. The prototype RCB star, R CrB, has shown such declines over the last $\sim$200 years. The dust clouds, made of amorphous carbon, are thought to be formed near the stellar atmosphere and accelerated away by radiation pressure. The composition of RCB atmospheres, being $\sim$98\% helium and $\sim$1\% carbon, indicates that they are currently passing through a late stage of stellar evolution. A detailed review of their characteristics has been written by \citet{1996PASP..108..225C}. RCB stars are also rare: only 51 have been discovered in our Galaxy \citep[][and references therein]{2008A&A...481..673T}, including 9 that are likely located inside the Galactic Bulge, 17 in the Large Magellanic Cloud \citep{2001ApJ...554..298A}, and 5 in the Small Magellanic Cloud \citep{2004A&A...424..245T}.
This rarity suggests that RCBs may correspond to a brief phase of stellar evolution or an uncommon evolutionary path.

Two major evolutionary scenarios have been suggested to explain their origin: the Double Degenerate (DD) scenario and the final Helium Shell Flash (FF) scenario \citep{1996ApJ...456..750I, 1990ASPC...11..549R}. The DD model involves the merger of a CO- and a He- white dwarfs. It was recently strongly supported by two observations: an $^{18}$O over-abundance in seven H-deficient carbon and cool RCB stars, that is not expected in the FF model \citep{2007ApJ...662.1220C}; and the large abundances of fluorine in the hotter RCBs' atmospheres \citep{2008ApJ...674.1068P}. Such a scenario for RCB stars would not only help to constrain simulations of low mass Double Degenerate merging events \citep{2008ASPC..391..335F,2008arXiv0811.4646D}, but their birthrates would also help us to understand better the rates of mergers for objects that could be supernovae type Ia progenitors \citep{1984ApJ...277..355W,2005ApJ...629..915B}. The FF model involves the expansion of a star, on the verge of becoming a white dwarf, to supergiant size. This outburst phenomenon has already been observed in three stars (Sakurai's object, V605 Aql and FG Sge), which transformed themselves into cool giant stars with spectral properties similar to those of RCB stars \citep{1997AJ....114.2679C, 1999A&A...343..507A,1998ApJS..114..133G}. However, \citet{2006ApJ...646L..69C} note that differences between FF star and RCB star light curve variations and abundance patterns (such as the $^{12}$C$/^{13}$C ratio) indicate that the former are unlikely to be the evolutionary precursors of the majority of the latter. Furthermore, we also stress these two peculiarities : V605 Aql, which formed dust in 1921, has been in a deep decline ever since - it even evolved back to being a hot ($\sim$ 95000 K) star - \citep{2006ApJ...646L..69C}, and the duration of the interval containing fading events observed in Sakurai's object is much shorter than the phase observed in R CrB ($>200$ yrs).

The systematic monitoring of millions of stars and the advanced light curve analysis techniques of microlensing surveys make these surveys ideal for discovering such rare variable stars. The MACHO and EROS-2 collaborations have thus discovered half of the known sample of RCB stars \citep{2001ApJ...554..298A,2004A&A...424..245T,2005AJ....130.2293Z,2008A&A...481..673T}. Note also that the OGLE-3 collaboration presents a useful real-time monitoring of known RCBs on their web site \citep{2008AcA....58..187U}. Further surveys such as VISTA \citep{2004SPIE.5489..638M}, SkyMapper \citep{2007PASA...24....1K} and LSST will generate an increase of the sample thanks to the important time-series measurements that are expected to be produced. A better understanding of the spatial distribution of RCBs will thus be possible as well as better constraints on their age.

This article is the third of the series that describe searches for RCB stars in the EROS-2 (Exp\'erience de Recherche d'Objets Sombres) database. The first two, \citet{2004A&A...424..245T,2008A&A...481..673T}, are hereafter related to as T04 and T08. We report here an analysis of the Magellanic EROS-2 database, which uses the same technique as that applied in T08. We looked for the main signature of RCB stars, some consequent, irregular and rapid fades. It is important to increase the number of known Magellanic RCB stars, as they enable a good determination of their absolute magnitude range.

DY Per type stars are thought to be a sub-class of RCBs, the coolest ones ($T\sim$3500 K). Only 8 examples are currently known: 4 in the LMC \citep{2001ApJ...554..298A}, 2 in the SMC \citep{2004A&A...424..245T}, and 2 others in our Galaxy \citep{2008A&A...481..673T}, including the prototype DY Persei \citep{1990IsSKZ..33...83A}. They are recognised among typical RCB stars by their slower declines in brightness and symmetric recoveries. A recent analysis of DY Persei by \citet{2007A&A...472..247Z} suggests a significant hydrogen deficiency and high carbon abundance, compared to classic known carbon stars, but does not confirm a significant enhancement of $^{13}$C as found by \citet{2001ApJ...554..298A} in the atmosphere of 4 LMC DY Per-like stars.  As stated in \citet{2007A&A...472..247Z}, the coolest RCB stars should be much more numerous than the warmest RCBs, and more observations and analysis need to be done as it is not clear whether the DYPer stars are related to either RCB stars or more classical carbon stars \citep{2001ApJ...554..298A}.

This article reports on the discovery of 6 new Magellanic RCB and 7 new DYPer stars and discusses their optical, infrared and mid-infrared properties. Some new candidates are also listed: 3 RCBs and 14 DYPers. The photometric and spectroscopic data used are presented in Section~\ref{sec_obs}. A discussion of the previously known RCB stars in and near the area monitored by EROS-2 is given in Section~\ref{sec_knownRCB}. The detection techniques, light curve analysies and catalogue mining are presented in Section~\ref{sec_mining}. The general characteristics of the newly discovered RCBs are discussed in Section~\ref{sec_newRCB}.

\section{Observational data \label{sec_obs}}

The EROS-2 project used the 1-metre MARLY telescope at ESO La Silla Observatory, Chile. The primary purpose of the project was to search for microlensing events \citep{1986ApJ...304....1P} due to baryonic dark matter in the halo \citep{2007A&A...469..387T} or to ordinary stars in the Galactic plane \citep{2006A&A...454..185H,2009arXiv0901.1325R}. The observations were performed between July 1996 and February 2003 with two wide-field cameras (0.69$\degr$ in right ascension $\times$ 1.38$\degr$ in declination, thus a sky area of $\sim$0.95 deg$^2$) behind a dichroic cube splitting the light beam into two broad passbands. The so-called ``blue'' channel (420-720 nm, hereafter magnitudes $B_\mathrm{E}$) overlapped the $V$ and $R$ standard bands, while the ``red'' channel (620-920 nm, hereafter magnitudes $R_\mathrm{E}$) roughly matched the mean wavelength of the Cousins $I$ band. Each camera had a mosaic of eight 2048 $\times$ 2048 CCDs with a pixel size of 0.6$\arcsec$ on the sky.

The photometric calibration was obtained by matching our star catalogues with those of \citet{2002AJ....123..855Z} for the SMC and \citet{2004AJ....128.1606Z} for the LMC. To a precision of 0.1 mag, we found the following transformations with the standard V Johnson and I Cousins broadband:
\begin{equation}
$$ R_E = I, \hspace{3 mm} B_E = V - 0.4(V - I)$$.
\label{eq.photcalib}
\end{equation}
The calibration has been refined for each individual field and CCD, with special care applied to the LMC central area. The Magellanic fields are designated ``ooxxxqq'', where ``oo'' is the object (lm or sm), ``xxx'' the field number (001 to 088 for the LMC and 001 to 010 for the SMC), and ``qq'' the CCD quadrant. The sly areas covered by these fields are $\sim$84 ($\sim$9) deg$^2$  for the LMC (SMC). They are represented on Figure~\ref{map_fig}. The photometry of individual images and the reconstruction of the light curves were processed using the PEIDA package which was specifically developed for the EROS experiment \citep{1996VA.....40..519A}.

The template images were formed with 15 good seeing co-added images. They were used to detect sources and then form the initial object catalogue. The exposure time ranged from 180 to 900 seconds. More detailed information on the Magellanic EROS-2 data can be found in \citet{2007A&A...469..387T}.

Spectroscopy of RCB candidates was performed with the Dual-Beam Spectrograph (DBS) \citep{1988PASP..100..626R} attached to the ANU\footnote{Australian National University} 2.3m diameter telescope at Siding Spring Observatory. The DBS is a general purpose optical spectrograph, permanently mounted at the Nasmyth A focus. The visible wave-band is split by a dichroic at around 600 nm and feeds two essentially similar spectrographs, with red and blue optimised detectors. The full slit length is 6.7 arcmin. Observations are presented, with a 2-pixel resolution of 2 \AA{}. They were all obtained in 2008; individual epochs are listed in Table~\ref{tab.RCB.GenInfo}.

\section{Previously known RCB stars \label{sec_knownRCB}}

Table~\ref{tab.KnownRCB.GenInfo} reviews the EROS-2 status of the 22 confirmed Magellanic RCB and DYPer stars (17 in the LMC \citep{2001ApJ...554..298A} and 5 in the SMC\footnote{Note that we have renamed these 5 stars to correspond to the new EROS-2 RCBs and DYPers designation, as in T08 and this article} \citep{2004A&A...424..245T}) as well as 5 RCB candidates selected either from their optical spectra \citep{2003MNRAS.344..325M}\footnote{We have not listed in Table~\ref{tab.KnownRCB.GenInfo} the RCB candidates RAW 21 and KDM 2492 presented by \citet{2003MNRAS.344..325M} as they are already confirmed RCB stars, namely EROS2-SMC-RCB-1 \citep{2004A&A...424..245T} and HV 5637 \citep{1972MNRAS.158P..11F}.} or their mid-infrared featureless spectra \citep{2005ApJ...631L.147K,2006AJ....132.1890B}. The EROS-2 light curves of all these objects, if catalogued, are available at the following URL: \texttt{http://eros.in2p3.fr/Variables/RCB/RCB-LMC.html} 

We catalogued and obtained photometric time-series for 16 of the 17 confirmed LMC RCB and DYPer stars presented by \citet{2001ApJ...554..298A}; only HV 12842, located in the extreme north part of the LMC (outside our fields) is missing. Large declines are observed during the 6.7 years of our monitoring in 11 cases.

Of the 5 RCB candidates, only one, namely MSX-SMC-014, was not catalogued being too faint on our reference image. We also did not find any brightness variation during the 6.7 years EROS-2 observations with a careful inspection of all available images, but we note that for the first time its light curve is available, thanks to the OGLE-3 work \citep{2008AcA....58..187U}. Two of the other four candidates, KDM 2373 and KDM 7101, will be discussed further in Section~\ref{sec_newRCB} as they are considered as new EROS-2 Magellanic RCB stars. Of the remaining two, MSX-LMC-775's light curve has large oscillations in brightness, but not fast fading typical of RCB stars. We estimate a colour $V-I\sim$1.3. KDM 5651 shows oscillations and two small symmetric drops of $\sim$0.3 mag (its MACHO light curve presents another drop of $\sim$0.5 mag at about JD$\sim$2450220 and no fading phase is observed in its OGLE-3 light curve). KDM 5651 is a good RCB candidate; in Section~\ref{sbsec_midir} we compare its mid-infrared properties with other RCB and DYPer stars.

\begin{table*} 
\caption{Known confirmed and candidate Magellanic RCB stars located in the EROS-2 fields; coordinates from 2MASS 
\label{tab.KnownRCB.GenInfo}}
\medskip
\centering
\begin{tabular}{llcllc}
\hline
\hline
Star name & EROS-2 star & Coordinates ($J_{2000}$) & Other identifier & Large decline\\
 & identifier & from 2MASS & & seen by EROS-2?\\
\hline
\hline
& & Confirmed RCB & & &\\
& & or DYPer stars & & &\\
\hline
HV 5637$^{a}$ & lm0181l22619 & 05:11:31.37 -67:55:50.6 & KDM 2492 & no  \\
 & & & MACHO-20.5036.12 & \\
W Men$^{a}$ & lm0584k5872 & 05:26:24.52 -71:11:11.8 & MACHO-21.7407.7 & yes \\
HV 12842$^{a}$ &  & 05:45:02.87 -64:24:22.8 & &  \\
MACHO-11.8632.2507$^{b}$ & lm0027m19359 & 05:33:48.94 -70:13:23.4 & HV 2671 & yes  \\
MACHO-81.8394.1358$^{b}$ & lm0025k33192 & 05:32:13.36 -69:55:57.8 & & yes  \\
MACHO-6.6575.13$^{c}$ & lm0016k29468 & 05:20:48.21 -70:12:12.5 & & yes, remained faint most of the time\\
MACHO-6.6696.60$^{c}$ & lm0016m10260 & 05:21:47.98 -70:09:56.9 & HV 942 & yes \\
MACHO-12.10803.56$^{c}$ & lm0601k18043 & 05:46:47.74 -70:38:13.5 & & yes  \\
MACHO-16.5641.22$^{c}$ & lm0190l20117 & 05:14:46.20 -67:55:47.4 & HV 2379 & yes  \\
MACHO-18.3325.148$^{c}$ & lm0111k3287 & 05:01:00.36 -69:03:43.2 & HV 12524 & no  \\
MACHO-79.5743.15$^{c}$ & lm0091k19526 & 05:15:51.79 -69:10:08.6 & & yes  \\
MACHO-80.6956.207$^{c}$ & lm0206l17380 & 05:22:57.37 -68:58:18.9 & & yes  \\
MACHO-80.7559.28$^{c}$ & lm0011m13457 & 05:26:33.91 -69:07:33.4 & & yes  \\
MACHO-2.5871.1759$^{c}$ & lm0196m8601 & 05:16:51.96 -68:45:16.8 & & no  \\
MACHO-10.3800.35$^{c}$ & lm0102l12314 & 05:03:44.65 -69:38:11.7 & & no  \\
MACHO-15.10675.10$^{c}$ & lm0603l29693 & 05:46:13.27 -71:07:40.4 & & yes  \\
MACHO-78.6460.7$^{c}$ & lm0014k7770 & 05:19:55.93 -69:48:05.9 & & no \\
EROS2-SMC-RCB-1$^{d}$ & sm0102l20592 & 00:37:47.11 -73:39:02.3 & RAW-21 &  yes   \\
EROS2-SMC-RCB-2$^{d}$ & sm0014k11612 & 00:48:22.96 -73:41:04.7 & RAW-476 &  yes  \\
EROS2-SMC-RCB-3$^{d}$ & sm0067m28134 & 00:57:18.15 -72:42:35.2 & MACHO-207.16426.1662 & yes  \\
 & & & MSX-SMC-155 & \\
EROS2-SMC-DYPer-1$^{d}$ & sm0077k11497 & 00:44:07.50 -72:44:16.4 & RAW-233 & yes  \\
 & & & MACHO-208.15571.60 & \\
EROS2-SMC-DYPer-2$^{d}$ & sm0106m19412 & 00:40:14.72 -74:11:21.6 & [MH95]-431 & yes \\
%Attention, Maybe good to change the number of the known SMC RCB, = order as appeared in SMC paper.
\hline
& & RCB candidates & & \\
\hline
KDM 2373$^{e}$ & lm0105m6306 & 05:10:28.52 -69:47:04.4 & & Confirmed RCB in this article \\
KDM 5651$^{e}$ & lm0593m31181 & 05:41:23.51 -70:58:01.7 & MACHO-15.9830.5 & Oscillations and small \\
 & & & & $\sim$0.3 mag drops observed\\
KDM 7101$^{e}$ & lm0745l21924 & 06:04:05.46 -72:51:22.8 & & Confirmed RCB in this article  \\
MSX-SMC-014$^{f}$ &  & 00:46:16.33 -74:11:13.6 & monitored by OGLE-3 & Too faint to be catalogued \\
MSX-LMC-755$^{g}$ & lm0213l10369 & 05:32:56.18 -68:12:49.0 & MACHO-8.8541.68 & Large oscillation observed \\
\hline
\multicolumn{6}{l}{$^{a}$ \citet{1972MNRAS.158P..11F,1996PASP..108..225C}, $^{b}$ \citet{1996ApJ...470..583A}, $^{c}$ \citet{2001ApJ...554..298A}, $^{d}$ \citet{2004A&A...424..245T}, }\\ 
\multicolumn{6}{l}{$^{e}$ \citet{2003MNRAS.344..325M}, $^{f}$ \citet{2005ApJ...631L.147K}, $^{g}$ \citet{2006AJ....132.1890B}}\\
\end{tabular}
\end{table*}

\section{Mining the EROS database \label{sec_mining}}

As in T08, we used an extended version of the EROS-2 Magellanic catalogue produced by \citet{2007A&A...469..387T} for the microlensing analysis. The new catalogue has grown in objects detected in only one filter, but has also been enriched with objects whose light curves are affected by diverse optical and electronic artefacts. About 62 (8) million different objects have then been catalogued and analysed in the LMC (SMC). We used two different strategies to search for RCB or DYPer stars in the EROS-2 database. In both cases, we searched for the main signature, a rapid drop in luminosity of $\sim$2 to 8 mag, but also paid attention to smaller declines.

The first and main technique is based on a series of selection cuts applied to each light curve, both filters being considered separately, to select variations larger than 2.5 mag and to reject variable stars. The analysis and its detection efficiency are as described in T08 and will not be discussed further here. 13 new RCB candidates were selected for spectroscopy follow-up with this technique. We note that 10 (5) of the 16 (5) known LMC (SMC) RCB stars located in our fields were detected. The missed ones simply did not show large enough variations. All 13 stars were considered as strong RCB or DYPer candidates.

The second strategy is a visual inspection of the light curves of all EROS-2 objects that matched at least one of the three following criteria: (a) catalogued as a carbon star, using \citet{2001A&A...369..932K} catalogue for the LMC and  \citet{1993A&AS...97..603R}, \citet{1995A&AS..113..539M}, \citet{2007yCat....102023S} catalogues for the SMC; (b) has a high infrared excess in the 2MASS database \citep{2006AJ....131.1163S} and a correlation factor between the red and blue light curves higher than 0.5; or (c) has a mid-infrared emission characteristic of a dust shell, thanks to the public Spitzer-SAGE data for the LMC \citep{2006AJ....132.2268M} and the Spitzer-S$^3$MC data for the SMC \citep{2007ApJ...655..212B}. The selection limits in the 2MASS (J-H, H-K plane) and the Spitzer ([8.0], [3.6]-[8.0] plane) database are presented in Figure~\ref{infra_red}. We note also that the MACHO light curves were used, if available, to study and reject selected interesting candidates.

This last analysis is deployed to find possible candidates missed by the first technique but also to recover stars
that would present smaller declines but with a shape characteristic of RCBs or DYPers. These stars, if confirmed as carbon stars, would then help us to carry out more follow-up studies on a possible evolutionary connection between RCBs, DYPers and more ordinary carbon stars presenting fading events. We inspected 5882, 1157 and 7064 light curves selected respectively by criteria (a), (b) and (c) and kept 8, 2 and 13 candidates for spectroscopy follow-up. Among these new 23 candidates, only two were considered as strong DYPer candidates (EROS2-LMC-DYPer-4 and -5), the others (mostly DYPer type) show small non-periodic declines (lower than 2.0 mag) but with a shape that reminds us of RCB or DYPer stars.

Finally, we note that we have also analysed the 8 million light curves from the EROS-1 LMC database (27 deg$^2$ monitored from 1990 to 1994 with photographic plates, and digitised). Unfortunately, we did not find any new RCB or DYPer stars, but obtained light curves of 12 known LMC RCB or DYPer stars. The EROS-1 measurements are also available at the URL mentioned in Section\ref{sec_knownRCB}.

\subsection{Spectroscopic selection}
\label{spect_sect}
If a well-sampled light curve is available, identification with the RCB class can be made with fairly high confidence because of the distinct nature of the RCB brightness drops. Of the 36 candidates, only 5 showed one or multiple drops of more than 4 magnitudes, therefore spectroscopic information was necessary.

\begin{figure} %figure Spectra
\centering
\includegraphics[scale=0.45]{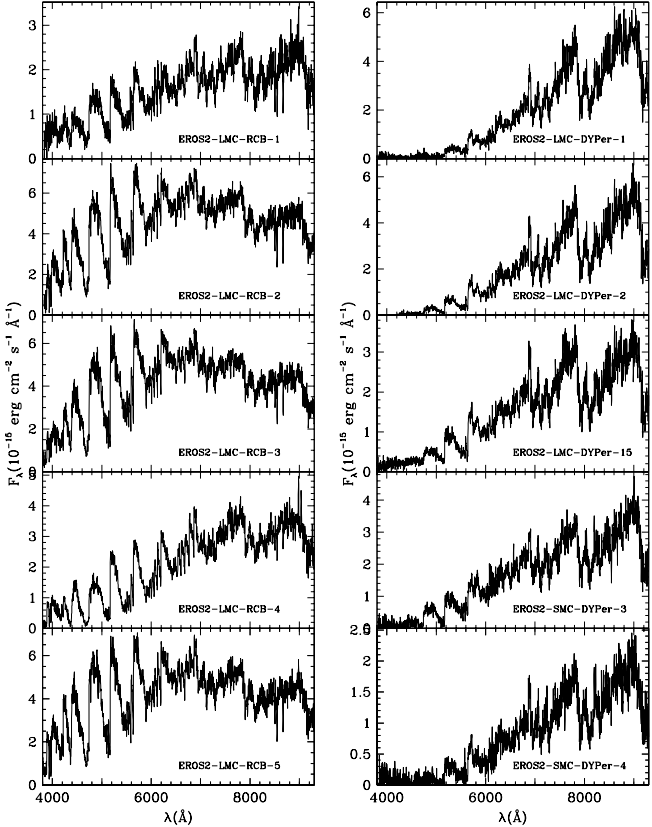}
\caption{Optical spectra at 2 \AA{} resolution of 5 RCB (left) and 5 DYPer (right) stars.}
\label{spectra_fig}
\end{figure}

We obtained spectra for 32 of the 36 candidates, 4 being too faint, namely EROS2-LMC-RCB-6, -7, -8, and EROS2-SMC-RCB-4. We consider nevertheless that EROS2-LMC-RCB-6 should be classified as a confirmed RCB star due to its multiple fast-fading phases that occurred during the EROS-2 observations (see Figure~\ref{lc}; EROS2-LMC-RCB-6 was brighter than $R_E<18$ only 30$\%$ of the time). The last three will be listed only as RCB star candidates. Of the 32 spectra obtained, 26 present a spectrum with carbon features due to $C_2$ and/or CN molecules (see Fig.~\ref{spectra_fig}). We note that only 10 of these 26 stars were already listed in a carbon star catalogue. The 6 candidates rejected are listed in Table~\ref{tab.RejectedCand}: two are hot stars, included a Wolf-Rayet type, and three others present strong TiO features that indicate a possible link with RV Tauri type stars. The remaining one, EROS2-lm0231l24809, is interestingly classified as a planetary nebulae (PN) in the SIMBAD database and we confirmed its classical oxygen rich PN type spectrum. Only a small drop of $\sim$1 mag is visible in its EROS-2 light curve.

Among all candidates tested, we confirm 6 new RCBs (EROS2-LMC-RCB-6 included) and 7 new DYPers and consider the 14 remaining carbon stars only as RCB or DYPer candidates, their fading amplitudes being too small (less than 2 mag). All new confirmed and candidate RCB or DYPer stars are listed in Table~\ref{tab.RCB.GenInfo}. A sample of the spectra is presented in Figure~\ref{spectra_fig}.

We searched for the presence of the isotope $^{13}$C in the atmosphere of the RCB and DYPer stars observed. We used the spectral atlas of carbon stars compiled by \citet{1996ApJS..105..419B} to identify isotopic $C_2$ and CN bands, respectively at 6100 and 6260 $\AA{}$ and compared each spectrum to the one of MACHO-78.6460.7, a DYPer star known to have no $^{13}C$ in its atmosphere \citep{2001ApJ...554..298A}. The results are summarized in Table~\ref{tab.RCB.MagInfo}. We found no trace of $^{13}$C in the atmospheres of the 5 RCB stars observed and a large range of abundances in those of DYPer stars, that we classified in three groups : none, weak and strong presence. This result confirms the one of \citet{2001ApJ...554..298A}, which shows that DYPers, unlike RCBs, have a significant amount of $^{13}$C in their atmosphere. Furthermore, we also confirm the presence of $^{13}$C in the atmospheres of 2 known LMC DYPers : MACHO-2.5871.1759 and MACHO-15.10675.10. An empirical analysis of the strength of the Ca II triplet absorption lines confirms that all RCBs have stronger lines than confirmed and candidate DYPers. The intensity of these lines is, as shown by \citet{1971ApJ...167..521R}, a good indicator of carbon star temperature: the cooler the temperature, the weaker the lines. 

We did not find any strong indication of a presence of hydrogen in any of our spectra, except in the case of four DYPers, where H$\alpha$ was found in weak emission :  EROS2-SMC-DYPer-3, EROS2-LMC-DYPer-14, -6 and -15 in decreasing order of H$\alpha$ strength.

\begin{table*}
\caption{General information on the new confirmed and candidate EROS-2 Magellanic RCB and DYPer stars.
\label{tab.RCB.GenInfo}}
\medskip
\centering
\begin{tabular}{llll}
\hline
\hline
EROS-2 Magellanic & EROS-2 star & Coordinates ($J_{2000}$) & Other identifier  \\
RCB name & identifier &  & \\
\hline
 &  &  & \\
 &  &  New confirmed RCB stars &  \\
\hline
EROS2-LMC-RCB-1$^{S2}$  &  lm0094n12459  &  05:14:40.17 -69:58:40.06  & [KDM2001]-2831 ; MACHO-5.5489.623\\
EROS2-LMC-RCB-2$^{S2}$  &  lm0105m6306  &  05:10:28.50 -69:47:04.54  & [KDM2001]-2373 ; SP77-39-12$^a$ ; MACHO-5.4887.14 \\
EROS2-LMC-RCB-3$^{S1}$  &  lm0174k9881  &  04:59:35.78 -68:24:44.68  & [KDM2001]-1296 ; SP77-30-16$^a$ \\
EROS2-LMC-RCB-4$^{S6+S7}$  &  lm0720k11917  &  05:39:36.97 -71:55:46.42  & MACHO-27.9574.93 \\
EROS2-LMC-RCB-5$^{S1}$  &  lm0745l21924  &  06:04:05.42 -72:51:22.73  & [KDM2001]-7101 ; SP77-65-2$^a$ \\
EROS2-LMC-RCB-6  &  lm0864m16909  &  06:12:10.48 -74:05:10.16  & \\
\hline
 &  &  & \\
 &  & New confirmed DYPer stars &  \\
\hline
EROS2-LMC-DYPer-1$^{S2}$ & lm0112l16201 & 04:56:28.47 -69:39:12.57 & [KDM2001]-1033 ; MACHO-17.2590.220\\
EROS2-LMC-DYPer-2$^{S1}$ & lm0405m10336 & 06:18:46.59 -67:00:59.59 & [KDM2001]-7544 \\
EROS2-LMC-DYPer-3$^{S1}$ & lm0485n10275 & 05:44:04.05 -65:46:00.84 & [KDM2001]-5900 \\
EROS2-LMC-DYPer-4$^{S1}$ & lm0013n33996 & 05:25:54.95 -69:43:54.03 & MACHO-77.7429.64 ; OGLE2-LMC-SC4-450098\\
EROS2-LMC-DYPer-5$^{S1}$ & lm0730k16380  &  05:50:15.47 -71:57:39.66  & [KDM2001]-6448 ; MACHO-51.11267.9 \\
EROS2-LMC-DYPer-6$^{S4}$ & lm0020n28519 & 05:29:27.19 -69:22:32.61 & MACHO-82.7918.2694 \\
EROS2-SMC-DYPer-3$^{S2}$ & sm0067m2668  &  00:55:54.97 -72:35:12.27  & RAW-961$^e$ ; OGLE2-SMC-SC7-368043 \\
\hline
\hline
 &  &  & \\
 &  &  New RCB candidates &  \\
\hline
EROS2-LMC-RCB-7 & lm0577n19651 & 05:24:43.68 -71:47:26.40 & MACHO-21.7156.850; MSX-428$^d$ \\
EROS2-LMC-RCB-8 & lm0675k8409 & 04:58:48.37 -72:37:43.37 & MSX-184$^d$ \\
EROS2-SMC-RCB-4 & sm0053n9475 & 01:04:52.89 -72:04:02.64  & OGLE2-SMC SC10-107856 \\
\hline
 &  &  & \\
 &  &  New DYPer candidates &  \\
\hline
EROS2-LMC-DYPer-7$^{S1}$ & lm0456m10958 & 05:19:48.13 -65:58:22.55 & [KDM2001]-3340 ; MACHO-63.6396.1154\\
EROS2-LMC-DYPer-8$^{S1}$ & lm0593k30837 & 05:39:00.21 -70:58:01.25  & [KDM2001]-5419 ; MACHO-14.9467.9 \\
EROS2-LMC-DYPer-9$^{S1}$ & lm0121k19954  &  04:52:58.35 -69:10:27.42  & [KDM2001]-804 ; LMC-BM 5-9$^b$ \\
EROS2-LMC-DYPer-10$^{S1}$ & lm0257n7946  &  06:05:35.32 -68:54:54.05  & [KDM2001]-7151 \\
EROS2-LMC-DYPer-11$^{S7}$ & lm0026n10671 &  05:29:22.61  -70:19:35.65  & MACHO-7.7904.11 ;  [KDM2001]-4359\\
EROS2-LMC-DYPer-12$^{S7}$ & lm0335m9403 & 05:29:26.21  -67:00:22.64  & MACHO-60.7954.68 ; [KDM2001]-4370\\
EROS2-LMC-DYPer-13$^{S4}$ & lm0010k3834 & 05:20:22.92 -69:04:43.11 & MACHO-80.6471.2523 \\
EROS2-LMC-DYPer-14$^{S4}$ & lm0020k5163 & 05:27:32.33 -69:05:35.48 & \\
EROS2-LMC-DYPer-15$^{S5}$ & lm0022n15676 & 05:30:43.25 -69:38:51.84 & MACHO-77.8156.16 ; BSDL2075$^c$\\
EROS2-LMC-DYPer-16$^{S4}$ & lm0034l8100 & 05:35:15.49 -69:57:43.84 & MACHO-81.8877.24 \\
EROS2-LMC-DYPer-17$^{S5}$ & lm0341m22605 & 05:35:24.08 -66:24:15.17 & [KDM2001]-5039\\
EROS2-SMC-DYPer-4$^{S2}$ & sm0061m7643 & 00:56:35.47 -71:32:32.66 & [MH95]-672 \\
EROS2-SMC-DYPer-5$^{S8}$ & sm0010l4971 & 00:47:41.71 -73:06:16.38 & MACHO-212.15793.25 ; RAW-421 \\
EROS2-SMC-DYPer-6$^{S8}$ & sm0101n8018 & 00:44:56.40 -73:12:25.02 & MACHO-212.15621.153 \\
\hline
\multicolumn{4}{l}{$^a$ \citet{1979ApJ...230..724R}, $^b$\citet{1990AJ....100..674B}, $^c$\citet{1999AJ....117..238B}, $^d$\citet{2001AJ....122.1844E}, $^e$\citet{1993A&AS...97..603R}}\\
\multicolumn{4}{l}{ Spectra epochs (JD-2450000): $^{S1}$= 4510.0, $^{S2}$= 4511.0 , $^{S3}$= 4645.0, $^{S4}$= 4646.0, $^{S5}$= 4647.0, $^{S6}$= 4747.0, $^{S7}$= 4820.0, $^{S8}$= 4821.0 }\\
\end{tabular}
\end{table*}

\begin{table*} % table: rejected candidate
\caption{General information on the rejected candidates, for which fading events were observed.
\label{tab.RejectedCand}}
\medskip
\centering
\begin{tabular}{lllll}
\hline
\hline
EROS-2 star & Coordinates ($J_{2000}$) & Classification & Other identifier & Information \\
identifier & from 2MASS & & & \\
\hline
lm0560k-21075$^{S1}$ & 05:07:56.45 -70:34:53.7 & Wolf Rayet & MACHO-9.4391.25 & $\Delta$mag $\sim$2.0\\
lm0127n-6759$^{S3+S7}$  & 04:55:15.75 -70:17:42.14 & A0 supergiant & MACHO-41.2336.11 & $\Delta$mag $\sim$0.7 \\
lm0174m-8630$^{S4+S7}$ &	05:02:17.16 -68:24:05.75 & RVa Tau (R Sct type)? & MACHO-19.3577.6 & $\Delta$mag $\sim$1.0, Strong TiO bands observed \\
lm0337l-17621$^{S4+S7}$ & 05:27:12.62 -67:35:07.42 & RVa Tau (R Sct type)? &  & $\Delta$mag $\sim$1.5, Strong TiO bands observed\\
lm0231l-24809$^{S7}$ & 05:48:22.32 -67:58:53.23 & Planetary nebula & MACHO-68.11085.71 & $\Delta$mag $\sim$1.0\\
sm0101n-16084$^{S8}$ & 00:44:54.02 -73:15:30.02 & RVa Tau (R Sct type)? & MACHO-212.15620.713 & $\Delta$mag $\sim$1.5, Strong TiO bands observed\\
\hline
\multicolumn{4}{l}{ Spectra epochs (JD-2450000): $^{S1}$= 4510.0, $^{S3}$= 4645.0, $^{S4}$= 4646.0, $^{S7}$= 4820.0, $^{S8}$= 4821.0 }\\
\end{tabular}
\end{table*}

\begin{figure*} %figure RA DEC
\includegraphics[scale=0.8]{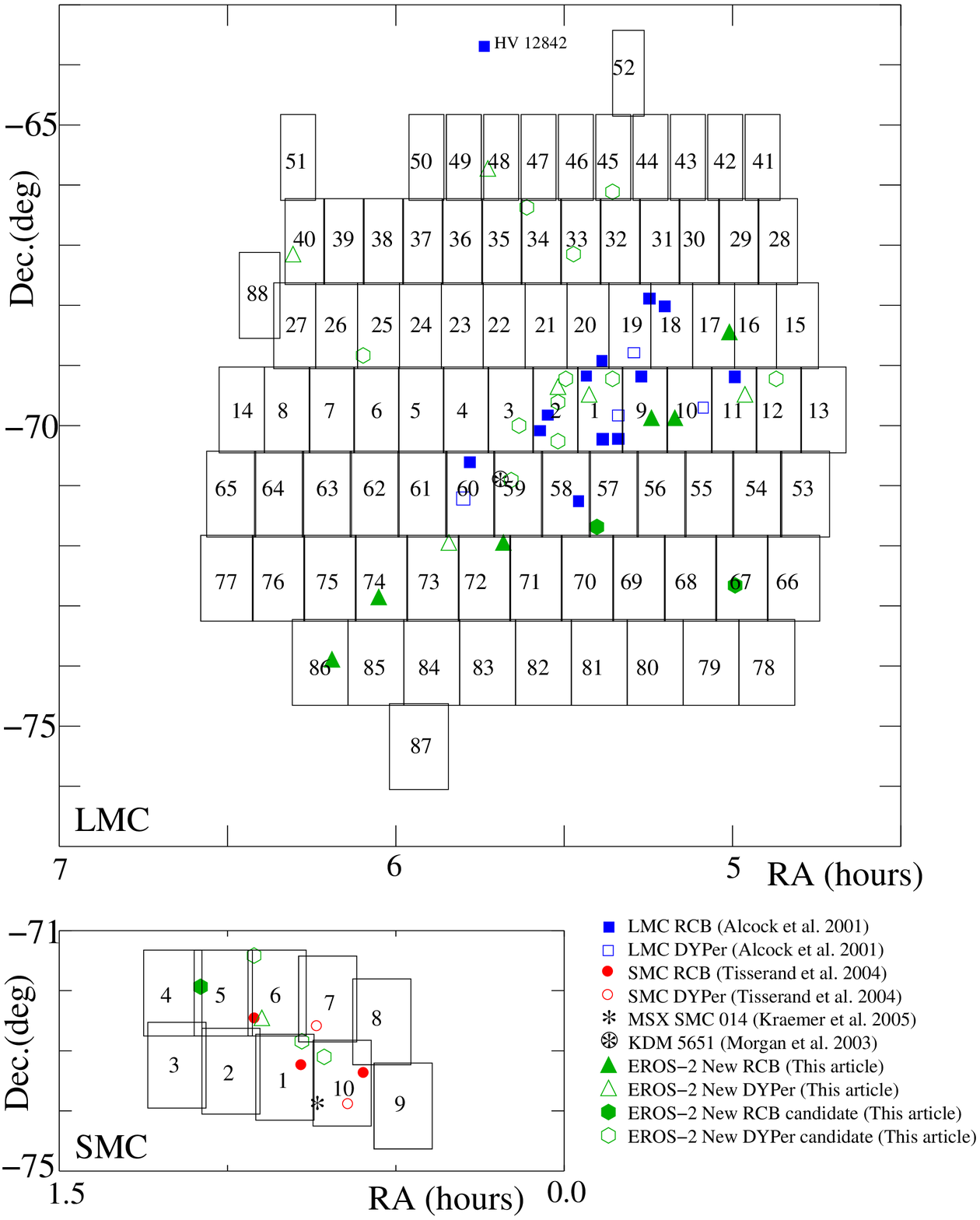}
\caption{Spatial distribution of all known and new Magellanic RCBs and DYPers (confirmed and candidates). The 98 Magellanic EROS-2 $\sim$1 deg$^2$ fields are represented with their respective numbers.}
\label{map_fig}
\end{figure*}

\section{The new EROS-2 Magellanic RCB and DYPer stars \label{sec_newRCB}}

We investigate in this section the optical, near-infrared and mid-infrared characteristics of the 13 newly confirmed and 17 candidate RCB and DYPer stars listed in Table~\ref{tab.RCB.GenInfo}. Their spatial distribution is shown in Figure~\ref{map_fig} together with previously known ones. Note that LMC RCB stars are distributed preferentially along the LMC bar. Their light curves are presented in Figures \ref{lc} to \ref{lc_end} and their charts in Figures \ref{chartA} and \ref{chartB}. We note that a MACHO light curve is available for 20 of them, as mentioned in Table~\ref{tab.RCB.GenInfo}. EROS-2 measurements of all objects listed in this article are available at the URL indicated in Section~\ref{sec_knownRCB}.

\subsection{Colour Magnitude Diagram \label{sbsec_cmd}}

The maximum optical magnitudes, average fading rates and maximum drop amplitudes recorded during the EROS-2 observations of the newly discovered RCBs and DYPers  are presented in Table~\ref{tab.RCB.MagInfo}. We used the photometric calibration defined in Equation~\ref{eq.photcalib}. We remark that the main differences in the classifications of RCB and DYPer stars are not only the shape of the drops observed, but also the rate of fading. There is a clear separation in the fading rates presented in Table~\ref{tab.RCB.GenInfo} between RCB and DYPer stars, DYPer rates being slower by on average a factor of four ($\sim$0.01 compared to $\sim$0.04 mag.day$^{-1}$).

\begin{figure} %figure CMD V vs V-I
\centering
%\sidecaption
\includegraphics[width=8.25cm]{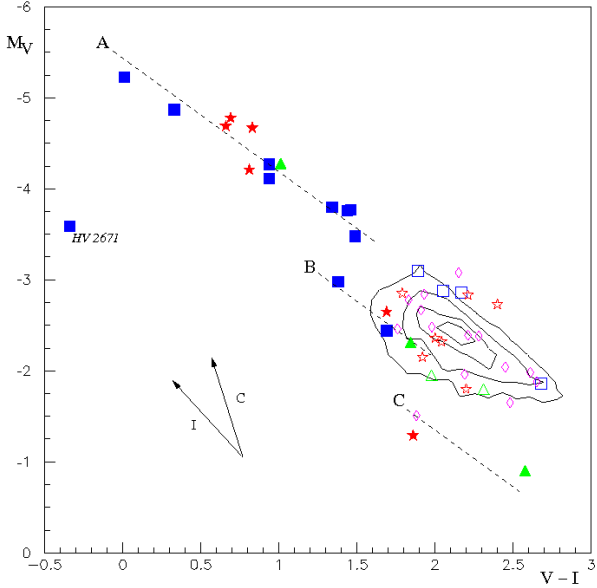}
\caption{Absolute magnitude at maximum, $M_V$, vs. $V-I$; full symbols represent RCB stars, open symbols the DYPers. The 10 LMC RCBs, plus 4 DY Per-like RCBs, from \citet{2001ApJ...554..298A} are indicated with blue squares. 2 SMC RCBs, plus 2 SMC DYPers, from \citet{2004A&A...424..245T} are shown with green triangles. The new confirmed RCBs and DYPers are indicated with red stars, and the candidate DYPers with purple diamonds. All data are from the EROS-2 photometry. All magnitudes have been corrected for their respective interstellar extinctions. The vectors represent the reddening correction directions due to the interstellar medium (I) and a carbon shell (C). Dashed lines and contours are described in the text.}
\label{cmd}
\end{figure}

We present in Figure~\ref{cmd} the colour magnitude diagram $M_V$ vs $V-I$ of the known and newly discovered (confirmed and candidate) Magellanic RCBs or DYPers. The magnitudes correspond to the average maximum magnitude found in each star's light curve.  We used a distance modulus of 18.5 mag for the LMC and 18.9 mag for the SMC. We also corrected the LMC and SMC magnitudes for the total reddening (Galactic foreground + intrinsic dust), corresponding to an average of $E(B-V)_{LMC}\sim$0.17 and $A_{V,LMC}\sim$0.5 mag for the LMC \citep{2001ApJ...554..298A}, and $E(B-V)_{SMC}\sim$0.06 and $A_{V,SMC}\sim$0.17 mag for the SMC \citep{2002AJ....123..855Z}.

RCB stars (plain symbols) in Figure~\ref{cmd} can be classified in three different groups. With the hot RCB star HV 2671 being on its own \citep{2002AJ....123.3387D}, we can separate the RCB stars hotter and cooler than T$\sim$4200 K ($V-I\sim$1.6). Below this temperature, we are not confident that the real maximum magnitude was observed during the EROS-2 observations. The maximum magnitude lasts only $\sim$50 days and a high fading activity was observed. 

We note that the three dashed lines indicated in Fig.~\ref{cmd} (A, B and C) are separated by only $\sim$1 magnitude from each other considering an extinction due to carbon dust. With a 1 mag correction, the four RCBs close to line B would have temperatures and magnitudes corresponding to the 13 brightest common RCBs close to line A, for which we are confident of their maximum magnitude. The two remaining faintest RCBs, close to line C, are EROS2-SMC-RCB-3 and EROS2-LMC-RCB-6. The latter would also have a more common temperature and magnitude if we applied a 2 magnitude correction, but with such a correction EROS2-SMC-RCB-3 would still be considered as the coolest RCB (with $V-I\sim$2.3 and $M_V\sim-2.6$).

Overall, we can affirm that we found a confident RCB absolute magnitude $M_V$ range between $\sim-5.2$ and $\sim-3.4$. The fainter limit can be extended: if we suppose a reasonable carbon extinction correction of 2 magnitudes to EROS2-SMC-RCB-3 (that would make it lies on line A), we find a conservative lower limit as faint as $M_{V_{inf}}\sim-2.6$. The DYPer (open symbols) absolute magnitude range is fainter: between $\sim-3$ and $\sim-1.8$.

The distribution of classical LMC carbon stars from \citet{2001A&A...369..932K} found in the EROS-2 database is also represented on Figure~\ref{cmd} (solid line contours). Interestingly, we can clearly observe that the distributions of all confirmed and candidate DYPer stars match very well the classical carbon star distribution. We note that DYPer stars' spectra are similar to more classical N carbon stars \citep{2003MNRAS.344..325M}.

\subsection{Near-infrared properties \label{sbsec_ir}}

\begin{figure*} %figure ALL IR diags
\centering
%\sidecaption
\includegraphics[scale=0.5]{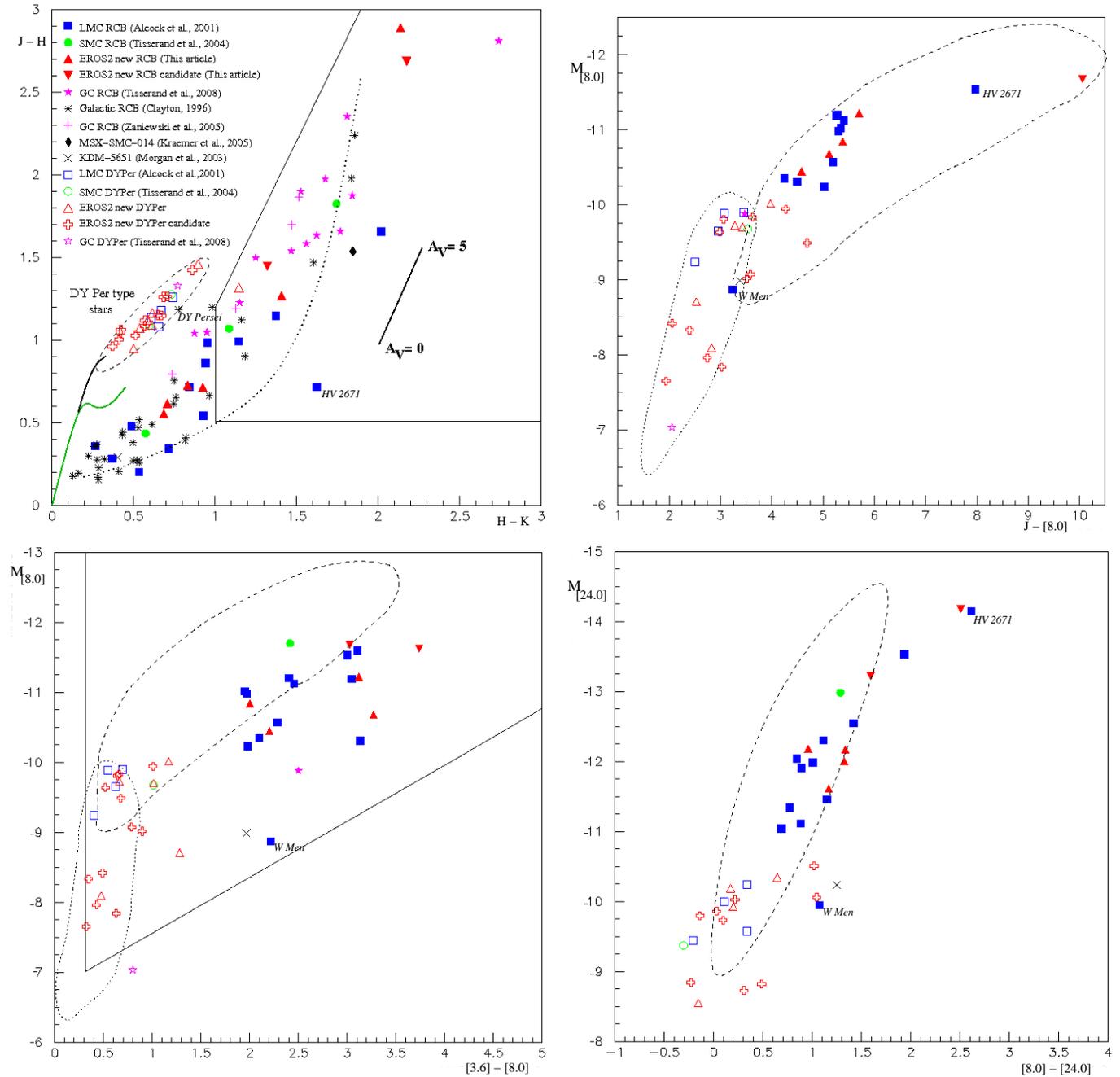}
\caption{Top-left: $J - H$ versus $H - K$ colour diagram. The line on the right side represents the reddening vector from \citet{1985ApJ...288..618R} and the dotted curve corresponds to the combination of blackbodies consisting of 
a 5500 K star and a 1000 K dust shell in various proportions ranging from all 'star' to all 'shell' \citep[from][]{1997MNRAS.285..339F}. Also shown are the expected positions (lines in the bottom-left) for common 
dwarf (green) and giant (black) stars from \citet{1988PASP..100.1134B}. All known and newly discovered RCBs and 
DYPers are represented. Their symbols are identical in all four diagrams: RCBs are represented 
with filled symbols and DYPers with open ones. DYPer stars are delimited by a dashed ellipse in the colour-colour
near-infrared diagram, which also encloses the area occupied by most ordinary carbon-rich stars. The black stars 
are the 31 confirmed Galactic RCBs listed by \citet{1996PASP..108..225C} plus ES Aql and V2552 Oph \citep{2003PASP..115.1301H}.
The solid lines in both left side diagrams delimit the selection zones in the 2MASS and Spitzer- SAGE or 
S$^3$MC databases where EROS-2 objects light curves were visually inspected (see Section~\ref{sec_mining}). The contours in the three mid-infrared colour magnitude diagrams represent the distribution envelopes of "Extreme-AGB" stars (dashed lines) and carbon rich "C-stars" (dotted lines) estimated from \citet[fig.4, 5 \& 6]{2006AJ....132.2034B}. The 2MASS J magnitude in the top-right diagram was corrected for extinction during fading events and interstellar extinction (as explained in Section~\ref{sbsec_sed}). The Spitzer magnitudes $M_{[8.0]}$ and $M_{[24.0]}$ are absolute magnitudes; we used distance moduli of 18.5, 18.9 and 14.4 for the LMC, SMC and Galactic center (GC) respectively.}
\label{infra_red}
\end{figure*}

We list in Table~\ref{tab.RCB.IRInfo} the J, H and K magnitudes of the newly discovered RCBs and DYPers observed by the 2MASS project as well as by the DENIS collaboration, and indicate also the RCB star phases during each epoch. In a $J-H$ versus $H-K$ diagram (see Figure~\ref{infra_red}), we note that most RCB stars have a near-IR excess with values distributed in a range between 0 and 3 in both colours. This wide range simply represents the fact that the 2MASS measurements, taken during the EROS-2 observations (i.e. when most newly discovered RCBs showed fading activity), happen to occur during different phases of obscuration by carbon dust clouds. See \citet{1997MNRAS.285..339F} for a more precise discussion.

As indicated in T08, we note also a clear separation between the DYPers (confirmed and candidate) and RCB stars. The first occupy the area where classical carbon stars are found.

Note that the 2 new RCB star candidates EROS2-LMC-RCB-7 and -SMC-RCB-4 have a high infra-red excess ($J-K>2$) certainly due to circumstellar dust extinction.

\begin{table*} 
\caption{Near-IR photometry.
\label{tab.RCB.IRInfo}}
\medskip
\centering
\begin{tabular}{llrrrlrrr}
\hline
\hline
Name & JD Epoch 2MASS & $J_{\mathrm{2MASS}}$ & $H_{\mathrm{2MASS}}$ & $K_{\mathrm{2MASS}}$ & JD Epoch DENIS & $I_{\mathrm{DENIS}}$ & $J_{\mathrm{DENIS}}$ & $K_{\mathrm{DENIS}}$  \\
\hline
& & & RCB stars  &    &    &  &  &  \\
\hline
EROS2-LMC-RCB-1 & 2451112.7481$^\Diamond$ & 12.796 & 12.181 & 11.475 &      2450366.85661$^\star$ & 16.980 & 15.211 & 12.445 \\
EROS2-LMC-RCB-2 & 2451112.7289$^\vee$ & 14.558 & 13.831 & 12.999 &      2450438.67682$^\Diamond$ & 13.219 & 12.619 & 11.741 \\
EROS2-LMC-RCB-3 & 2451111.7875$^\star$ & 15.205 & 14.491 & 13.567 &      2450492.60428$^\Diamond$ & 13.169 & 12.657 & 12.038 \\
EROS2-LMC-RCB-4 & 2451151.6263$^\wedge$ & 14.583 & 13.314 & 11.906 &      2451497.78554$^\star$ &  &  & 13.593 \\
EROS2-LMC-RCB-5 & 2451151.7192$^\wedge$ & 13.438 & 12.883 & 12.198 &      2450510.59451$^\Diamond$ & 13.253 & 12.708 & 11.968 \\
EROS2-LMC-RCB-6 & 2451151.7527$^\star$ & 18.313 & 15.422 & 13.286 &      2451133.78876$^\star$ &  &  & 12.984 \\
&  &  &  & &     2451135.818206$^\star$  &  &  &  13.396  \\
&  &  &  & &     2451134.826748$^\star$  &  &  &  13.292  \\
EROS2-LMC-RCB-7  & 2451583.6023$^\star$ & 16.881 & 14.195 & 12.021 &  2450103.598553$^\star$  &    &    &  12.393  \\
&  &  &  & &     2450417.691493$^\star$  &    &    &  12.008  \\
EROS2-LMC-RCB-8 & 2451149.7282$^\wedge$ & 13.826 & 12.250 & 10.957 & 2450492.610428$^\wedge$ & 17.167 & 14.200 & 10.764\\
&  &  &  & &     2450413.754525$^\wedge$ & 16.828 & 14.270 & 10.927 \\
EROS2-SMC-RCB-4 & 2451034.7855$^\wedge$ & 14.345 & 12.899 & 11.580 &      2451127.59750$^\wedge$ & 17.278 & 14.813 & 11.974 \\
\hline
& & & DYPer stars  &    &    &  &  &  \\
\hline
EROS2-LMC-DYPer-1 & 2451111.7728$^\star$ & 13.851 & 12.535 & 11.389 &      2450413.67902$^\Diamond$ & 13.365 & 11.503 & 9.793 \\
EROS2-LMC-DYPer-2 & 2451155.7431$^\Diamond$ & 11.985 & 10.864 & 10.278 &     2451148.82308$^\Diamond$ & 13.705 & 11.933 & 10.182 \\
EROS2-LMC-DYPer-3 & 2451157.7545$^\wedge$ & 13.461 & 12.001 & 11.104 &     2450119.56135$^?$ & 14.374 & 12.537 & 10.579 \\
EROS2-LMC-DYPer-4 & 2451583.6045$^\vee$ & 12.745 & 11.583 & 10.967 &      2450417.74903$^\Diamond$ & 14.122 & 12.292 & 10.552 \\
EROS2-LMC-DYPer-5 & 2451151.6726$^\Diamond$ & 12.274 & 11.325 & 10.826 &     2451127.82639$^\Diamond$ & 13.800 & 12.172 & 10.748 \\
&  &  &  & &     2450416.79795$^\Diamond$ & 14.015 & 12.414 & 10.703 \\
EROS2-LMC-DYPer-6 & 2451625.5271$^\wedge$ & 13.232 & 12.141 & 11.527 &  2450439.734606$^\Diamond$  &  14.922  &  13.255  &  11.525  \\
EROS2-LMC-DYPer-7 & 2451167.7629$^\Diamond$ & 12.299 & 11.236 & 10.814 &     2450386.81200$^\Diamond$ & 13.879 & 12.313 & 10.754 \\
EROS2-LMC-DYPer-8 & 2450891.5653$^\Diamond$ & 12.128 & 10.865 & 10.158 &     2450516.55506$^\wedge$ & 14.354 & 12.240 & 10.229 \\
&  &  &  & &     2451497.78409$^\Diamond$ & 13.825 & 11.895 & 10.068 \\
EROS2-LMC-DYPer-9 & 2451111.7589$^\Diamond$ & 12.949 & 11.987 & 11.618 &      2451140.73852$^\Diamond$ & 14.506 & 12.991 & 11.553 \\
&  &  &  & &     2450509.53213$^\Diamond$ & 14.727 & 13.043 & 11.719 \\
EROS2-LMC-DYPer-10 & 2451112.8178$^\Diamond$ & 12.582 & 11.580 & 11.169 &      &  &  &  \\
EROS2-LMC-DYPer-11 & 2451625.5278$^\Diamond$	& 11.758 &  10.560 & 9.913  &  2450439.736076$^\star$  &  14.366  &  12.286  &  10.186  \\
EROS2-LMC-DYPer-12 & 2451625.5253$^\Diamond$  & 12.290 & 11.046 & 10.318 &  2450439.731169$^\wedge$  &  14.865  &  12.626  &  10.373  \\
&  &  &  & &     2450417.691493$^\wedge$  &    &    &  12.008  \\
% Spitzer cand:
EROS2-LMC-DYPer-13 & 2451167.7606$^\Diamond$ & 13.689 & 12.571 & 12.001 &  2450386.816690$^\Diamond$  &  15.212  &  13.653  &  12.112  \\
&  &  &  & &     2450412.719178$^\Diamond$  &  15.348  &  13.607  &  12.120  \\
EROS2-LMC-DYPer-14 & 2451583.6101$^\star$ & 13.704 & 12.281 & 11.421 &  2450417.748032$^\Diamond$  &  14.218  &  12.365  &  10.379  \\
&  &  &  & &     2450509.585324$^\Diamond$  &  13.864  &  12.068  &  10.250  \\
EROS2-LMC-DYPer-15 & 2451602.5509$^\wedge$ & 12.565 & 11.537 & 11.026 &  2450440.624398$^\Diamond$  &  14.085  &  12.481  &  10.991  \\
EROS2-LMC-DYPer-16 & 2451580.5851$^\wedge$ & 13.280 & 12.235 & 11.815 &  2450419.724514$^\Diamond$  &  14.381  &  12.563  &  11.212  \\
&  &  &  & &     2450503.596481$^\Diamond$  &  14.288  &  12.677  &  11.298  \\
EROS2-LMC-DYPer-17 & 2451580.5877$^\wedge$ & 13.000 & 11.853 & 11.183 &  2450503.591030$^\star$  &  14.703  &  12.949  &  11.213  \\
&  &  &  & &     2450419.719317$^\Diamond$  &  14.259  &  12.405  &  10.758  \\
EROS2-SMC-DYPer-3 & 2451034.7496$^\Diamond$ & 13.139 & 12.066 & 11.528 &      2450432.56433$^\star$ & 17.011 & 15.242 & 12.791 \\
&  &  &  & &     2450431.571458$^\star$  &  17.247  &  15.025  &  13.195  \\
EROS2-SMC-DYPer-4 & 2451107.5892$^\Diamond$ & 13.253 & 12.166 & 11.598 &      2450433.56201$^\Diamond$ & 14.830 & 12.977 & 11.314 \\
EROS2-SMC-DYPer-5 & 2451034.7209$^\star$ & 13.412 & 12.261 & 11.613 &  2451039.799132$^\star$  &  15.011  &  13.524  &  11.579  \\
EROS2-SMC-DYPer-6 & 2451034.7111$^\wedge$ & 13.236 & 11.974 & 11.290 &  2451048.775845$^\wedge$  &  15.453  &  13.222  &  11.216  \\
&  &  &  & &     2450418.552407$^\wedge$  &  15.872  &  13.305  &  11.285  \\
%sm0101n-16084 & 2451034.7111$^\wedge$ & 12.133 & 11.102 & 10.665 &  2451048.776100$^\wedge$  &  13.849  &  11.963  &  10.577  \\
%&  &  &  & &     2450418.552650$^\Diamond$  &  13.330  &  11.779  &  10.435  \\
%lm0231l-24809  & 2450893.5624$^\vee$	& 14.572 &  13.085 & 11.730 &  2450416.792025$^\wedge$  &  16.799  &  14.904  &  12.098 \\
\hline
\multicolumn{9}{l}{$\star$: during a faint phase, $\Diamond$: during a bright phase, $\vee$ and $\wedge$: during a dimming or recovering phase, ?: phase unknown. }\\
\end{tabular}
\end{table*}

\begin{table*}
\caption{Properties of the new Magellanic RCB and DYPer stars, including derived absolute magnitudes and intrinsic colours.
\label{tab.RCB.MagInfo}}
\medskip
\centering
\begin{tabular}{lccccccccc}
\hline
\hline
Name & $^{13}C$ & $R_{E,max}$ & $(dR_E/dt)_{max}$ & Drop $\Delta R_E$ & $B_{E,max}$ & $(dB_E/dt)_{max}$ & Drop $\Delta B_E$ &  $M_V$ & ($V-I$)$_0$\\
& & & $mag.day^{-1}$ & & & $mag.day^{-1}$ & & &\\
\hline
& & & & RCB stars &    &    &  &  &  \\
\hline
EROS2-LMC-RCB-1 & None & 13.71 & 0.039 & 5.37 & 14.36 & 0.033 & 4.57 & -4.21 & 0.81 \\
EROS2-LMC-RCB-2 & None & 13.38 & 0.004 + 0.071 & 3.88 & 13.93 & 0.005 + 0.060 & 3.28 & -4.69 & 0.66 \\
EROS2-LMC-RCB-3 & None & 13.27 & 0.027 & 4.65 & 13.84 & 0.028 & 4.78 & -4.78 & 0.69 \\
EROS2-LMC-RCB-4 & None & 14.39$^*$ & 0.028 & $>7.40$ & 15.57$^*$ & 0.026 & $>7.00$ & -2.65$^*$ & 1.69$^*$ \\
EROS2-LMC-RCB-5 & None & 13.23 & 0.047 & 4.06 & 13.89 & 0.042 & 5.50 & -4.67 & 0.83 \\
EROS2-LMC-RCB-6 & ... & 15.58$^*$ & 0.038 & $>5.60$ & 16.86$^*$ & 0.040 & $>5.20$ & -1.29$^*$ & 1.86$^*$ \\
EROS2-LMC-RCB-7 & ... & nd & nd & nd & nd & nd & nd & nd & nd \\
EROS2-LMC-RCB-8 & ... & 15.90$^*$  & 0.026 & 4.10 & 17.95$^*$ & 0.031 & $>4.10$ & 0.82$^*$ & 3.42$^*$ \\
EROS2-SMC-RCB-4 & ... & 18.05$^*$ & 0.037 & 2.75 & 19.61$^*$ & 0.035 & 2.00 & 1.58$^*$ & 2.51$^*$ \\
\hline
& & & DYPer stars  &    &    &  &  &  \\
\hline
EROS2-LMC-DYPer-1 & Strong & 13.60 & 0.009 & 2.98 & 15.20 & 0.009 & 3.10 & -2.73 & 2.40 \\
EROS2-LMC-DYPer-2 & Strong & 13.69 & nd & $>0.50$ & 15.18 & 0.016 & 2.00 & -2.83 & 2.21 \\
EROS2-LMC-DYPer-3 & Weak & 14.37 & 0.014 & 3.61 & 15.73 & 0.014 & 3.51 & -2.36 & 2.00 \\
EROS2-LMC-DYPer-4 & Weak & 14.09 & 0.009 & 1.97 & 15.32 & 0.009 & 1.94 & -2.85 & 1.79 \\
EROS2-LMC-DYPer-5 & Weak & 14.16 & 0.016 & 1.94 & 15.46 & 0.018 & 2.22 & -2.67 & 1.91 \\
EROS2-LMC-DYPer-6 & Strong & 14.73 & 0.013 & 1.67 & 16.21 & 0.018 & 2.19 & -1.30 & 2.47 \\
EROS2-LMC-DYPer-7 & None & 13.96 & 0.012 & 1.37 & 15.28 & 0.012 & 1.59 & -2.84 & 1.93 \\
EROS2-LMC-DYPer-8 & ... & 14.07 & nd & $>1.18$ & 15.60 & nd & $>1.48$ & -2.38 & 2.28 \\
EROS2-LMC-DYPer-9 & None & 14.66 & 0.003 & 0.48 & 15.98 & 0.003 & 0.55 & -2.15 & 1.92 \\
EROS2-LMC-DYPer-10 & Strong & 14.26 & 0.005 & 0.48 & 15.62 & 0.007 & 0.59 & -2.48 & 1.98 \\
EROS2-LMC-DYPer-11 & Weak &	13.50 &	nd & $>1.20$ & 14.95 & nd & $>1.45$ & -2.58 & 2.42 \\
EROS2-LMC-DYPer-12 & Strong &	14.20 &	nd & $>0.84$ & 15.95 & nd & $>1.04$ & -1.38 & 2.92 \\
EROS2-LMC-DYPer-13 & ... & 15.34 & 0.005 & 0.48 & 16.63 & 0.006 & 0.52 & -1.01 & 2.15 \\
EROS2-LMC-DYPer-14 & Weak & 14.14 & 0.008 & 1.70 & 15.87 & 0.009 & 1.80 & -1.48 & 2.88 \\
EROS2-LMC-DYPer-15 & None & 14.12 & 0.008 & 0.83 & 15.38 & 0.009 & 1.00 & -2.28 & 2.10 \\
EROS2-LMC-DYPer-16 & Weak & 14.51 & 0.003 & 0.88 & 15.73 & 0.004 & 0.99 & -1.96 & 2.03 \\
EROS2-LMC-DYPer-17 & Strong & 14.24 & 0.008 & 1.08 & 15.87 & 0.010 & 1.36 & -1.54 & 2.72 \\
EROS2-SMC-DYPer-3 & Strong & 14.61 & 0.018 & 2.50 & 15.89 & 0.019 & 2.66 & -2.32 & 2.04 \\
EROS2-SMC-DYPer-4 & Weak & 14.84 & 0.003 & $>1.85$ & 16.38 & 0.004 & 2.88 & -1.65 & 2.48 \\
EROS2-SMC-DYPer-5 & ... & 14.38 & 0.006 & 0.92 & 15.76 & 0.007 & 1.30 & -2.22 & 2.30 \\
EROS2-SMC-DYPer-6 & ... & 14.83 & nd & $>1.22$ & 16.20 & nd & $>1.96$ & -1.79 & 2.28 \\
%sm0101n-16084 &  & 13.55 & 0.007 & 1.34 & 15.18 & 0.011 & 1.90 & -2.22 & 2.72 \\
%lm0231l-24809 &  & 16.55 & 0.009 & 0.91 & 17.10 & 0.009 & 1.23 & -1.03 & 0.92 \\
\hline
\multicolumn{10}{l}{nd = not detected; * = not confident that real maximum magnitude reached during EROS-2 observations;}\\
\multicolumn{10}{l}{ ... = either no spectrum obtained or spectrum with too low signal-to-noise.}\\
\end{tabular}
\end{table*}

\begin{table*}
\caption{Spitzer -SAGE, -S$^3$MC, GLIMPSE II magnitudes and MSX A-band (8.3$\mu$m) magnitudes for RCB and DYPer stars.
\label{tab.RCB.SAGE}}
\medskip
\centering
\begin{tabular}{lccccclc}
\hline
\hline
Name & Mag[3.6] & Mag[4.5]  & Mag[5.8] & Mag[8.0] & Mag[24.0] & SAGE epoch & MSX A-band\\
& & & & & & IRAC / MIPS &  (8.3$\mu$m) \\
\hline
& & & & & & &\\
& & & & RCB stars & & &\\
\hline
EROS2-LMC-RCB-1 & 10.256 & 9.548 & 8.857 & 8.054 & 6.888 & 1 / 1+2 & \\
EROS2-LMC-RCB-2 & 11.090 & 10.149 & 9.033 & 7.821 & 6.498 & 1 / 1+2 &  \\
EROS2-LMC-RCB-3 & 10.398 & 9.328 & 8.307 & 7.278 & 6.319 & 1 / 1+2 & \\
EROS2-LMC-RCB-4 & 9.662 & 8.958 & 8.372 & 7.661 & 6.327 & 1 / 1+2 & 7.70\\
EROS2-LMC-RCB-5 & nd & nd & nd & nd & nd &  & \\
EROS2-LMC-RCB-6 & nd & nd & nd & nd & nd &   & \\
EROS2-LMC-RCB-7 & 9.850 & 8.739 & 7.787 & 6.823 &  4.317 & 2 / 1 & 6.81\\
EROS2-LMC-RCB-8 & nd & nd & nd & nd & 6.836 & / 2 & 7.38\\
EROS2-SMC-RCB-1 &  nd & nd & nd & nd & nd &   & 8.44\\
EROS2-SMC-RCB-2 &  nd & nd & nd & nd & 6.840 &   &  \\
EROS2-SMC-RCB-3 & 9.611 & 8.612 & 7.770 & 7.199 & 5.914 &  &  7.13 \\
EROS2-SMC-RCB-4 & 11.010 & 9.494 & 8.356 & 7.273 & 5.679 &  &  \\
HV-5637 & 11.326 & 10.363 & 9.316 & 8.193 & 7.044 & 1 / 1+2  & \\
W-Men & 11.847 & 11.155 & 10.489 & 9.630 & 8.556 & 1 / 1+2  & \\
HV-12842 & nd & nd & nd & nd & nd  &  & \\
MACHO-11.8632.2507 & 9.967 & 8.880 & 7.971 & 6.966 & 4.352 & 1 / 1+2 & 7.15\\
MACHO-81.8394.1358 & 10.245 & 9.586 & 8.936 & 8.266 & 7.384 & 1 / 1+2  & \\
MACHO-6.6575.13 & 10.005 & 8.975 & 7.964 & 6.902 & 4.966 & 1 / 1+2 & 6.74\\
MACHO-6.6696.60 & 9.830 & 9.019 & 8.236 & 7.376 & 5.955 & 1 / 1+2 & 7.27\\
MACHO-12.10803.56 & 10.250 & 9.541 & 8.859 & 8.151 & 7.461 & 1 / 1+2  & \\
MACHO-16.5641.22 & 9.700 & 8.830 & 8.062 & 7.300 & 6.455 & 1 / 1+2  & \\
MACHO-18.3325.148 & 10.358 & 9.226 & 8.221 & 7.311 & 6.196 & 1 / 1+2  & \\
MACHO-79.5743.15 & 9.432 & 8.779 & 8.134 & 7.482 & 6.590 & 1 / 1+2  & \\
MACHO-80.6956.207 & 10.216 & 9.357 & 8.686 & 7.931 & 7.159 & 1 / 1+2  & \\
MACHO-80.7559.28 & 9.489 & 8.802 & 8.126 & 7.519 & 6.514 & 1 / 1+2  & \\
KDM-5651$^a$ & 11.876 & 11.410 & 10.704 & 9.911 & 8.662 & 1 / 1+2  &  \\
MSX-SMC-014 & nd & nd & nd & nd & nd &  & 6.83 \\
EROS2-CG-RCB-1 & 5.388 & 4.645 & 3.765 & nd & nd  &  & \\
EROS2-CG-RCB-8 & 7.019 & 6.310 & 5.281 & 4.519 & nd  &  & \\
\hline
& & & & & & &\\
& & & & DYPer stars & & &\\
\hline
EROS2-LMC-DYPer-1 & 9.436 & 9.478 & 9.359 & 8.773 & 8.574 & 1 / 1+2  & \\
EROS2-LMC-DYPer-2 & nd & nd & nd & nd & nd &  \\
EROS2-LMC-DYPer-3 & 9.654 & 9.322 & 8.955 & 8.484 & 8.312 & 1 / 2  & \\
EROS2-LMC-DYPer-4 & 9.809 & 9.622 & 9.304 & 8.794 & 8.152 & 1 / 1+2 &  \\
EROS2-LMC-DYPer-5 & 10.232 & 10.242 & 10.037 & nd & nd & 1 /  &  \\
EROS2-LMC-DYPer-6 & 10.882 & 10.998 & 10.910 & 10.406 & nd & 1 /  &  \\
EROS2-LMC-DYPer-7 & 10.569 & 10.552 & 10.334 & 10.082 & 9.775 & 1 / 1+2  & \\
EROS2-LMC-DYPer-8 & 9.381 & 9.506 & 9.395 & 8.863 & 8.768 & 1 / 1+2  & \\
EROS2-LMC-DYPer-9 & 11.172 & 11.293 & 11.103 & 10.849 & nd & 1 /  &  \\
EROS2-LMC-DYPer-10 & nd & nd & nd & nd & 9.854 &  / 1+2  & \\
EROS2-LMC-DYPer-11 & 9.334  & 9.281 & 9.165 &  8.692 & 8.475 & 1+2 / 1+2 &  \\
EROS2-LMC-DYPer-12 & 9.325  & 9.327 & 9.169 &  8.667 & 8.638 & 1+2 / 1+2  & \\
EROS2-LMC-DYPer-13 & 11.288 & 11.272 & 11.099 & 10.660 & nd & 2 / & \\
EROS2-LMC-DYPer-14 & 9.690 & 9.682 & 9.572 & 9.011 &  7.993 & 2 / 1+2 & \\
EROS2-LMC-DYPer-15 & 10.516 & 10.724 & 10.608 & 10.167 & 9.680 & 2 / 1  & \\
EROS2-LMC-DYPer-16 & 10.965 & 11.004 & 10.798 & 10.538 & nd & 2 /  & \\
EROS2-LMC-DYPer-17 & 10.380 & 10.448 & 10.193 & 9.485 & 8.440 & 2 / 1  & \\
EROS2-SMC-DYPer-1 & 10.236 & 10.022 & 9.600 & 9.225 & 9.531  &  & \\
EROS2-SMC-DYPer-2 &  nd & nd & nd & nd & nd  &  & \\
EROS2-SMC-DYPer-3 & 11.473 & 10.994 & 10.502 & 10.190 & 10.346  &  & \\
EROS2-SMC-DYPer-4 &  nd & nd & nd & nd & nd  &  & \\
EROS2-SMC-DYPer-5 & 10.62  & 10.60 & 10.20 & 9.83 & 10.06  &  & \\
EROS2-SMC-DYPer-6 &  9.97  &  9.66 &  9.19 & 8.96 & 9.10  &  & \\
MACHO-2.5871.1759 & 9.663 & 9.775 & 9.637 & 9.259 & 8.920 & 1 / 1+2  & \\
MACHO-10.3800.35 & 9.292 & 9.318 & 9.177 & 8.598 & 8.258 & 1 / 1+2  & \\
MACHO-15.10675.10 & 9.472 & 9.454 & 9.213 & 8.847 & 9.057 & 1 / 1+2 &  \\
MACHO-78.6460.7 & 9.155 & 9.279 & 9.194 & 8.612 & 8.500 & 1 / 1+2  & \\
EROS2-CG-RCB-2 & 8.167 & 7.893 & 7.667 & 7.367 & nd  &  & \\
%sm0101n-16084 & 10.02  & 10.06 &  9.59 & 9.48 & 9.14  &  & \\
%lm0231l-24809 & 10.166 & 9.407 & 8.784 &  8.011 & 5.853 & 1+2 / 1+2 \\
\hline
\multicolumn{8}{l}{nd = not detected ;$^a$ RCB candidate from \citet{2003MNRAS.344..325M} }\\
\end{tabular}
\end{table*}

\begin{figure} %figure Dmag vs 8.0
\centering
%\sidecaption
\includegraphics[width=8.25cm]{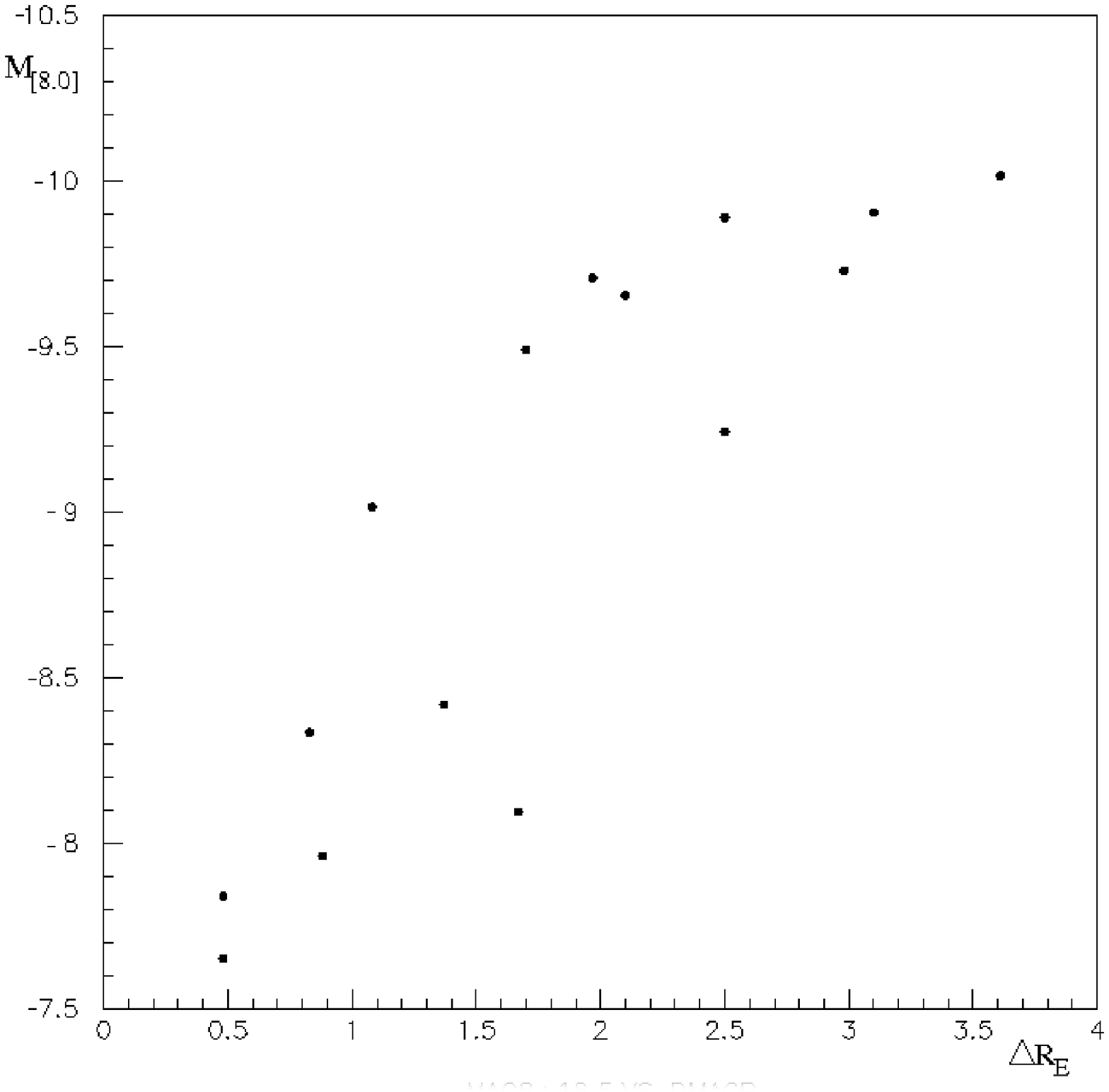}
\caption{Spitzer absolute magnitude $M_{[8.0]}$ versus the maximum drop in optical $R_E$ magnitude observed in LMC DYPer light curves (confirmed and candidates).}
\label{dmag_8.0}
\end{figure}

\subsection{Mid-infrared properties \label{sbsec_midir}}

RCB stars are known to have a dust shell made of amorphous carbon grains \citep{1991ApJ...367..635H} and recently, direct observations of this shell were made by \citet{2007AAS...211.9307C} for 4 RCBs using the Spitzer MIPS camera at 24, 70 and 160 microns. Studies of these circumstellar envelopes should help us to understand better the preceding stages of RCB stars. Here we present and discuss RCB and DYPer star's mid-infrared broadband magnitudes centred at 3.6, 4.5, 5.8, 8.0 and 24.0 $\mu$m obtained by the LMC SAGE \citep{2006AJ....132.2268M} and SMC $S^3$MC \citep{2007ApJ...655..212B} surveys, using the Spitzer IRAC and MIPS cameras. We list in Table~\ref{tab.RCB.SAGE} the magnitudes obtained for all known and newly discovered Magellanic RCB and DYPer stars. If more than one epoch were available, we averaged them. We also included magnitudes of 3 galactic RCBs found in T08 and detected by the Spitzer galactic survey GLIMPSE II; and, if available, the MSX A-band (8.3 $\mu$m) magnitudes \citep{2003yCat.5114....0E} (magnitude limit: $A_{lim}\sim$7.5). These last magnitudes all agree within 0.1 mag with the $[8.0]$ Spitzer magnitudes.

In the mid-infrared colour magnitude diagrams of Figure~\ref{infra_red}, we can observe a clear separation between RCBs and DYPers. DYPers have magnitudes generally bluer and fainter than RCB ones, and therefore have warmer shells. RCB's shells are very bright at 8.0 and 24 $\mu$m. Two interesting conclusions can be drawn when one compare the position of RCB and DYPer stars in these diagrams to that of other set of known stars. We used figures 4, 5 and 6 presented by \citet{2006AJ....132.2034B} for such a comparison. It appears first, that in a $[8.0]$ vs. $[3.6]-[8.0]$ diagram, RCB stars are located in a rather unpopulated area ($-10.2<M_{[8.0]}<-11.8$ and $2<[3.6]-[8.0]<4$). This fact could greatly simplify the search for RCB stars in the future. Their position indicates that their circumstellar shells are thinner, than the so-called "extreme AGB" stars presented in \citet{2006AJ....132.2034B} (these stars were selected on the basis of their extreme red $J-[3.6]$ colour indicating a dusty circumstellar envelope; they are either C- or M-stars). We note that this property may be explained by a clumpy model for RCB shell formation. Second, the DYPer positions correspond to those of the "C-stars" (classical carbon-rich stars selected from the 2MASS catalogue by \citet{2006A&A...448...77C}). As in the optical and near-infrared, we don't observe major differences in mid-infrared broadband photometry between classical carbon stars and DYPers. Nevertheless, in the diagrams of $[8.0]$ vs. $[3.6]-[8.0]$ and $J-[8.0]$, we note that DYPer stars are distributed preferentially in the brighter part of the "C-stars" sequence and at the beginning of the "extreme AGB" one. This seems to indicate an early mass loss phase. This last remark is also supported by the observed correlation between $[8.0]$ magnitudes and the maximum drop detected in optical $R_E$ band (see Figure~\ref{dmag_8.0}).

We note that the warm RCB star W-Men ($T_{eff}\sim$7000 K, \cite{1990MNRAS.245..119G}) has an $[8.0]$ magnitude about 2 mag. lower than classical RCBs, but an identical $[3.6]-[8.0]$ colour. Its shell should therefore be much thinner. We observe the same properties for the RCB candidate KDM-5651. We find also that the shell properties of the 3 new RCB candidates correspond to those of confirmed RCB stars, as they share the same positions in the 3 mid-infrared colour magnitude diagrams presented. This re-enforces our assignment of these stars as RCB stars. We note also that the only mid-infrared measurement for the RCB candidate MSX-SMC-014 comes from MSX A-band, which shows a shell as bright as EROS2-LMC-RCB-7 and MACHO-6.6575.13.

\subsection{Spectral energy distributions\label{sbsec_sed}}

\begin{figure*} %figure SED
\includegraphics[scale=0.85]{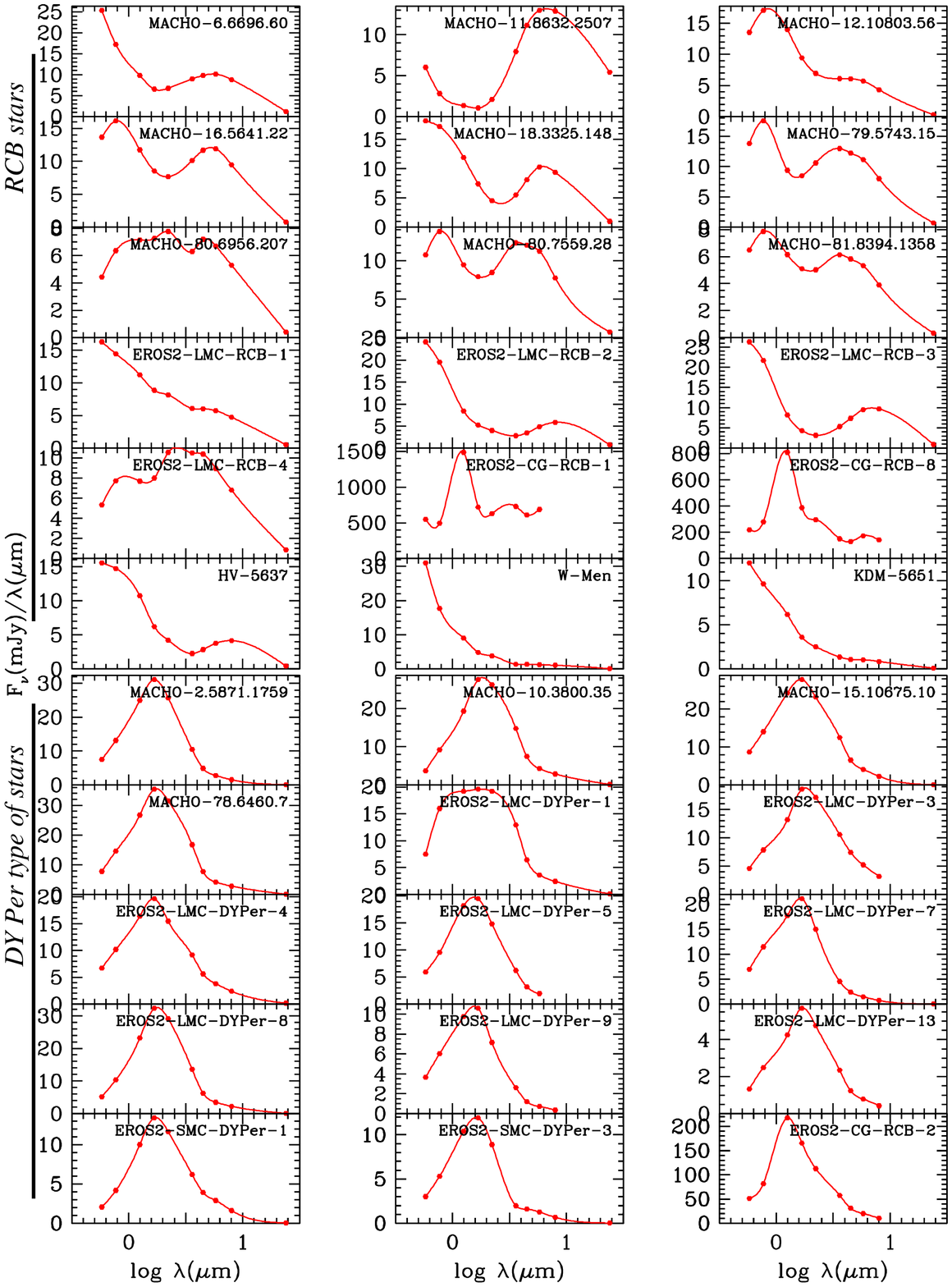}
\caption{Spectral energy distributions from optical to mid-infrared of confirmed and candidate RCBs (top) and DYPers (bottom).}
\label{sed_fig}
\end{figure*}

\begin{figure*} %figure Charts A
\centering
%\sidecaption
\includegraphics[scale=0.9]{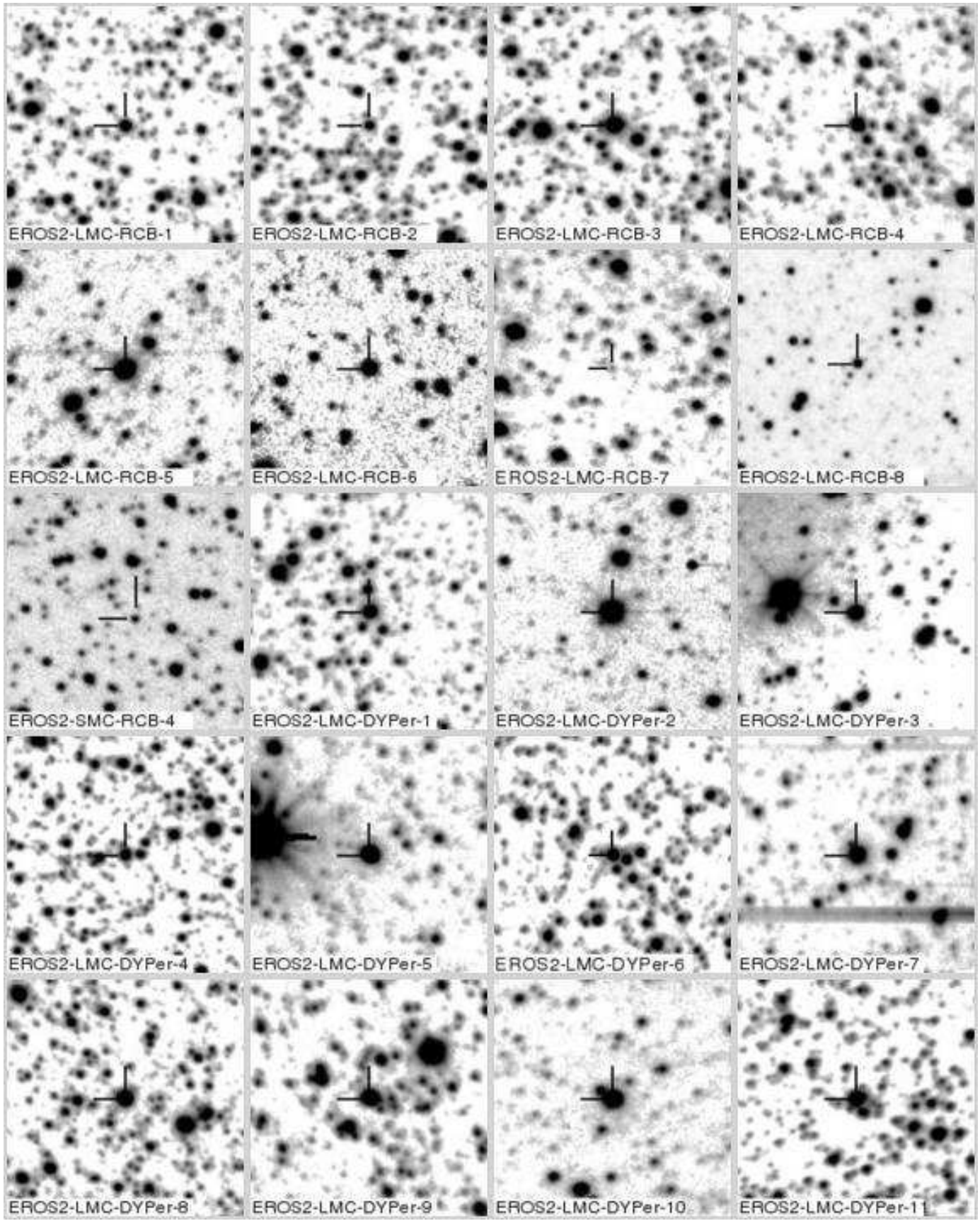}
\caption{Charts of the new Magellanic RCB and DYPer stars, confirmed and candidates (2'x2'). North is up, East is to the left.}
\label{chartA}
\end{figure*}

\begin{figure*} %figure Charts B
\centering
%\sidecaption
\includegraphics[scale=0.9]{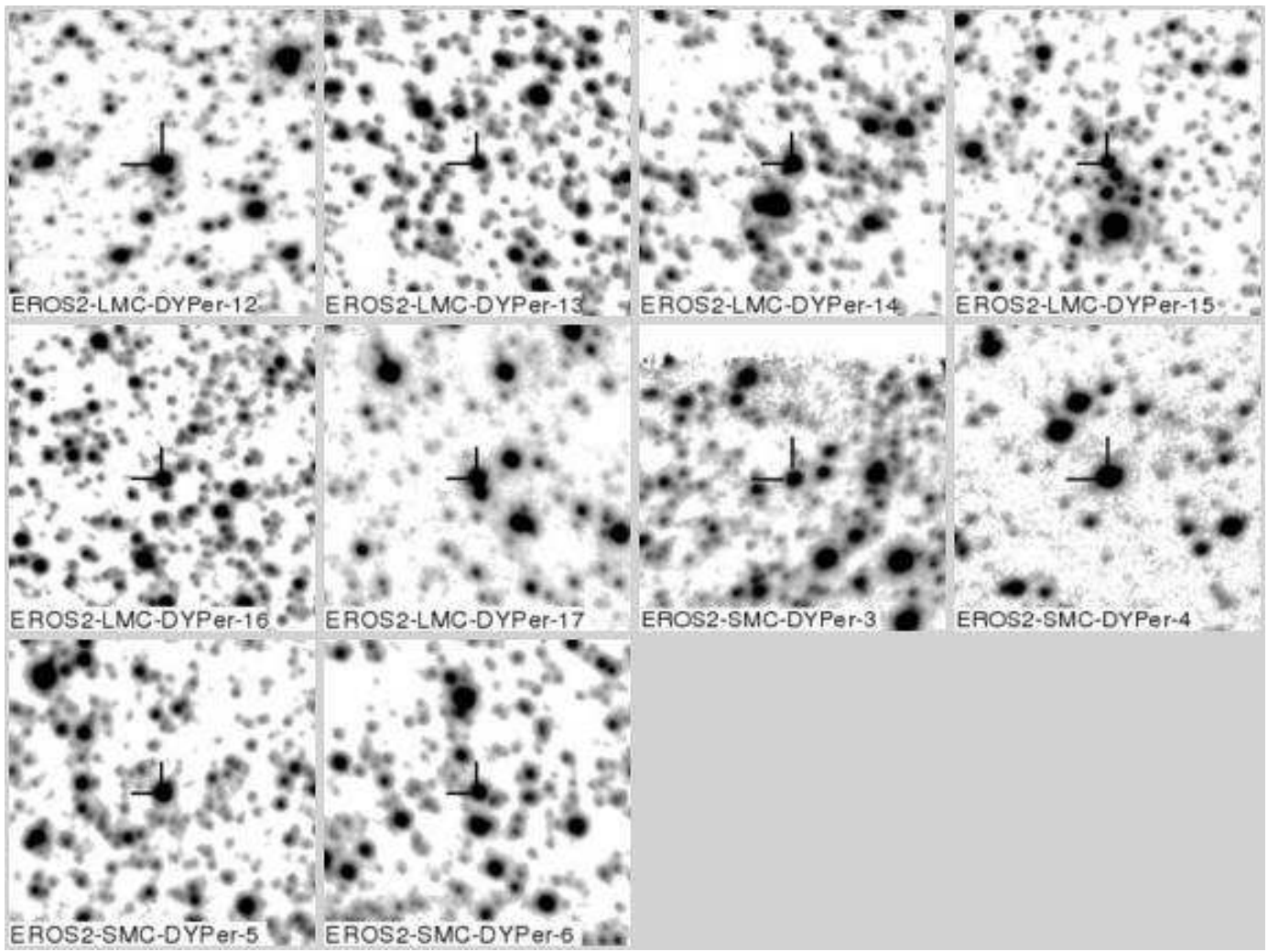}
\caption{Charts continued.}
\label{chartB}
\end{figure*}

Figure~\ref{sed_fig} shows the spectral energy distributions (SEDs) from optical to mid-infrared of RCBs and DYPers (confirmed or candidate) where a complete set of magnitudes are available (i.e. maximum magnitudes $B_E$ and $R_E$ observed in EROS-2 light curves, 2MASS JHK magnitudes and Spitzer magnitudes from IRAC and MIPS camera). The optical and near-infrared magnitudes were corrected for interstellar extinction using average Magellanic values as mentioned in Section~\ref{sbsec_cmd}. We also corrected the 2MASS JHK magnitudes for carbon dust extinction, if the epochs corresponded to a fading phase observed in the EROS-2 optical light curves. We used for such corrections the $\Delta R_E$ magnitude variation observed at that epoch and the absorption coefficients of pure amorphous carbon dust presented by \citet[fig.2]{1995A&A...293..463G} (note that the DYPer dust may not be made of amorphous carbon as observed in RCB stars).

We can clearly see from Figure~\ref{sed_fig} that most RCB stars' SEDs appear to be made of two distinct blackbodies, one from the stellar component and the other from the circumstellar shell. We can estimate that shell temperatures range between 360 and 600 K, which are cooler than previous estimates obtained by \citet{1973MNRAS.161..293F} using Galactic RCBs (800 - 1000 K). We observe almost no sign of a cool blackbody in the SED of two stars: the warm RCB W-Men and the RCB candidate KDM 5651.

The SEDs of DYPers are essentially identical: only one peak is observable, the stellar and the shell blackbody components being mixed. They all have a maximum intensity at wavelength $\lambda_{max}\sim$1.7 $\mu$m, which is also found in the SEDs of classical carbon stars by \citet{1995A&A...293..463G}: his fitted models of ordinary carbon stars favour a stellar component with $T_{eff}\sim$2000 - 2500 K and a circumstellar shell with  $T_{eff}\sim$800 - 1500 K (hotter than that of RCBs).
% Bulge RCB SED: weird.

\subsection{Discussion of individual stars}

\begin{figure*}
\centering
\includegraphics[scale=0.38]{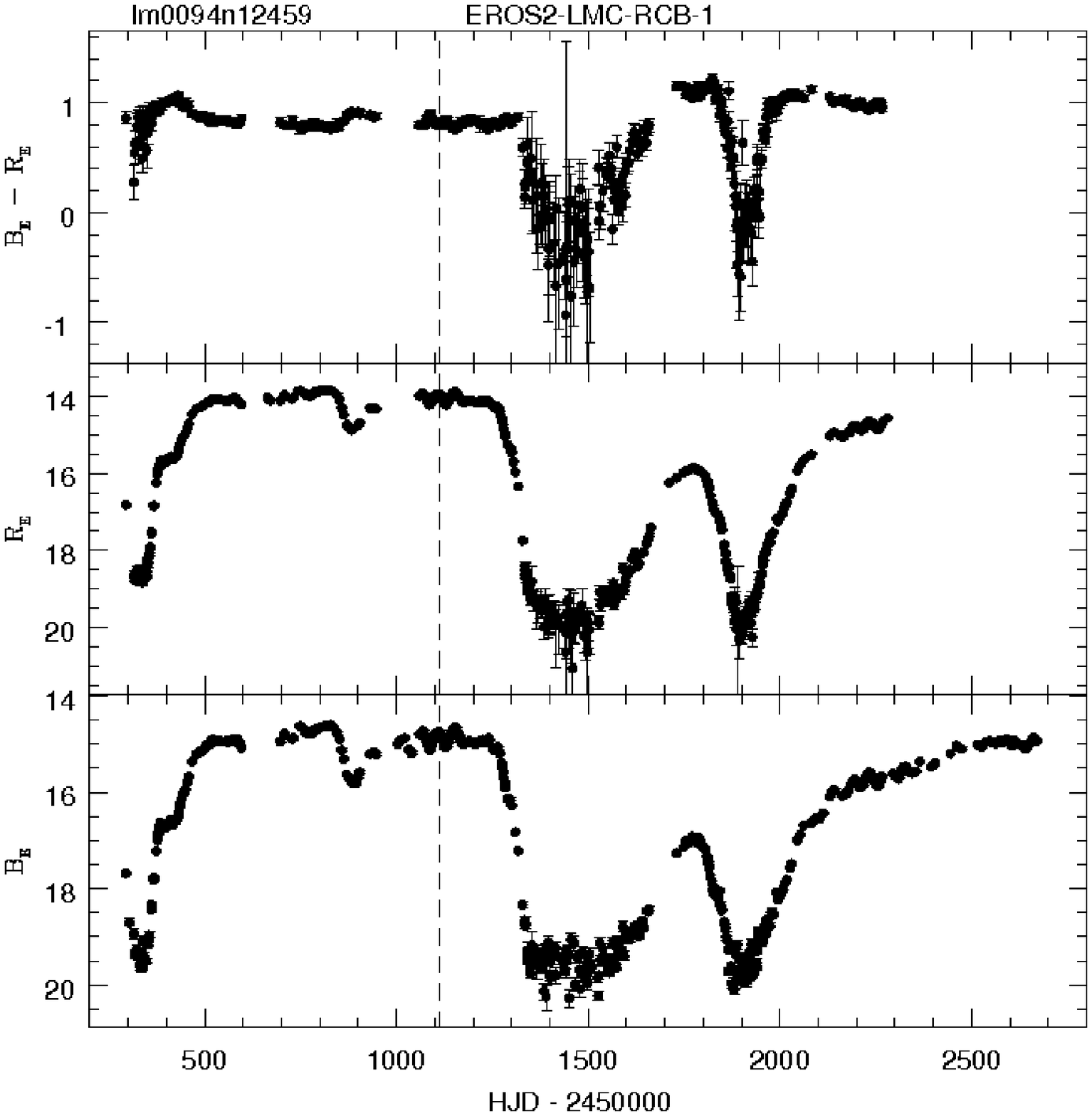}
\includegraphics[scale=0.38]{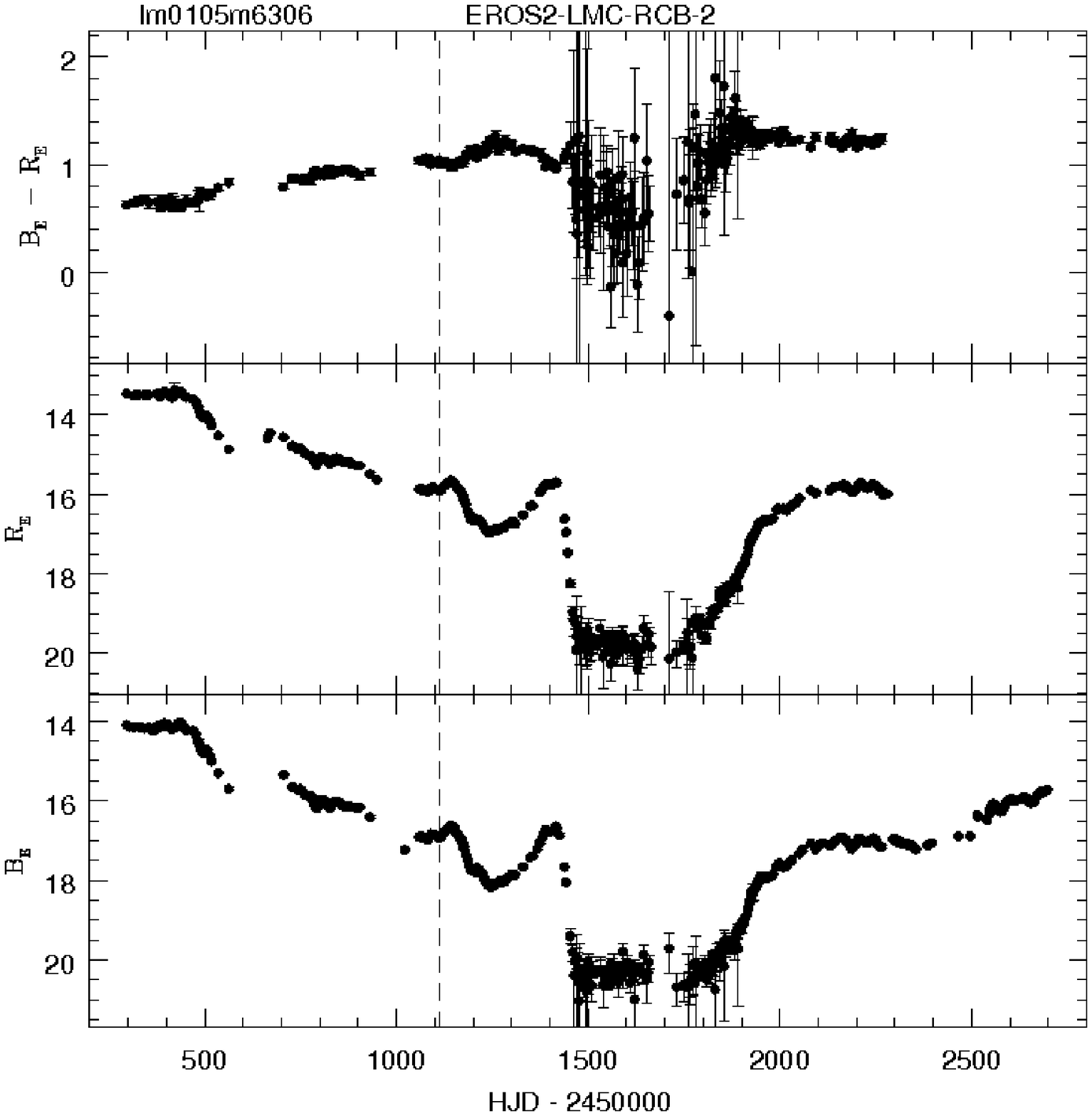}
\includegraphics[scale=0.38]{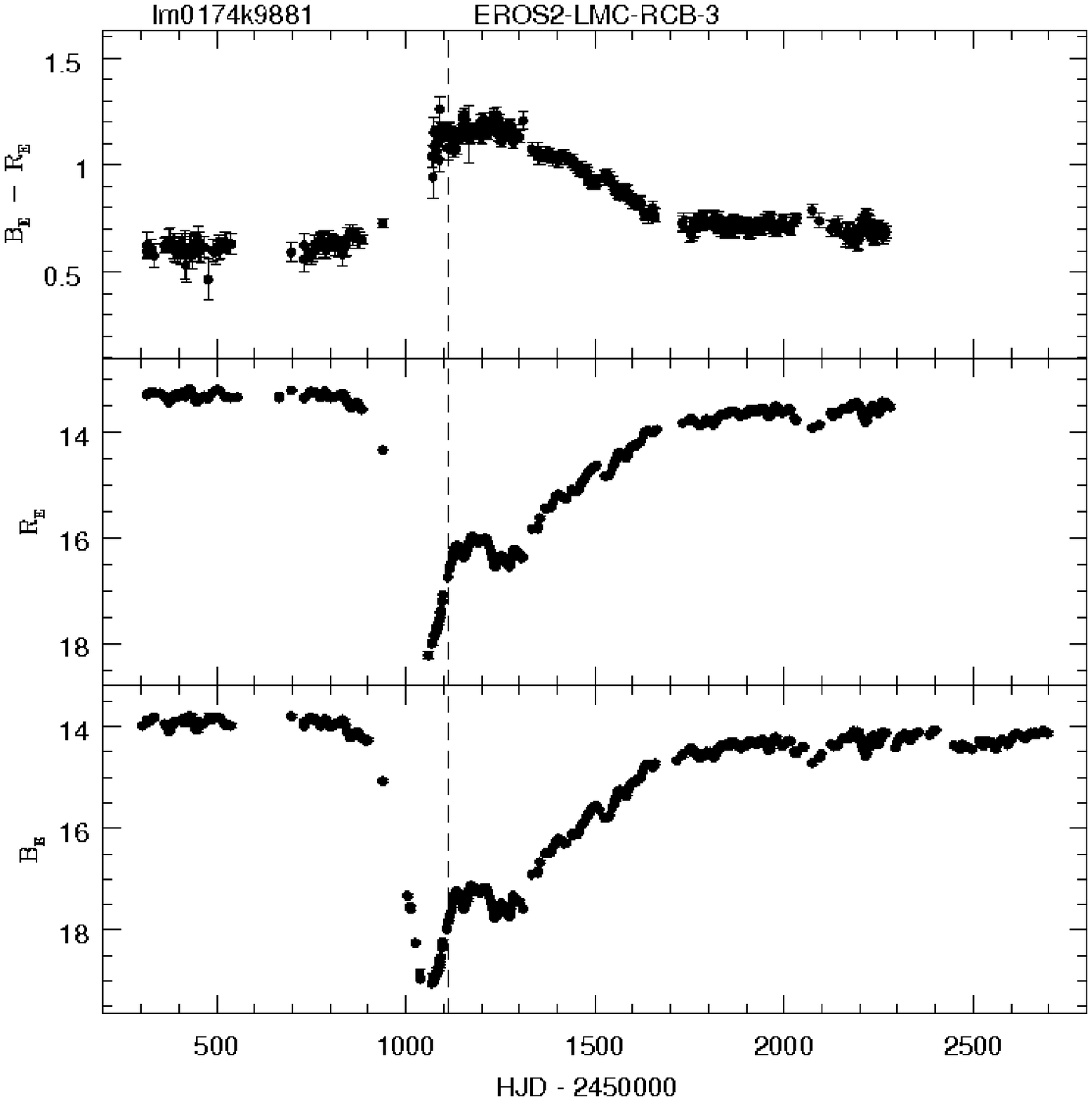}
\includegraphics[scale=0.38]{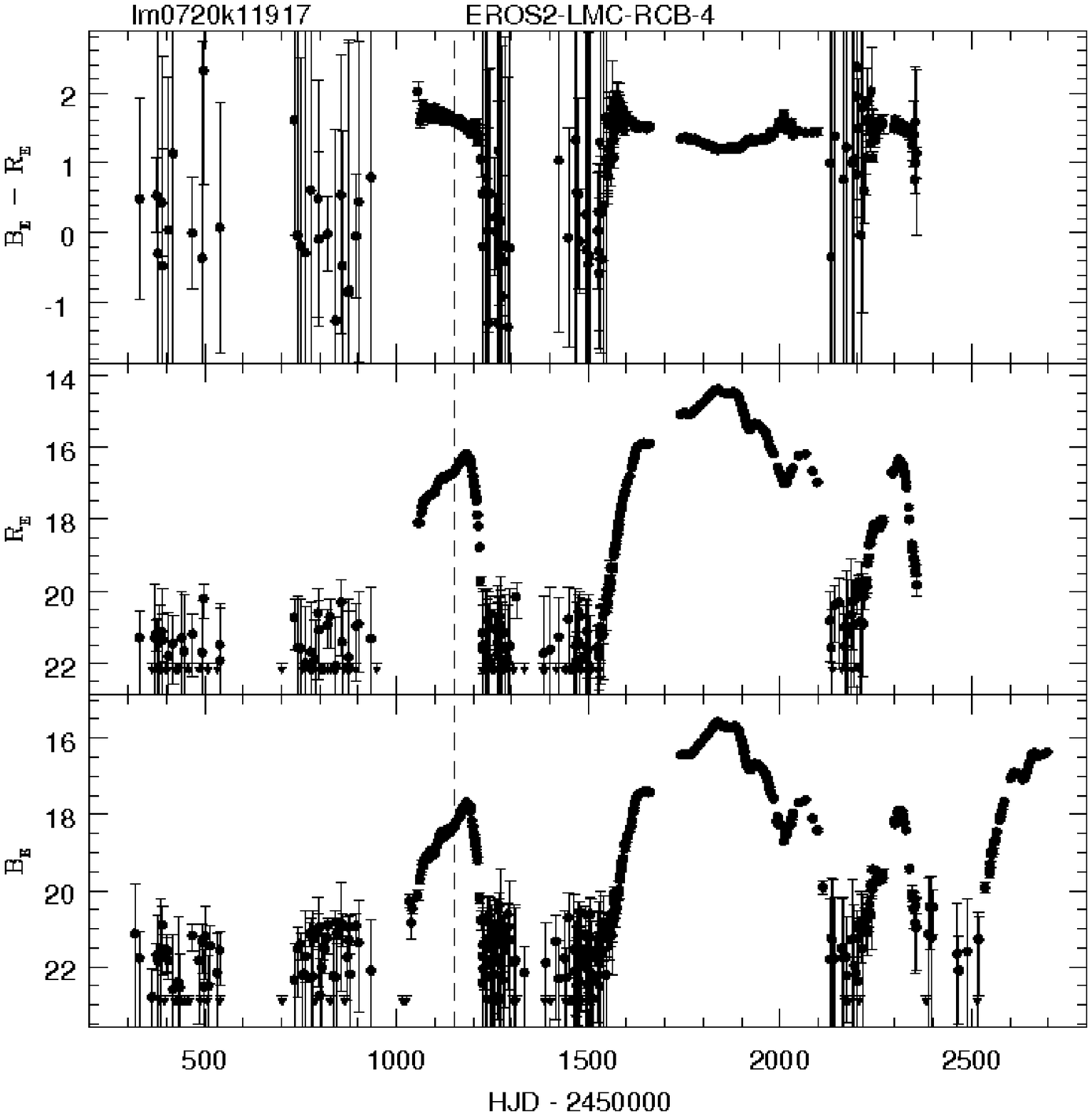}
\includegraphics[scale=0.38]{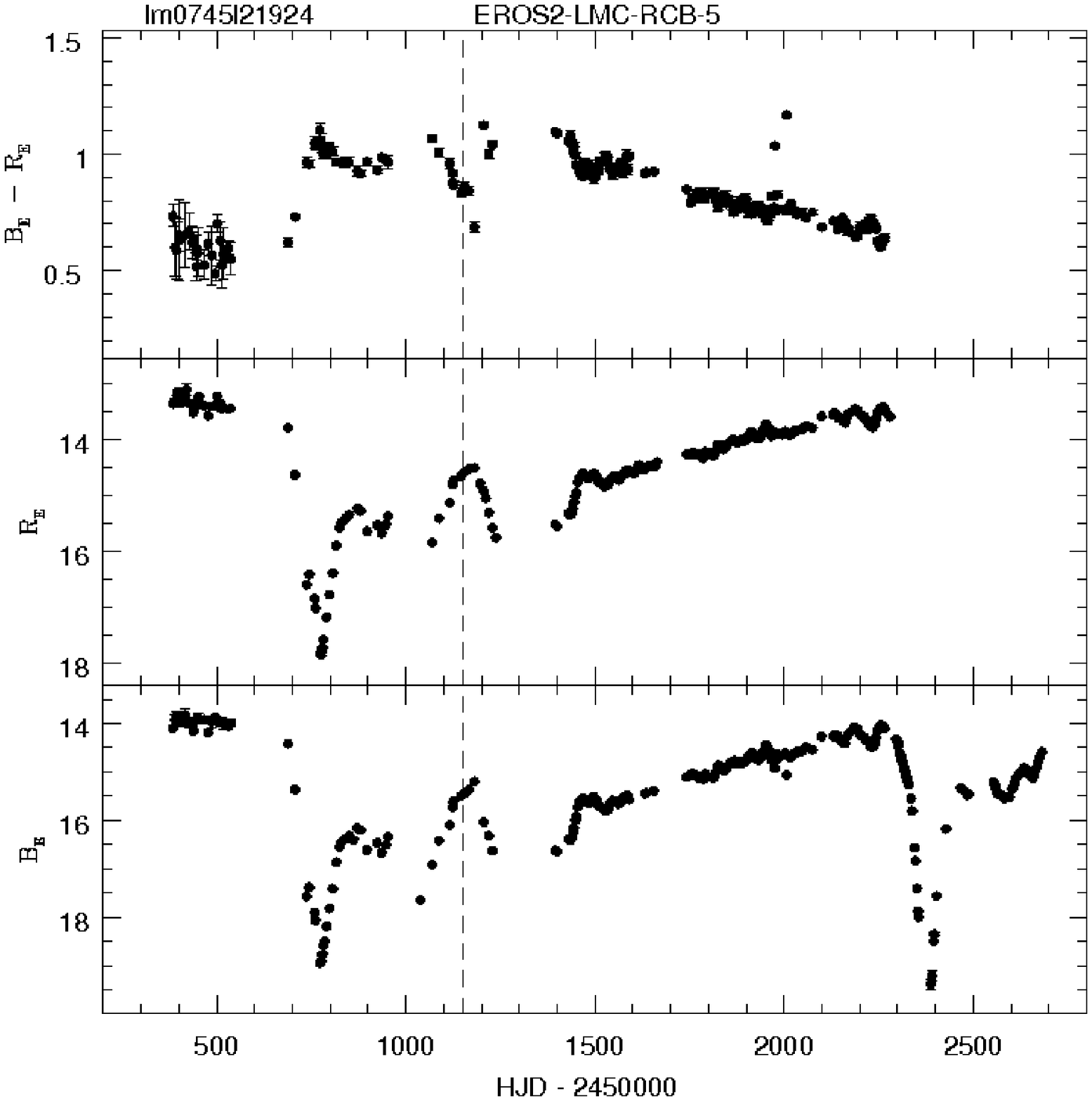}
\includegraphics[scale=0.38]{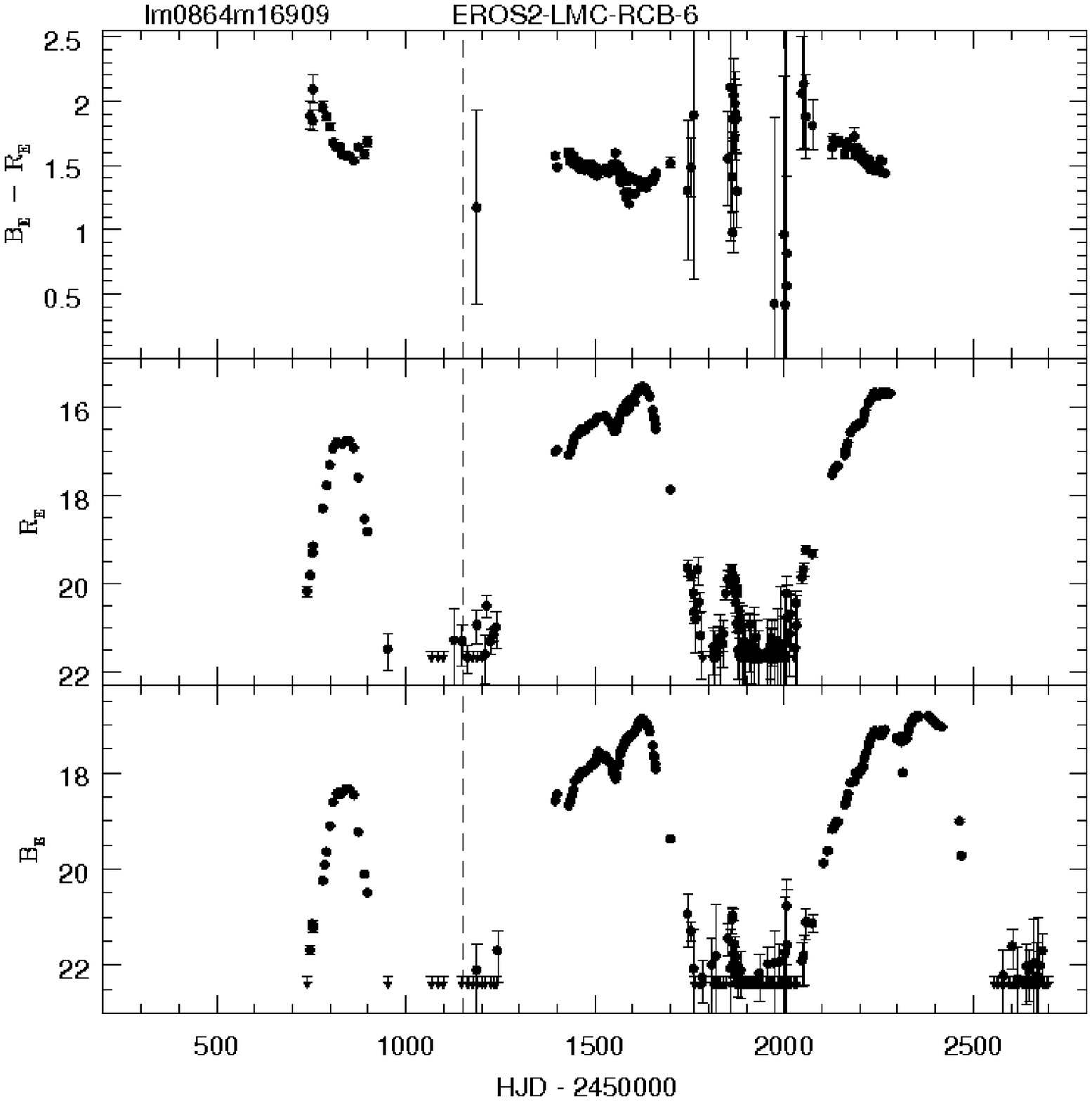}
\caption{Light curves of the new Magellanic RCB and DYPer stars (confirmed and candidates). The arrows represent our detection limits. The dashed vertical lines indicate the 2MASS epochs. For each group of three: top: $B_E - R_E$ colour vs. time; middle: $R_E$ light curve; bottom: $B_E$ light curve.}
\label{lc}
\end{figure*}

EROS2-LMC-RCB-1, -2, -3 and -5 are confirmed RCBs with temperatures in the T $\sim$6000 K range as determined from their $V-I$ colour indices. All four are catalogued as carbon star in \citet{2001A&A...369..932K}. We suspect that EROS2-LMC-RCB-1 has not been catalogued in the MACHO database, as the only objects found around EROS2-LMC-RCB-1's position have light curves with photometry influenced by a bright neighbour, and variations occurring during those of EROS2-LMC-RCB-1. We point out the atypical lightcurves of EROS2-LMC-RCB-2 (proposed as an RCB candidate by \citet{2003MNRAS.344..325M}) and EROS2-LMC-RCB-5. EROS2-LMC-RCB-2 has two fading phases (see Fig.~\ref{lc}), which is unusual for RCB type stars: a first slow fading ($\sim$0.004 mag. day$^{-1}$) followed by a fast one ($\sim$0.070 mag. day$^{-1}$). Its recovery phase is symmetric to the first fading event. EROS2-LMC-RCB-5 shows an overall remarkably slow and linear recovery phase ($\sim$0.002 mag. day$^{-1}$) that lasted for more than 3.5 years. Such recoveries are usually observed in DY Per type of stars.

EROS2-LMC-RCB-4 and -6 seem to have lower temperatures than the first four RCBs. From the strength of the Ca II triplet absorption lines, EROS2-LMC-RCB-4 is cooler than the four $\sim$6000 K RCBs and therefore its temperature is more likely to be in the 5000 K range. We are not sure that their real maximum optical brightness was reached during the EROS-2 observations, as it lasts less than 50 days in both cases. EROS2-LMC-RCB-4 and -6 are highly active RCBs as their respective EROS-2 and MACHO light curves show. EROS2-LMC-RCB-4 was faint during February 2008, but bright enough during October to obtain a spectrum which shows strong carbon features. EROS2-LMC-RCB-6 remained too faint during all of 2008, and we did not get any spectrum for it. We nevertheless consider EROS2-LMC-RCB-6 as a strong RCB star, due to the distinct signature of RCB-type drops observed in its light curve.

EROS2-LMC-RCB-7, -8 and EROS2-SMC-RCB-4 are only considered as RCB candidates. We did not obtain any spectrum for these stars as they remained too faint during 2008:

\begin{itemize}
\item EROS2-SMC-RCB-4 has a bright shell, with magnitudes and colours expected for RCB stars, as seen on Figure~\ref{infra_red}. Its EROS-2 light curve presents small variations but with a fast fading rate. We consider EROS2-SMC-RCB-4 as a strong RCB candidate.

\item EROS2-LMC-RCB-7 presents a unique variation of $\sim$1 mag. in the EROS-2 red band, and remained stable during all MACHO observations. It is therefore a weak RCB candidate, but we decided to keep this star in the RCB candidate list for three reasons. First, EROS2-LMC-RCB-7 has known a brighter phase in the past : it has been catalogued as a very blue object ($B-V\sim$ -0.4) in the YB6 USNO catalogue with magnitudes 2.5 mag. brighter than the EROS-2 ones ($B_{USNO}\sim$ 17.22 and $V_{USNO}\sim$ 17.64). We note that its EROS-2 colour is also relatively blue ($V-I\sim1.0$), but became redder during the variation observed. This may be due to a blending effect; the better astrometric resolution of OGLE-3 should help to answer this ambiguity. Second, EROS2-LMC-RCB-7 has a thick and cold circumstellar shell, similar to the one of the hot RCB star HV 2671; they have almost identical [24] Spitzer magnitudes and $[8.0]-[24]$ colours. The infra-red observations indicate also a very high extinction phase (with $J-H\sim$2.7 and $H-K\sim$2.0). We note that EROS2-LMC-RCB-7 was classified as a OH/IR star in \citet{2001AJ....122.1844E} due to its large $K-A$ colour (the largest value of all the sample analysed). Finally, we did not find any other object in our catalogue analysis with a light curve similar to EROS2-LMC-RCB-7, which makes it really peculiar. There should exist RCB stars that remain hidden during a long phase of extinction. EROS2-LMC-RCB-7 may be such a star, a hot RCB star that undergoes actually an intense phase of dust production.

\item EROS2-LMC-RCB-8 presents unusually large oscillations for an RCB star, before its fading phase (drop of 4 mags and fast fading rate). The OGLE-3 EROS2-LMC-RCB-8's light curve will be really interesting to study. It will indicate if the large oscillations are due to intrinsic pulsation of the star or some fading events. We note that similar large variations were also observed with EROS2-CG-RCB-12 \citep{2008A&A...481..673T}, but with shorter periods. If EROS2-LMC-RCB-8 is an RCB, it would be the coolest ever found, even cooler than EROS2-SMC-RCB-3. A spectrum is therefore needed to confirm its nature. If we suppose a 2 magnitudes extinction correction of its maximum magnitude due to carbon dust, EROS2-LMC-RCB-8 would lie on a continuation of line A in Figure~\ref{cmd} with $V-I\sim2.6$ (i.e. T$\sim$3400 K) and $M_V\sim-2.3$. Finally we note that EROS2-LMC-RCB-8's circumstellar bright shell has been observed only with the Spitzer MIPS camera and the MSX experiment; no IRAC magnitudes are currently available.
\end{itemize}

All DYPers have carbon spectra and photometric broad-band magnitudes from optical to mid-infrared almost identical to those of classical carbon stars. One DYPer seems apart from the rest: EROS2-LMC-DYPer-1 is the only DYPer that does not lie in the expected position of carbon stars in the colour-colour infrared diagram of Figure~\ref{infra_red}. Even if its 2MASS measurements were taken during a fading phase, we expect this star to be $\sim$0.5 mag bluer. Its circumstellar shell seems classical compared to a carbon star, but its SED (Fig.~\ref{sed_fig}) indicates a hotter stellar component. A blending effect may explain this observation.

EROS2-LMC-DYPer-1 to -7  are DYPers that we consider as confirmed, since they present a drop higher than 2 magnitudes (empirical limit). Only one drop is observed for each of them in the EROS-2 light curves. The remaining 14 DYPer candidates show less than 2 magnitudes drops in their EROS-2 and MACHO light curves. We note that in the case of 4 DYPer candidates, EROS2-LMC-DYPer-8, -11, -12, and EROS2-SMC-DYPer-6,  we only observed an increase in brightness that may indicate a recovery phase from an ejection event. For EROS2-LMC-DYPer-8 and -11, a fading phase is observed in the MACHO light curve. It will be interesting to see how the two remaining stars behave during the OGLE-3 observations.

We stress that the cool RCB star candidate, KDM-5651, has mid-infrared magnitudes corresponding to the warm RCB star W-Men. Its circumstellar shell is therefore as thin as W-Men's, which may explain the relatively small magnitude drops observed in its light curve and therefore the low level of ejection activity. At the other extreme, the other RCB candidate MSX-SMC-014 \citep{2005ApJ...631L.147K} has a bright circumstellar shell, as bright as EROS2-SMC-RCB-7's, which in that case indicates strong ejection activity. Its OGLE-3 light curve shows multiple variations.

Finally, we note that EROS-SMC-RCB-3 (alias EROS2-sm0067m28134, MSX-SMC-0155) is not a DY Per type star as mentioned by \citet{2005ApJ...631L.147K}. It is a very cool RCB star that presents a fast fading rate and a bright, cool circumstellar shell typical of RCB stars.

\section{Summary}

Our search for new RCB and DYPer stars in the EROS Magellanic database has resulted in the discovery of 6 new RCBs and 7 new DYPers. The total number of confirmed Magellanic RCB (DYPer) stars is therefore now 23 (13). We have also presented a list of candidates for both types of stars: 3 RCB and 14 DYPer candidates. We define candidates as stars with fading drops lower than 2 magnitudes. EROS-2 has contributed to the discovery of more than half of the currently known Magellanic RCB and DYPer stars.

We note that the \citet{2003MNRAS.344..325M} technique to find RCB stars, based on carbon spectra with weak CN bands, is conclusive. Four of the six stars presented are now considered as RCB stars and another one, KDM 5651, should be considered as a strong candidate. 

We confirm the \citet{2001ApJ...554..298A} result on the difference in the $^{13}$C isotope abundance in the atmospheres of RCB and DYPer stars. No trace was found in the RCB spectra, unlike those of DYPers which show a significant amount of $^{13}$C in most cases.

We stress that we observe strong similarities between the SEDs of DYPers and those of classical carbon stars, from optical to mid-infrared wavelengths. This suggests that they are ordinary carbon stars with ejection events. However, more spectroscopic observations and abundance analysis will be necessary to really answer this question.

We observe an RCB absolute magnitude ($M_V$) range between $\sim$-5.2 and $\sim$-3.4, but note that the lower limit could be conservatively extended to $M_V\sim$-2.6. In a similar way, this last limit may also be extended to $\sim$-2.3 if the star EROS2-LMC-RCB-8 is confirmed as an RCB. A fainter absolute magnitude range is found for DYPers, between $\sim$-3 and $\sim$-1.8.

We have compared the publicly available broadbands mid-infrared magnitudes of RCB and DYPer stars. We observe that most RCB stars have a brighter and cooler circumstellar shell than DYPer stars and we estimate a range of temperature for Magellanic RCB shells between 360 and 600 K, cooler than first estimates obtained with bright Galactic RCB stars \citep{1973MNRAS.161..293F}.  Finally, it appears that in the colour-magnitude diagram, $M_{[8.0]}$ vs $[3.6]-[8.0]$, RCB stars are located in a rather unpopulated area. They seem to have thinner shells than most common AGB stars. This fact could greatly simplify the search for RCB stars in the future.

\begin{acknowledgements}
We thank Tony Martin-Jones for his careful reading and comments.
This publication makes use of data products from the Two Micron All Sky Survey, which is a joint project of the University of Massachusetts and the Infrared Processing and Analysis Centre, California Institute of Technology, funded by the National Aeronautics and Space Administration and the National Science Foundation. The DENIS data have also been used. DENIS is the result of a joint effort involving human and financial contributions of several institutes mostly located in Europe. It has been supported financially mainly by the French Institut National des Sciences de l'Univers, CNRS, and French Education Ministry, the European Southern Observatory, the State of Baden-Wuerttemberg, and the European Commission under networks of the SCIENCE and Human Capital and Mobility programs, the Landessternwarte, Heidelberg and Institut d'Astrophysique de Paris. JA acknowledges support from the Danish Natural Science Research Council. This work is also based in part on observations made with the Spitzer Space Telescope, which is operated by the Jet Propulsion Laboratory, California Institute of Technology under a contract with NASA.
\end{acknowledgements}

\clearpage 

\bibliographystyle{aa}
\bibliography{RCB_tisserand_MC}

\clearpage 

\begin{figure*}
\centering
\includegraphics[scale=0.38]{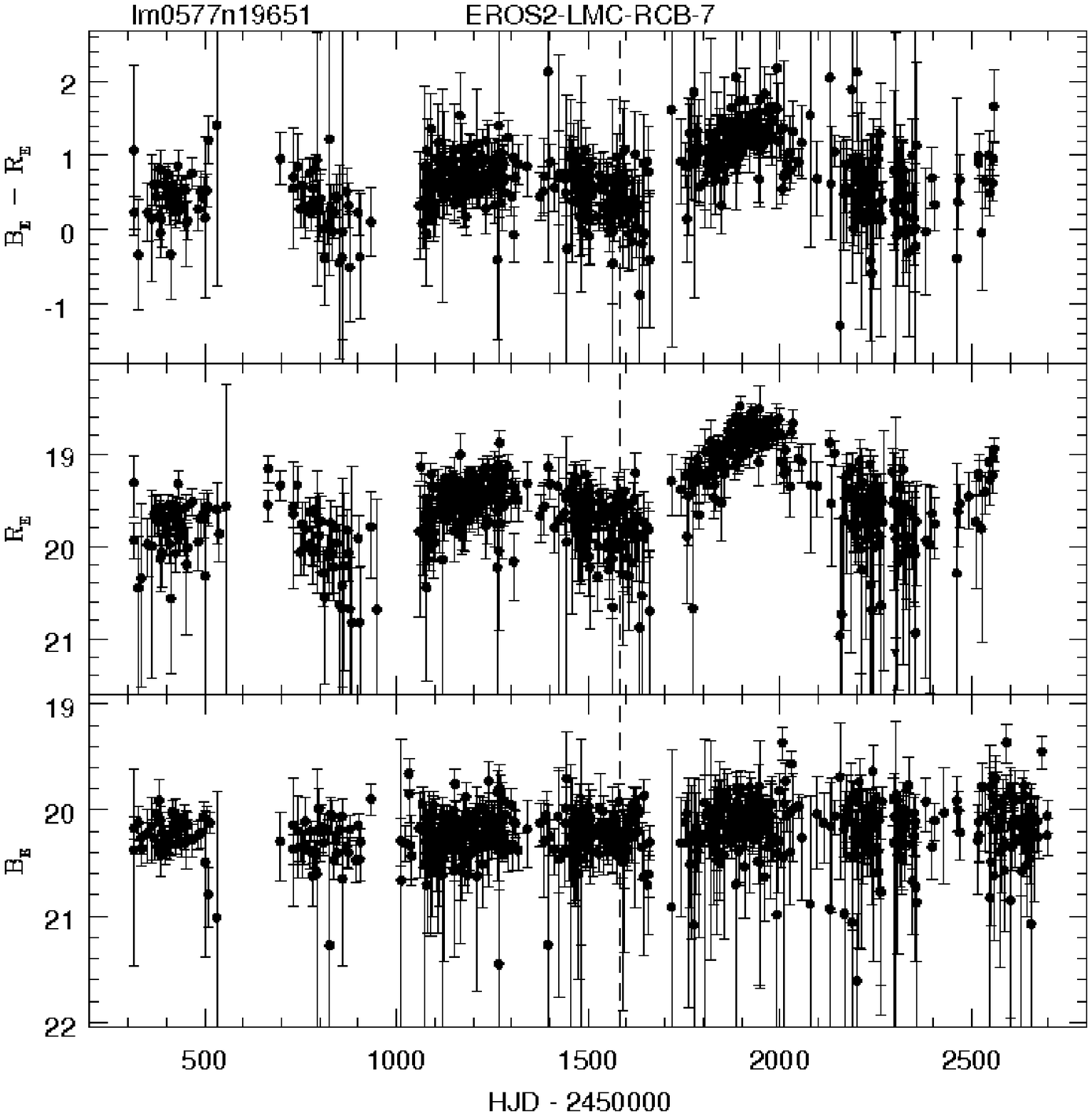}
\includegraphics[scale=0.38]{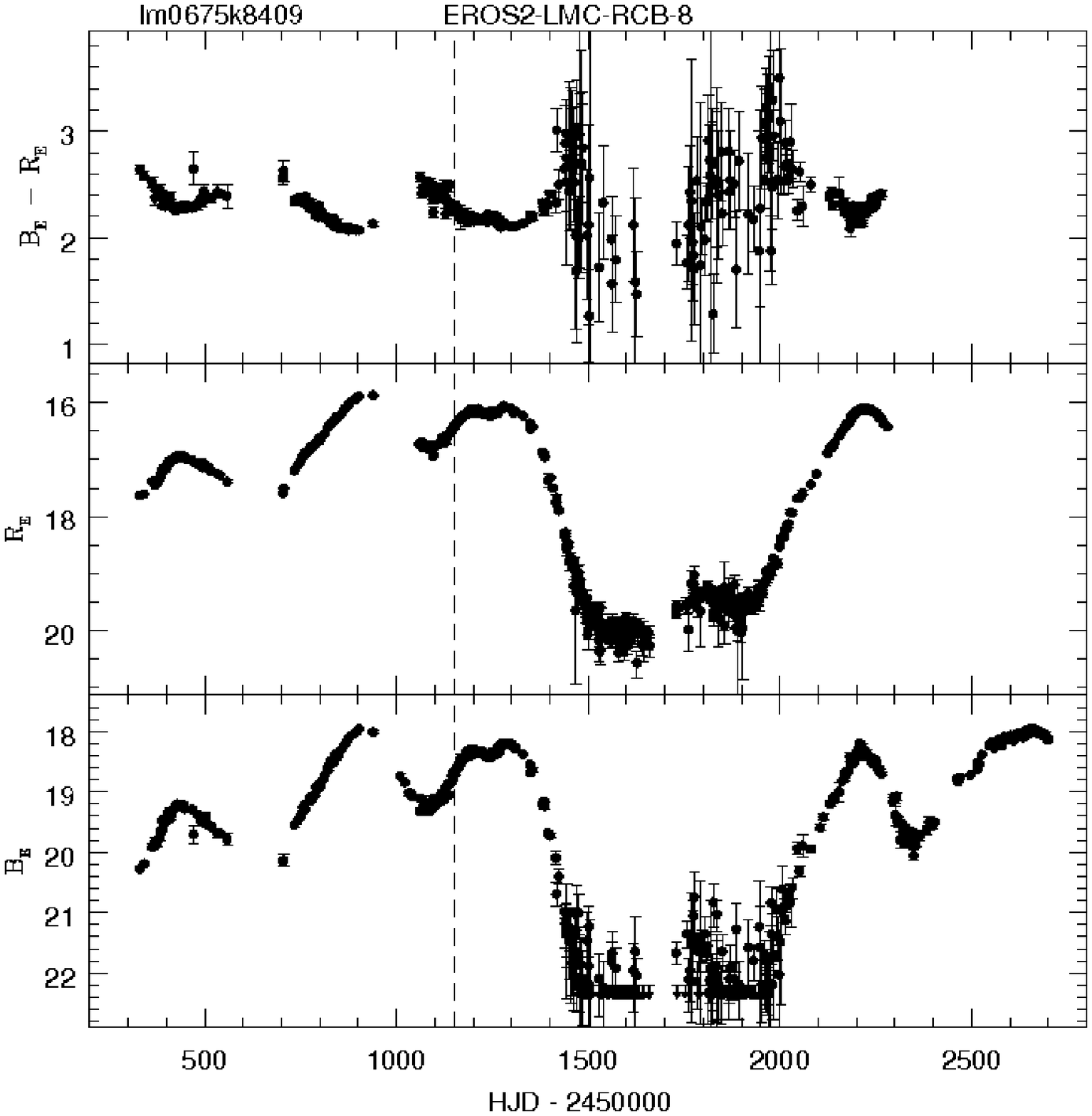}
\includegraphics[scale=0.38]{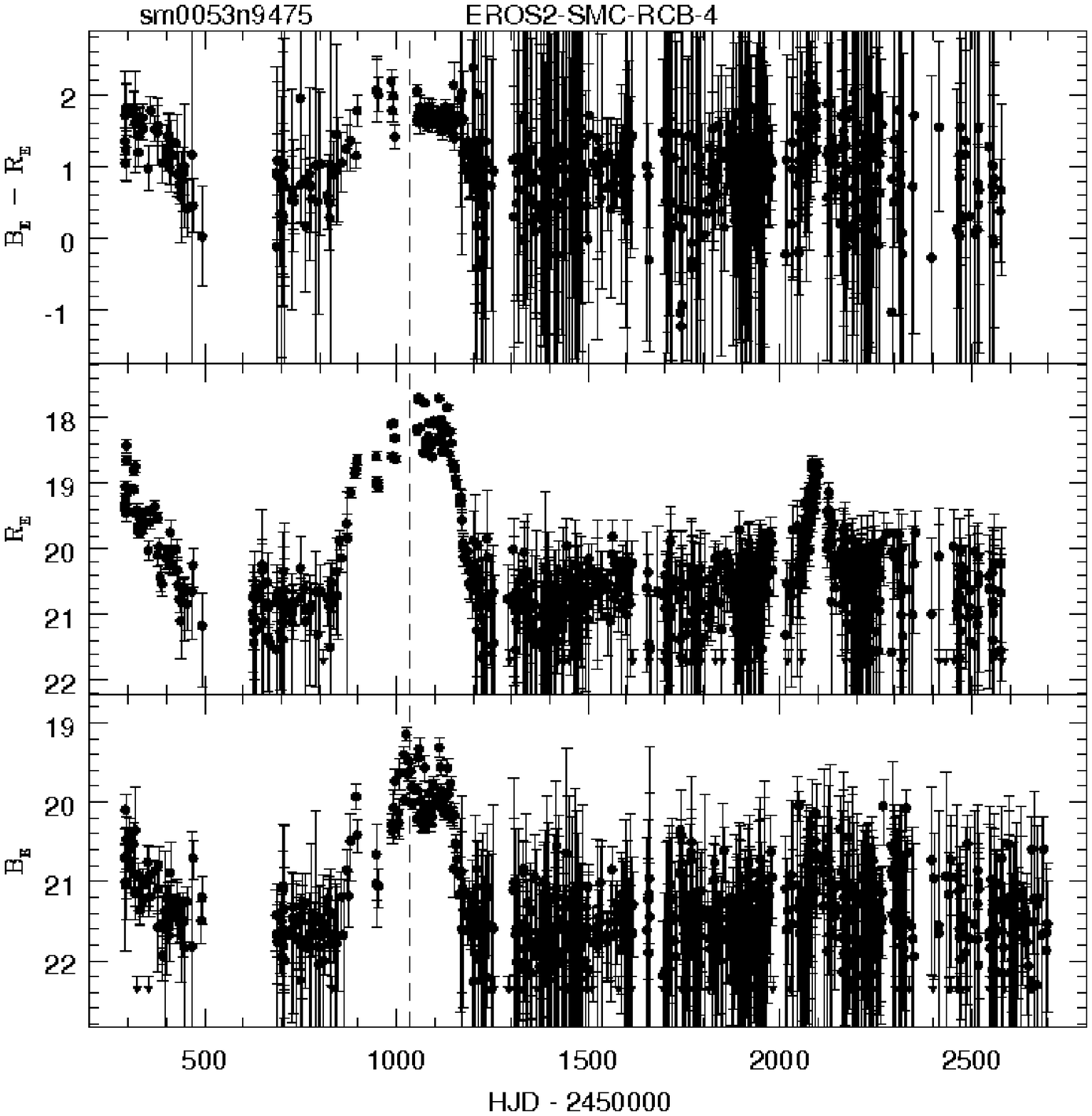}
\includegraphics[scale=0.38]{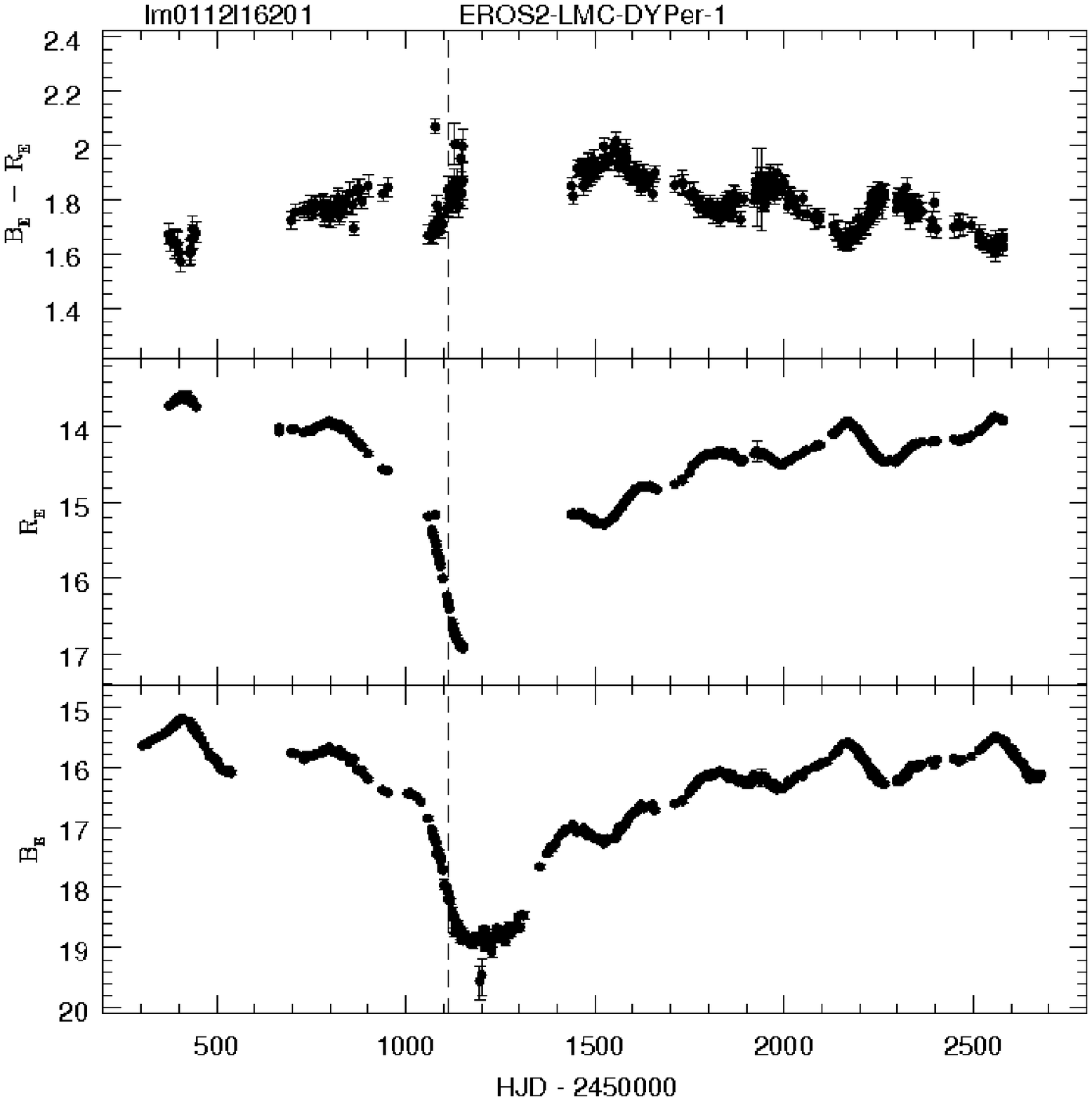}
\includegraphics[scale=0.38]{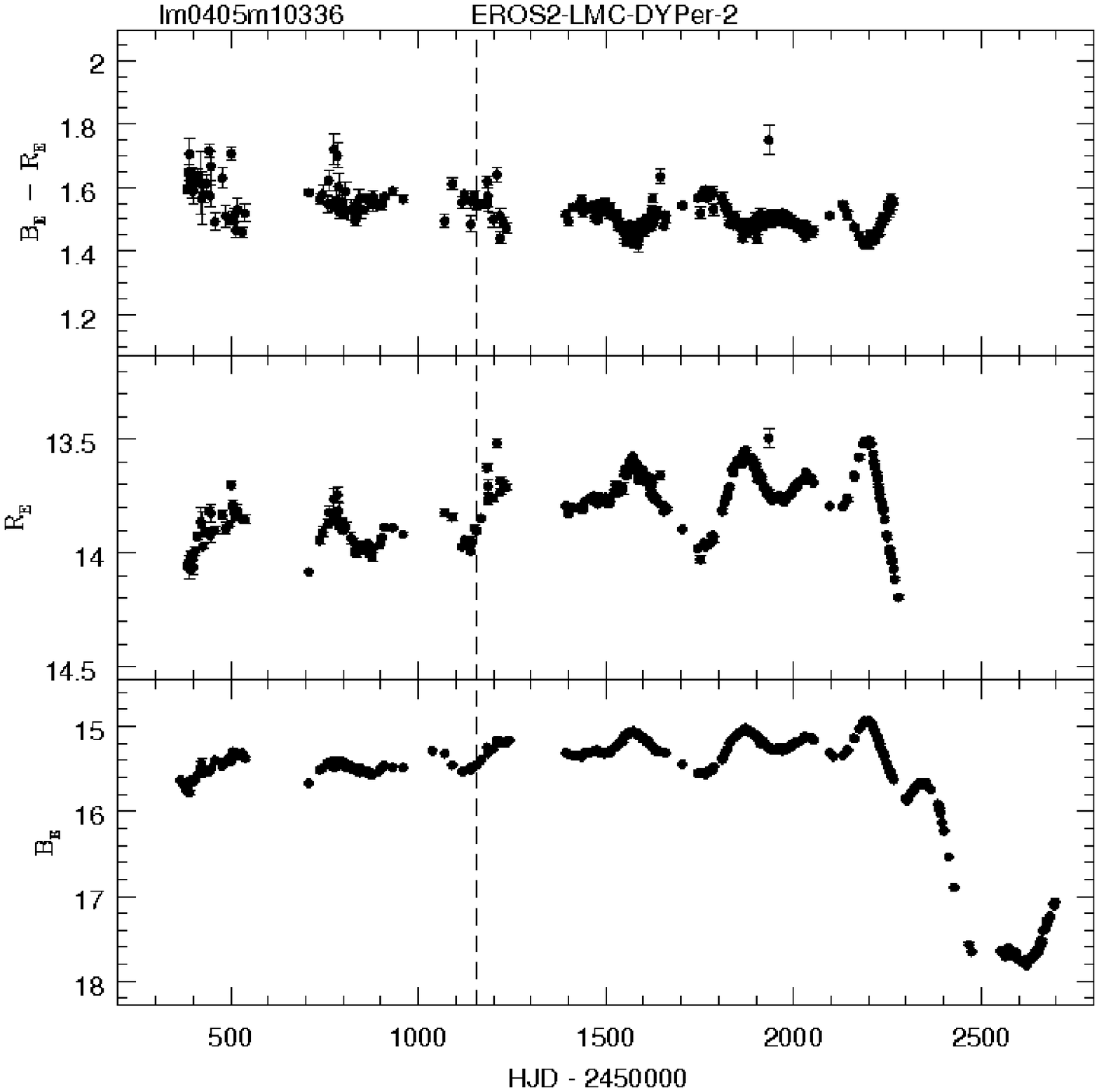}
\includegraphics[scale=0.38]{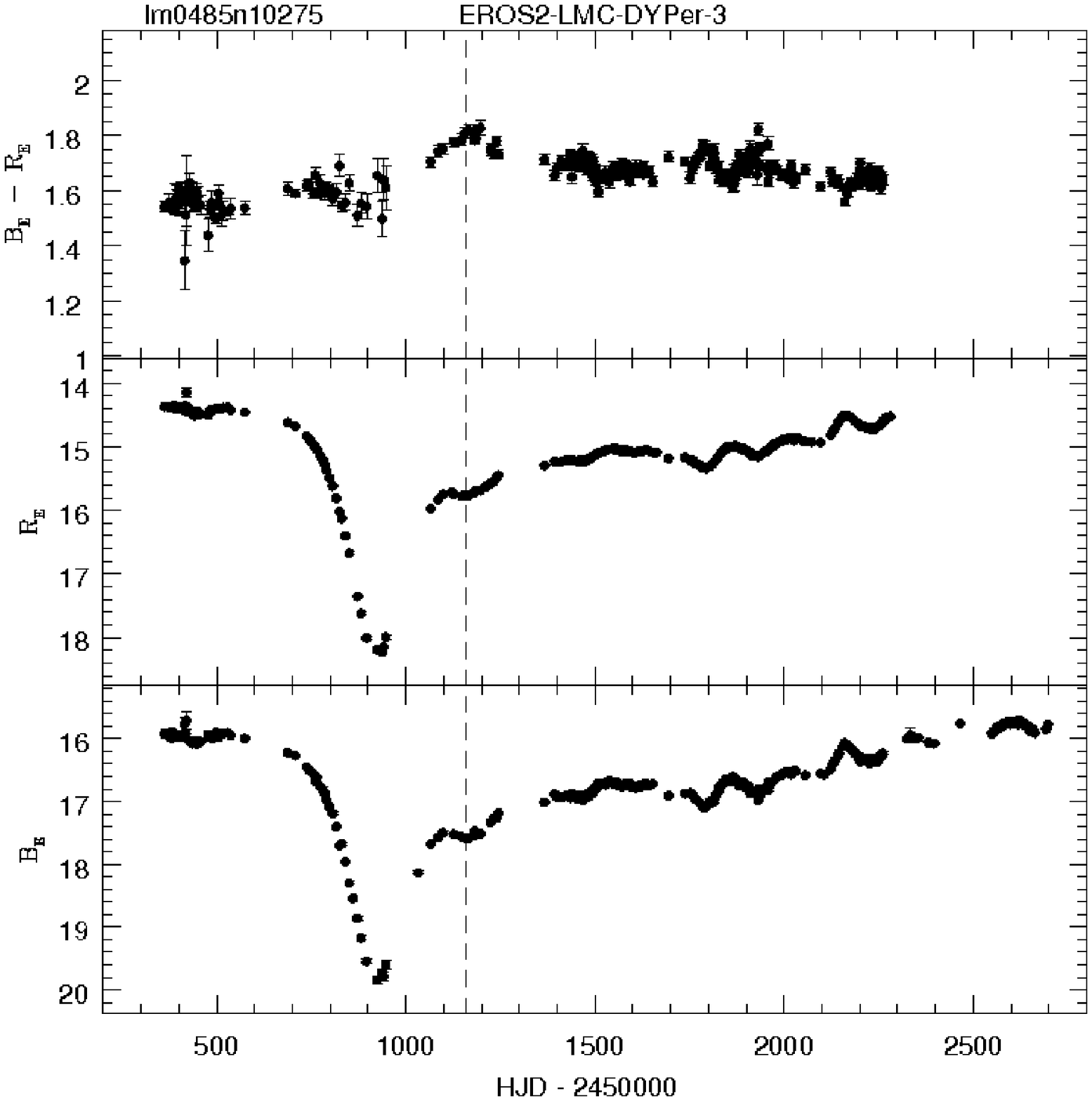}
\caption{Light curves of the new RCB and DYPer stars (continued).}
\end{figure*}

\begin{figure*}
\centering
\includegraphics[scale=0.38]{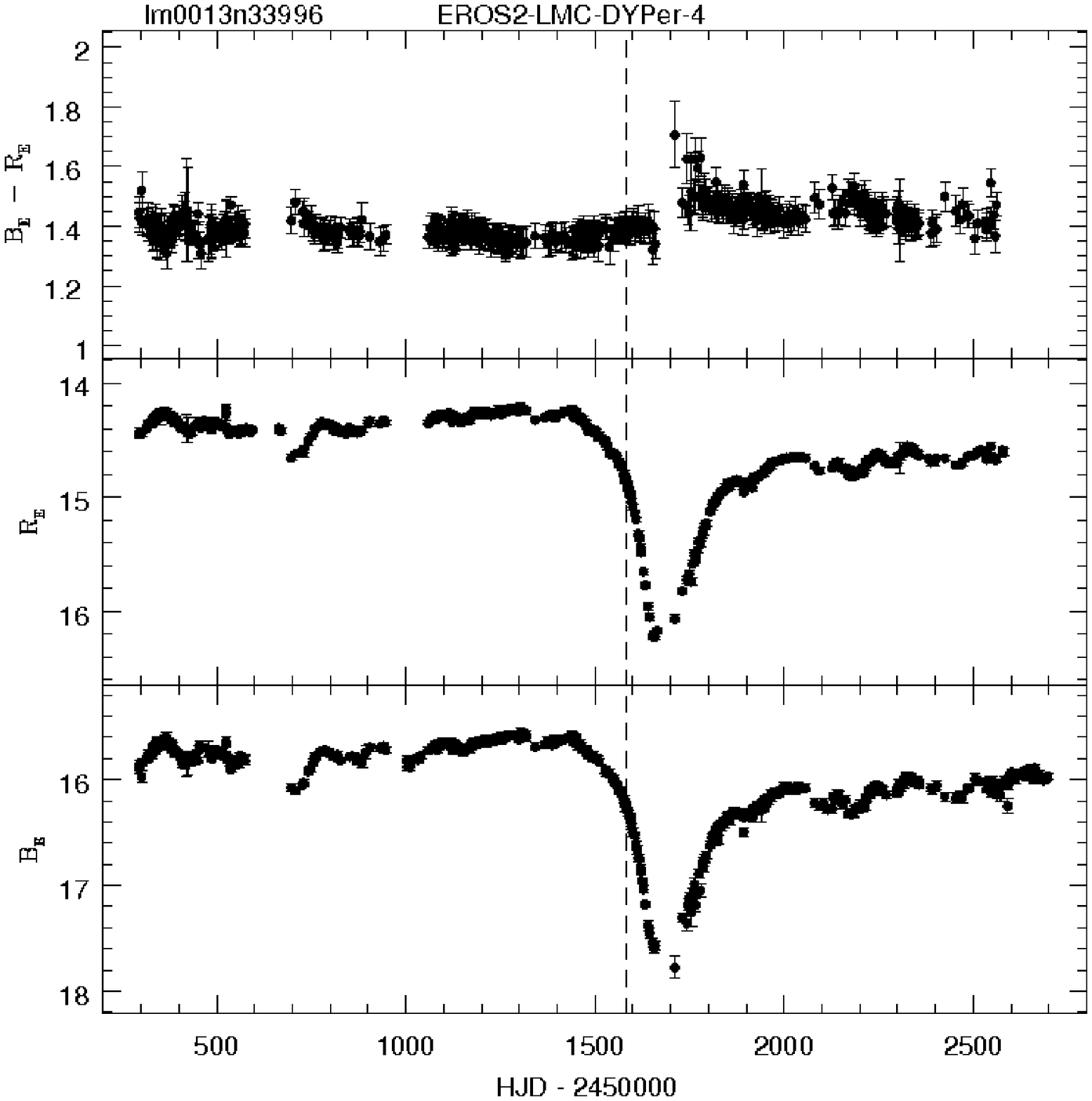}
\includegraphics[scale=0.38]{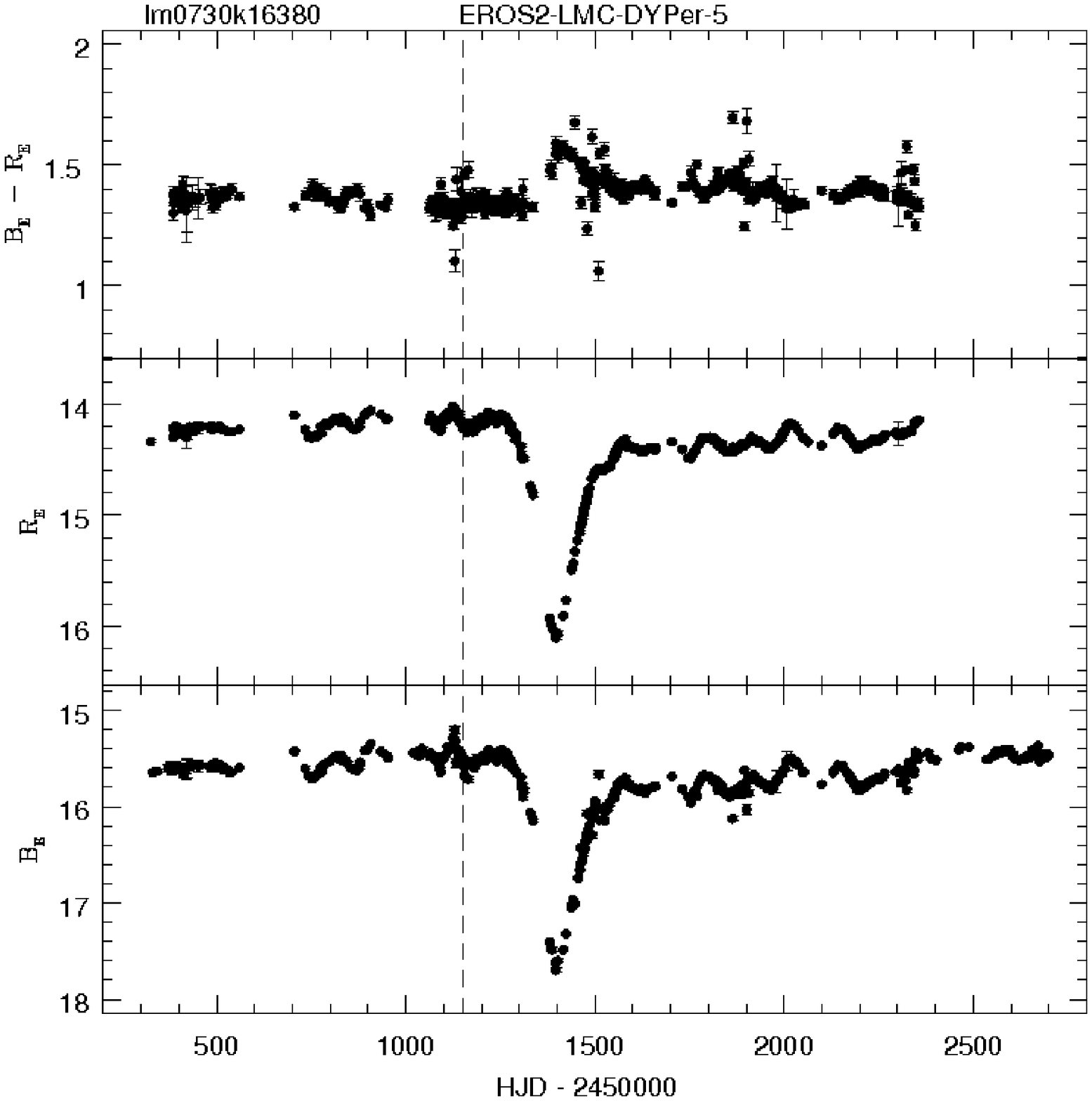}
\includegraphics[scale=0.38]{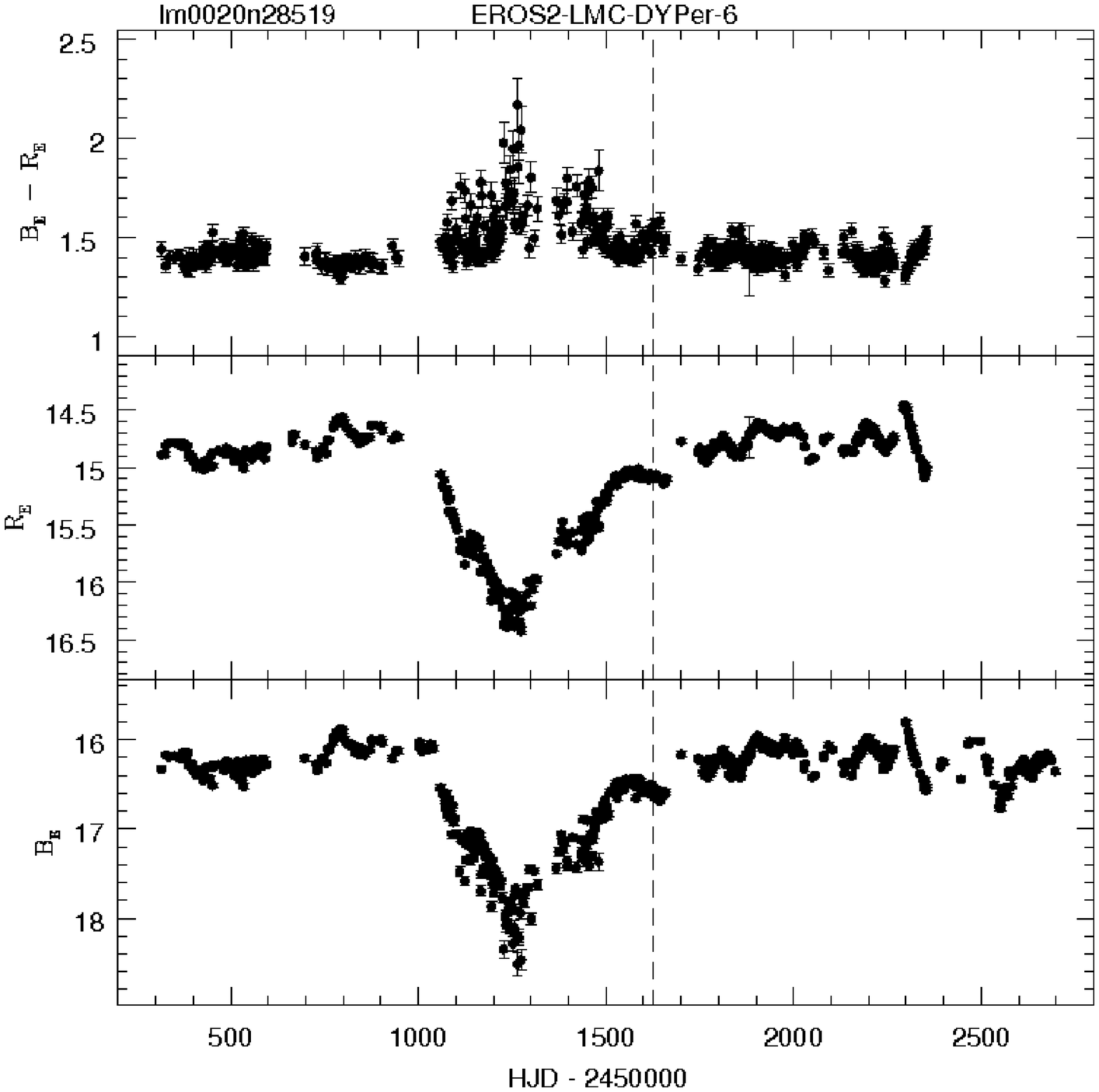}
\includegraphics[scale=0.38]{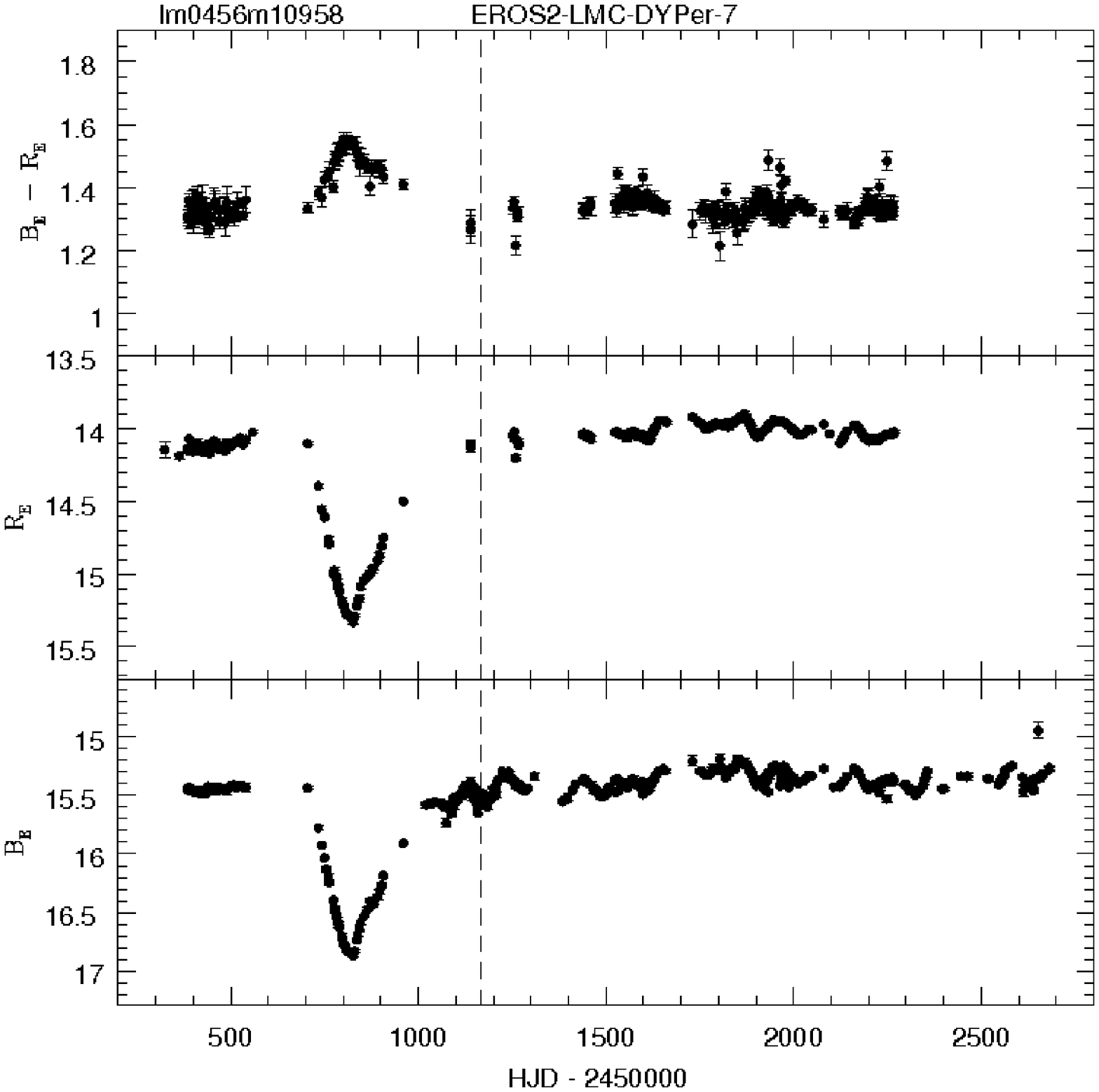}
\includegraphics[scale=0.38]{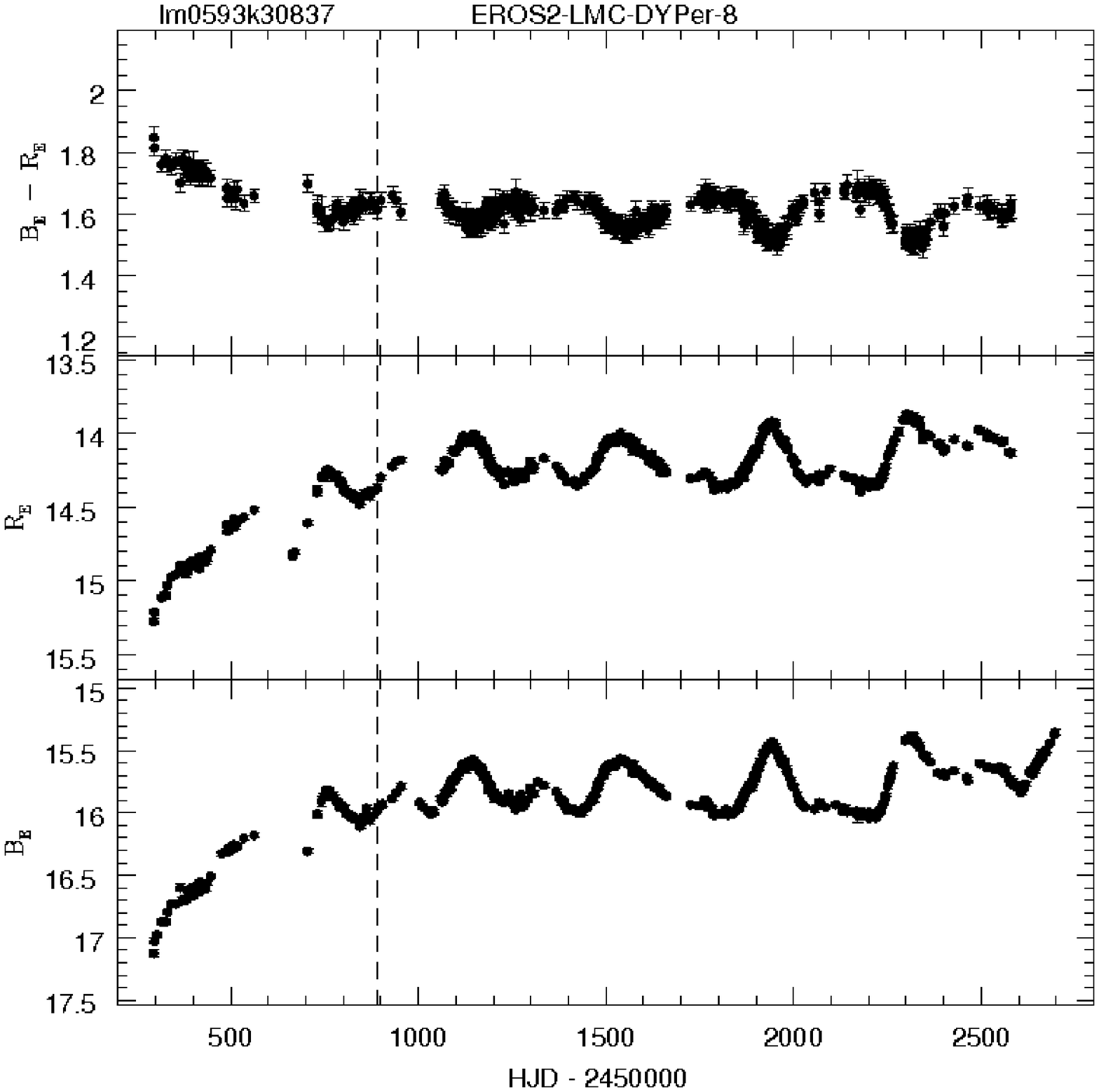}
\includegraphics[scale=0.38]{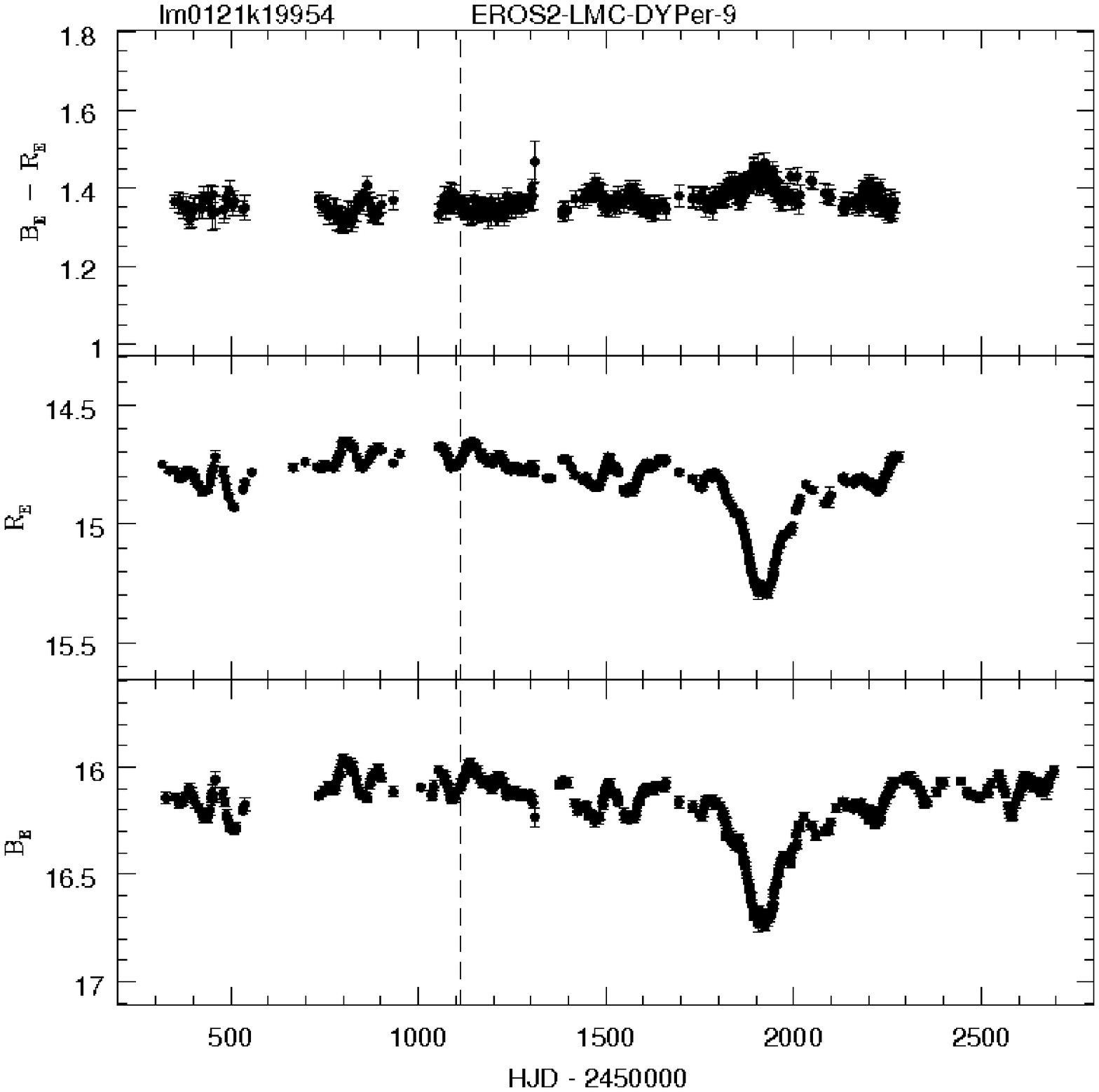}
\caption{Light curves of the new RCB and DYPer stars (continued).}
\end{figure*}

\begin{figure*}
\centering
\includegraphics[scale=0.38]{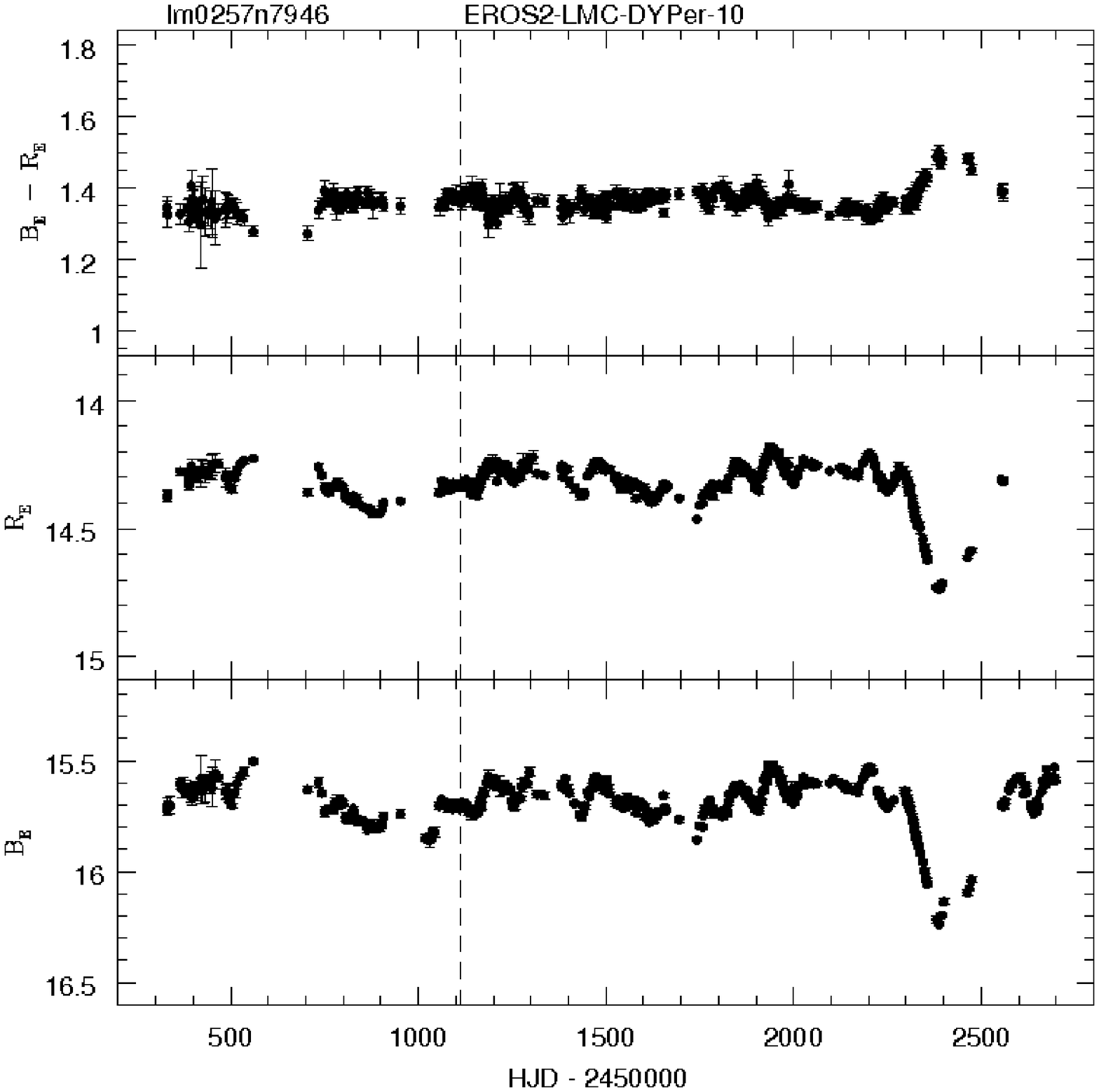}
\includegraphics[scale=0.38]{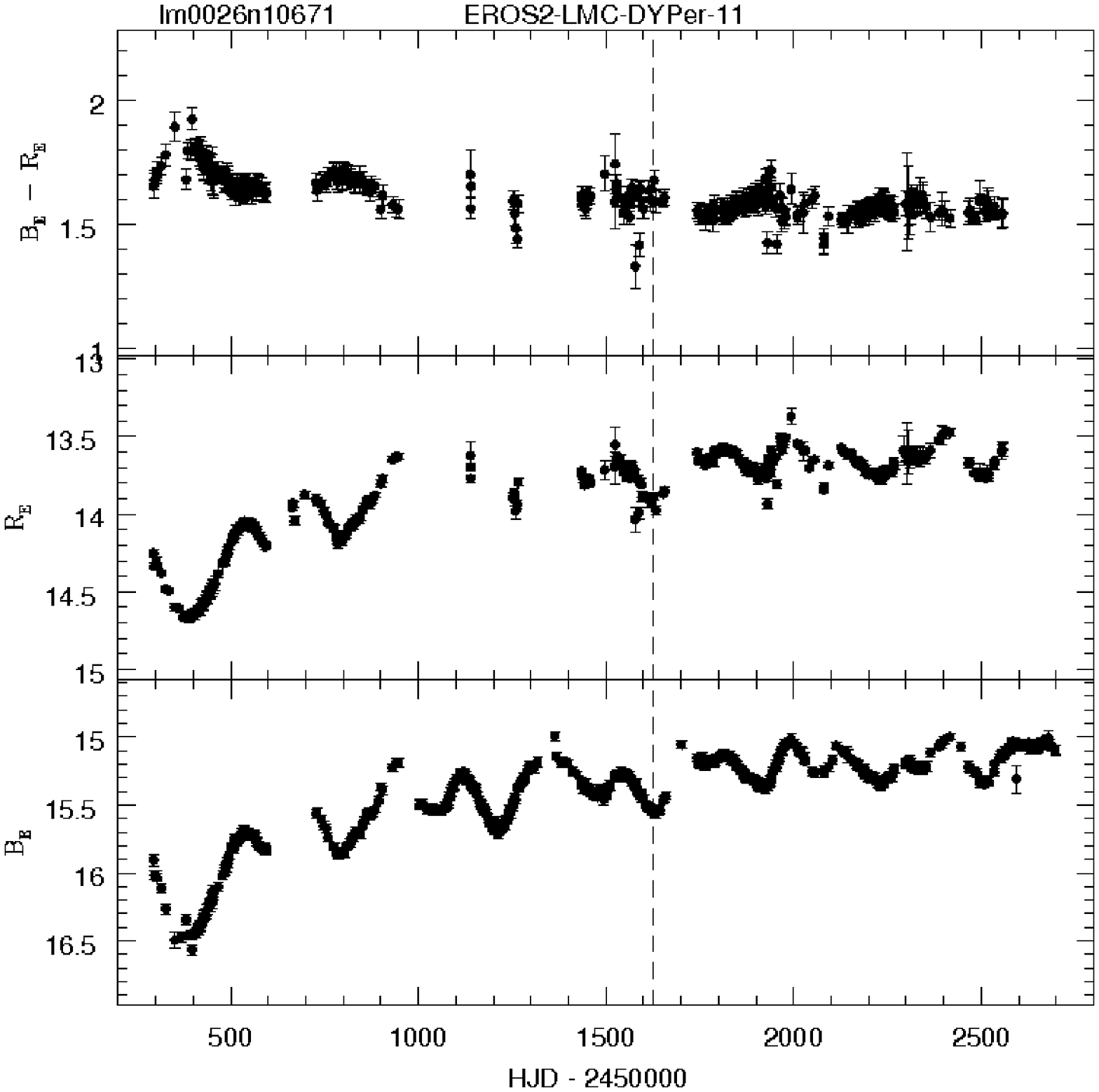}
\includegraphics[scale=0.38]{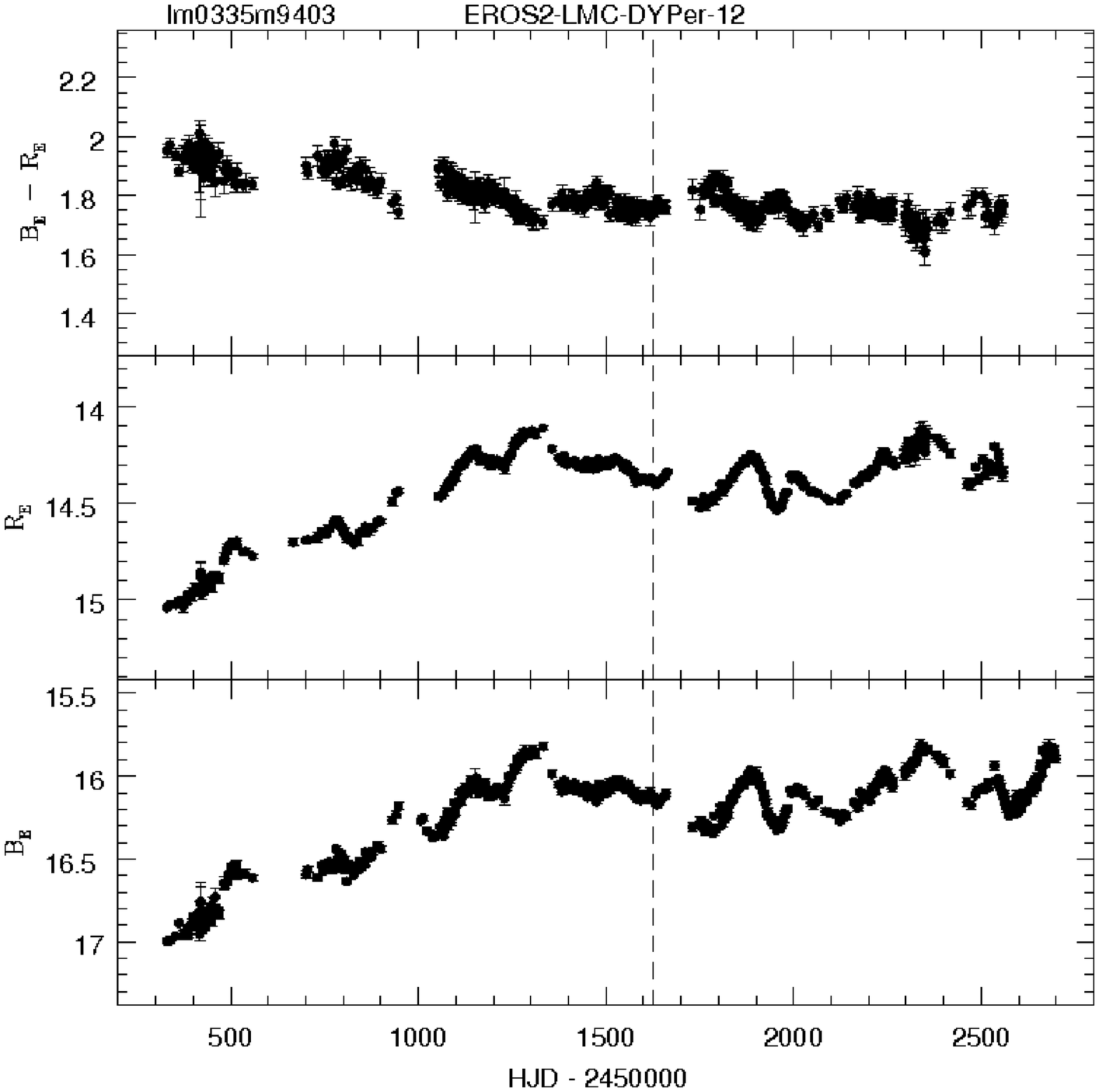}
\includegraphics[scale=0.38]{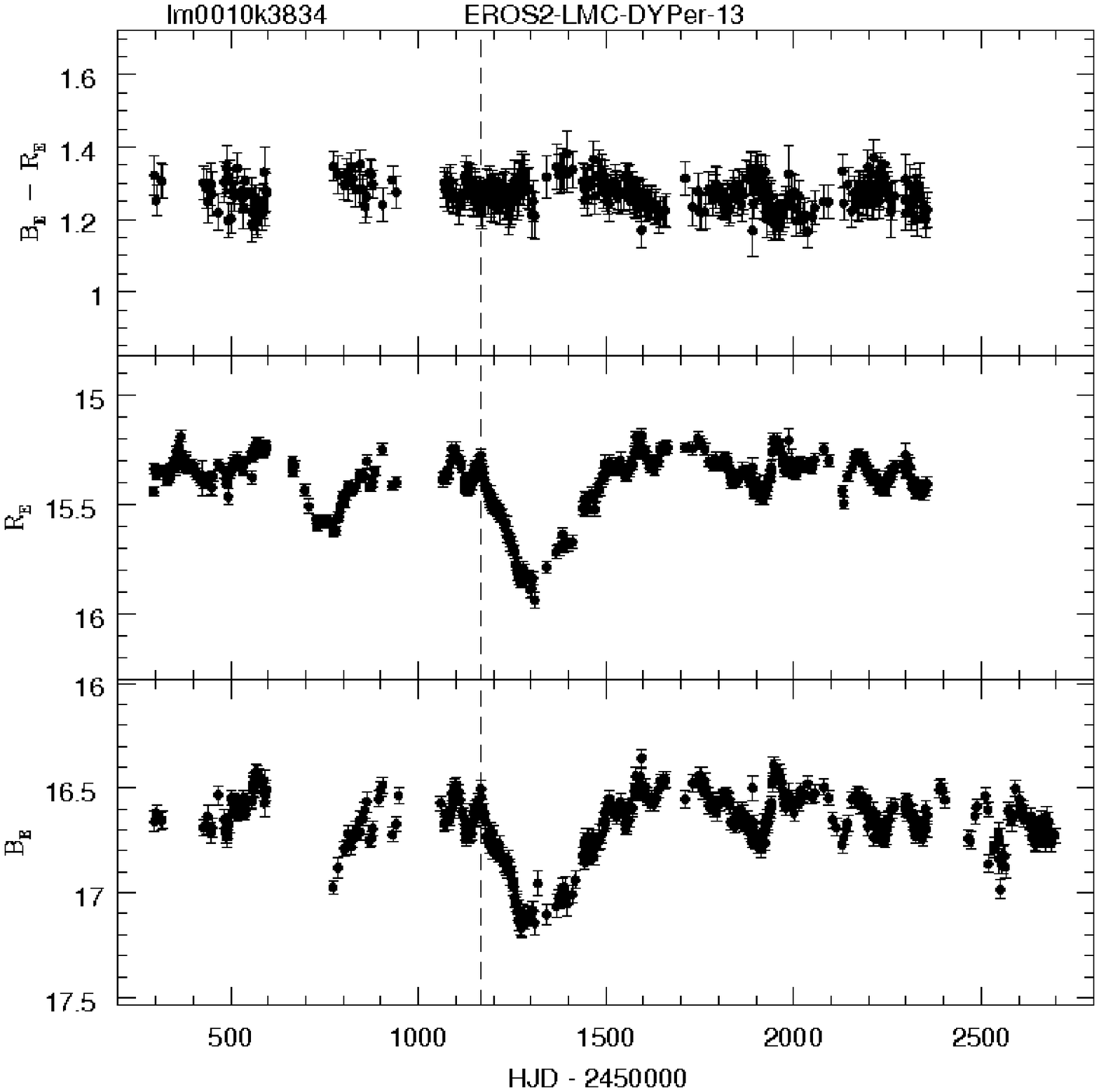}
\includegraphics[scale=0.38]{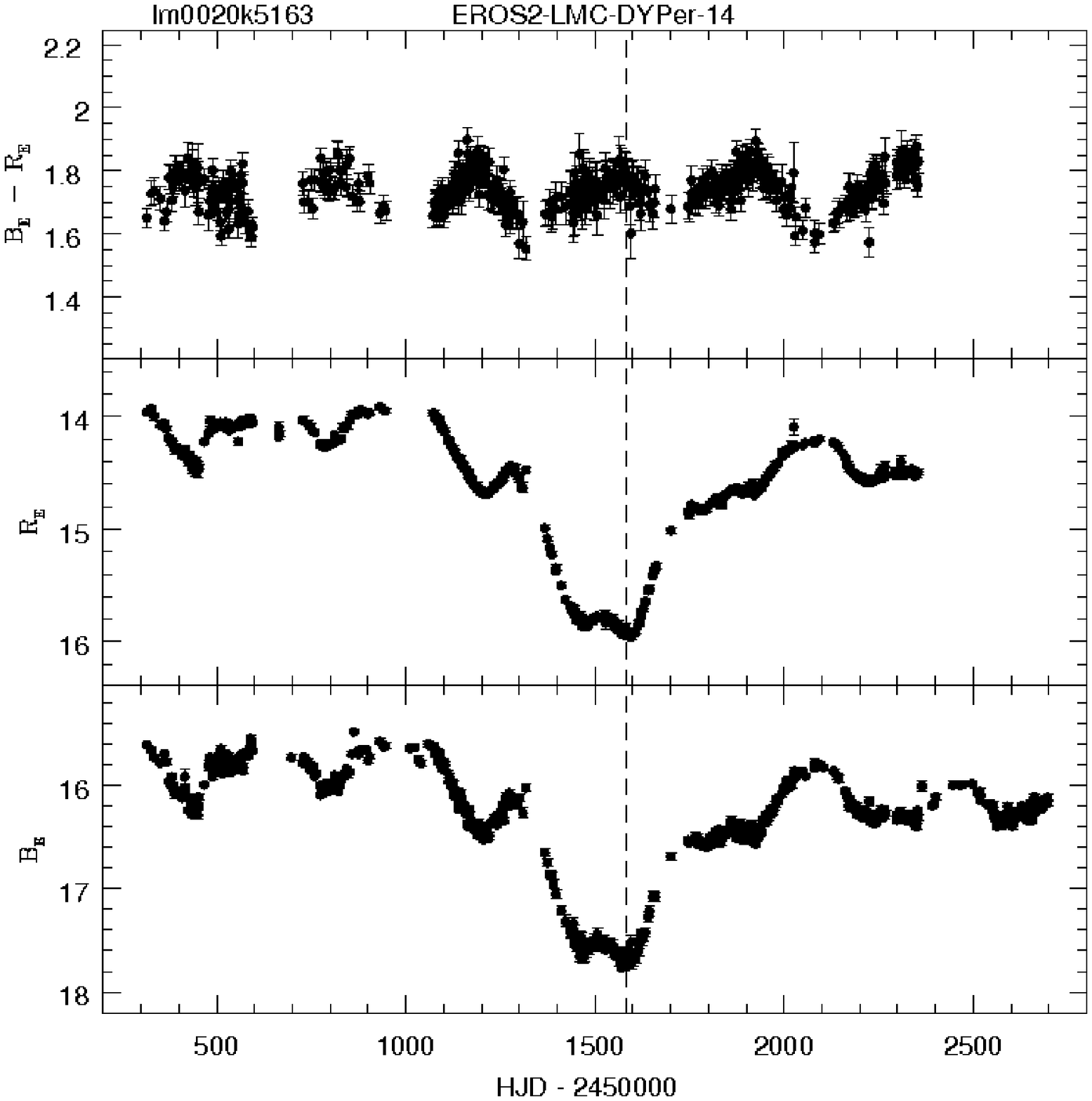}
\includegraphics[scale=0.38]{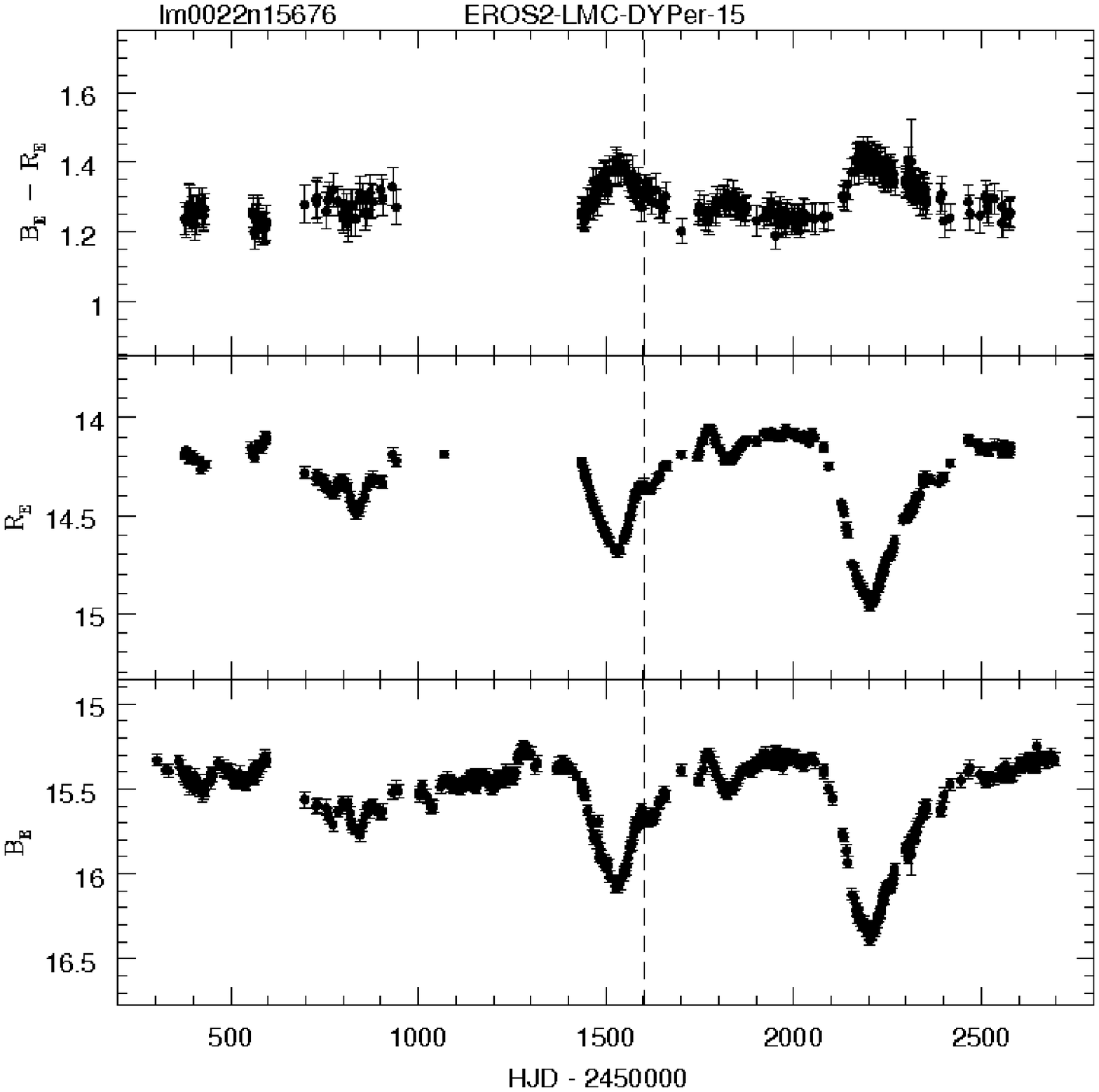}
\caption{Light curves of the new RCB and DYPer stars (continued).}
\end{figure*}

\begin{figure*}
\centering
\includegraphics[scale=0.38]{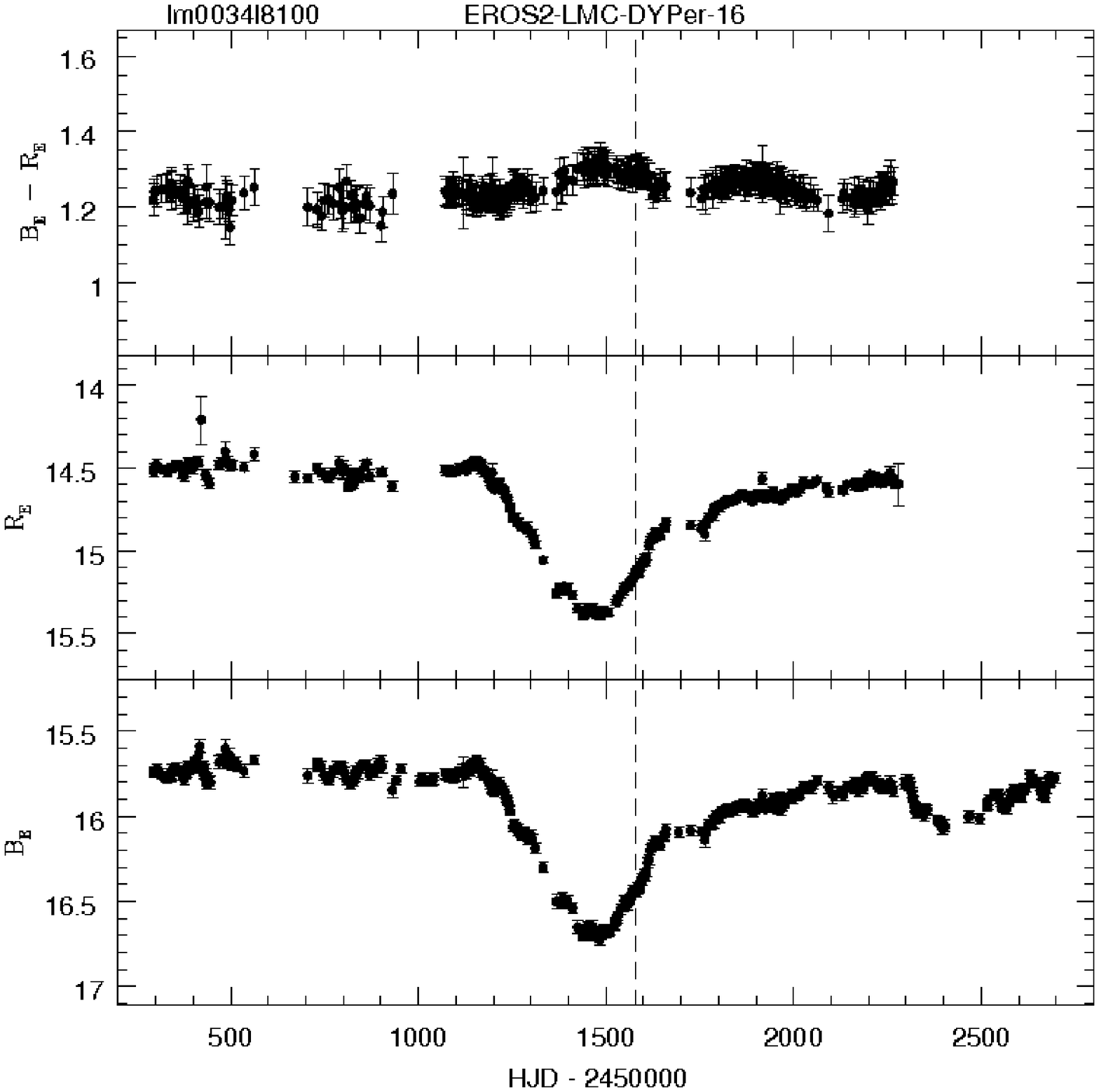}
\includegraphics[scale=0.38]{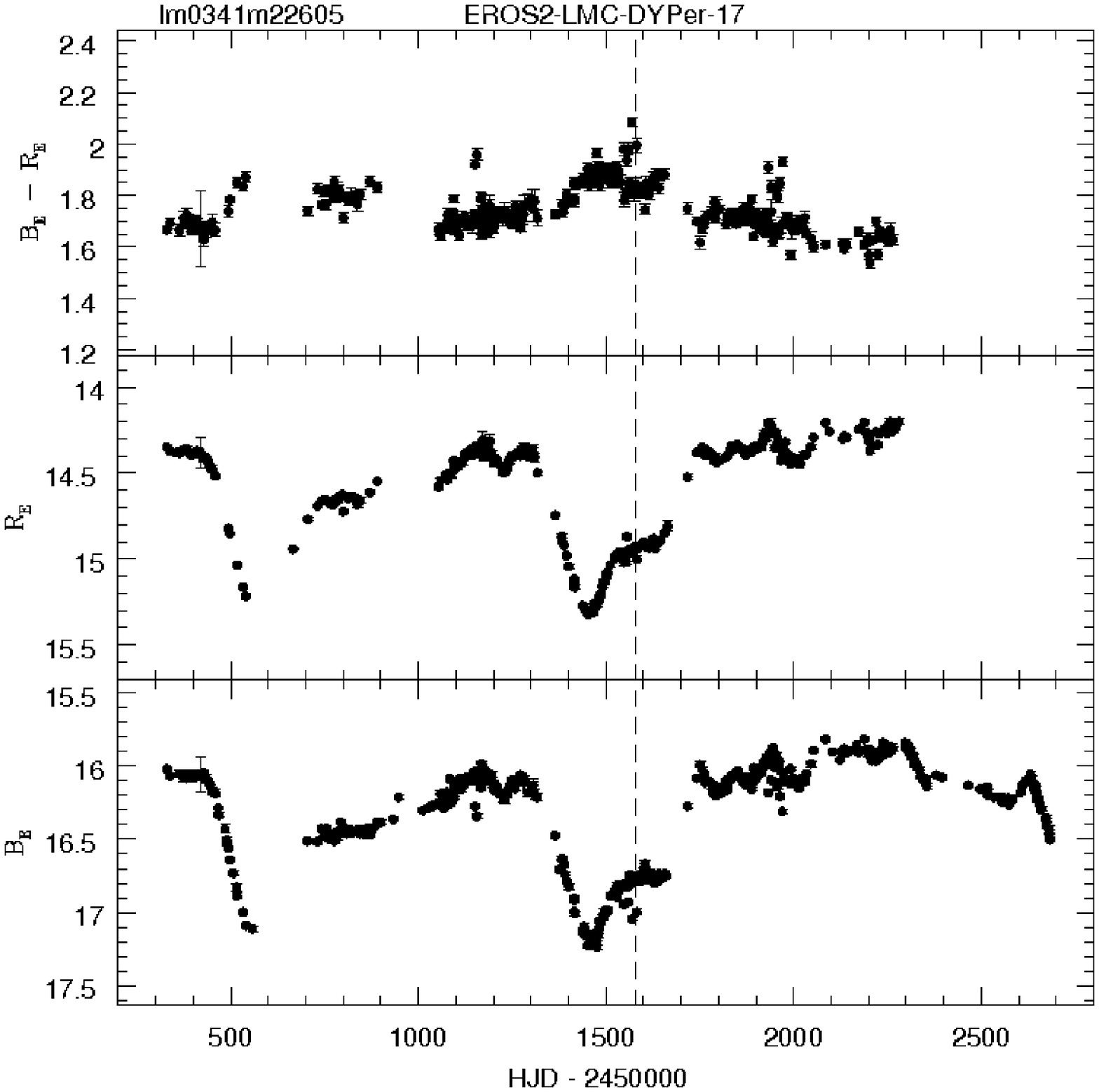}
\includegraphics[scale=0.38]{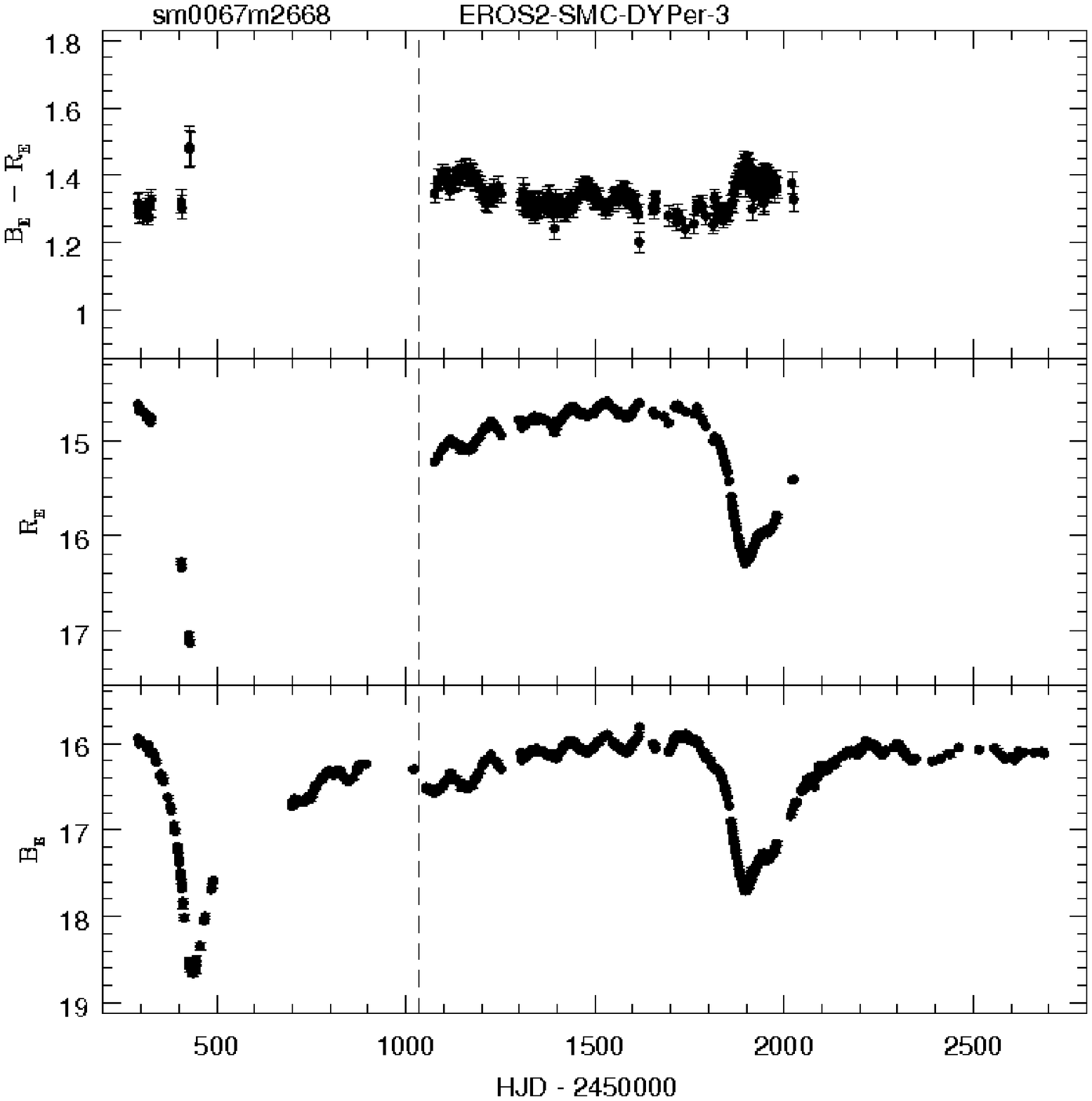}
\includegraphics[scale=0.38]{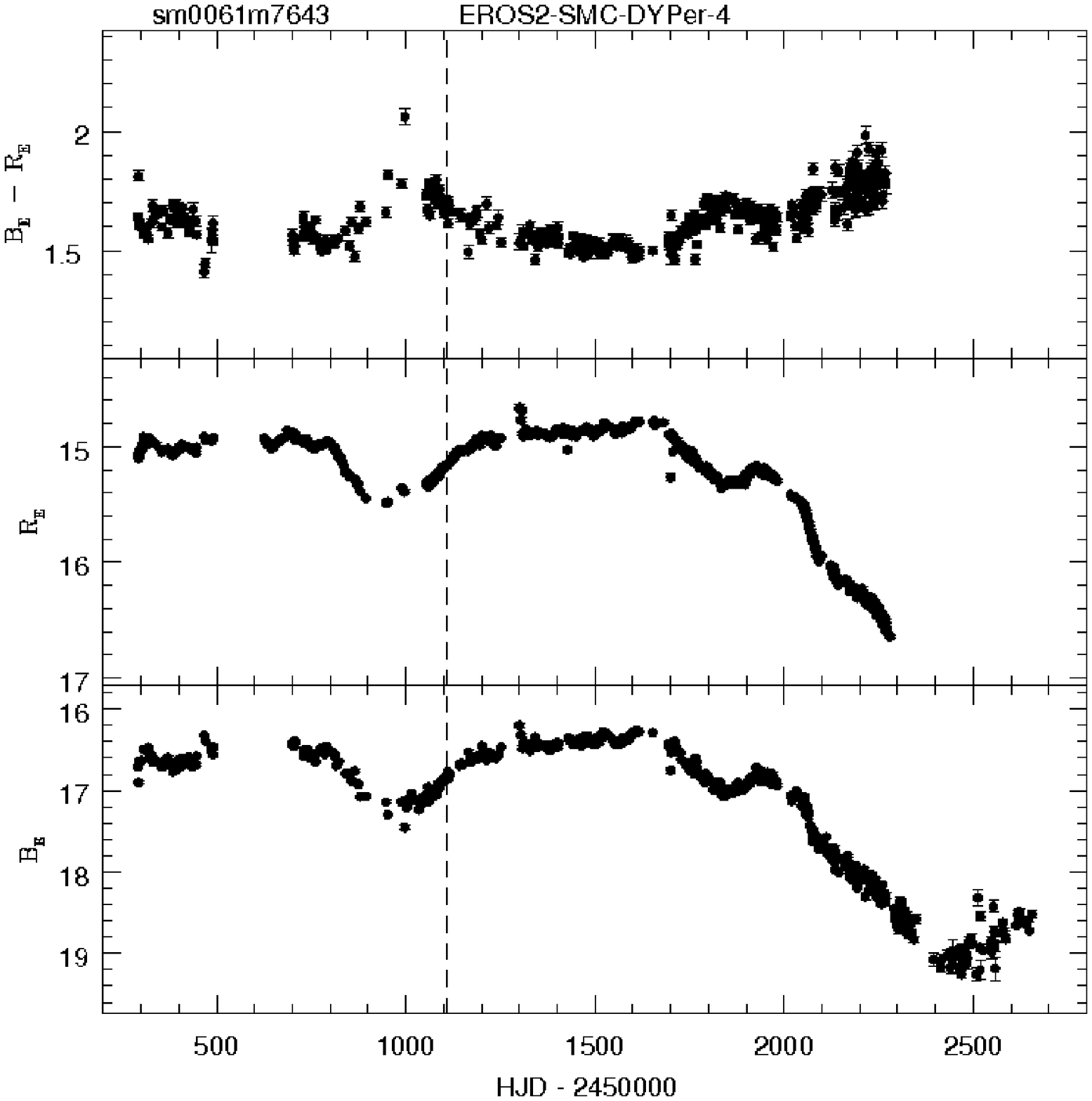}
\includegraphics[scale=0.38]{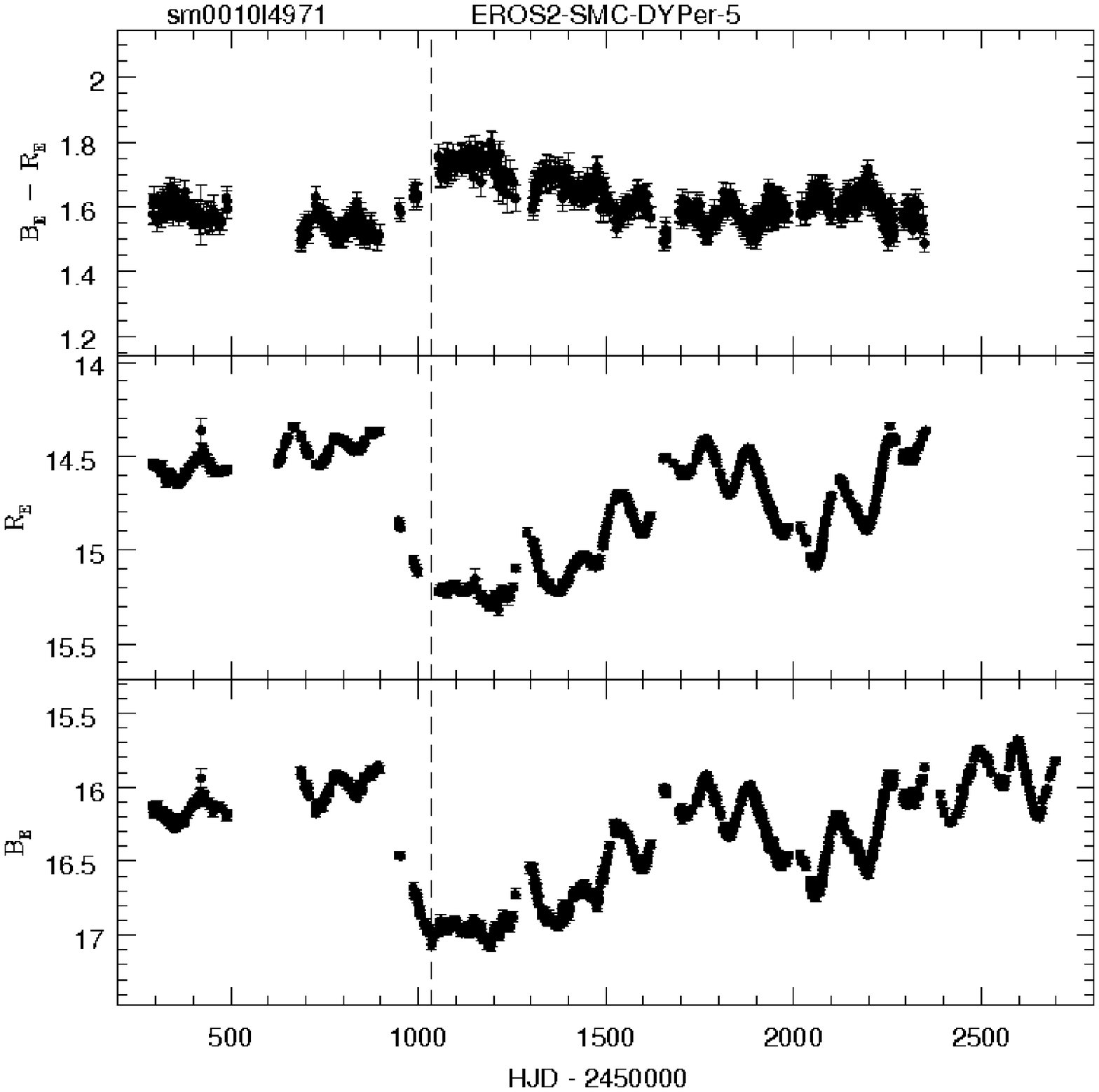}
\includegraphics[scale=0.38]{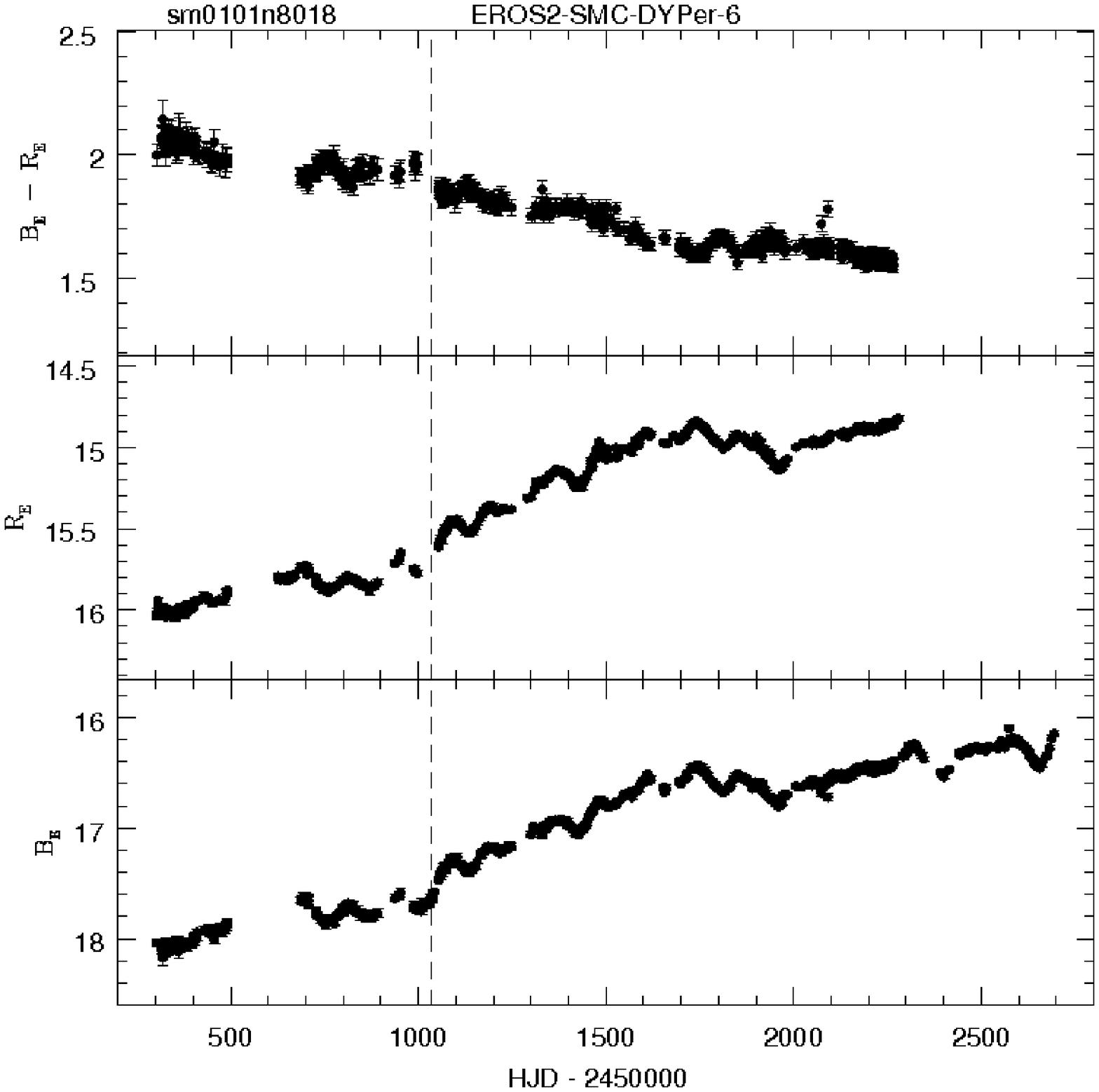}
\caption{Light curves of the new RCB and DYPer stars (continued).}
\label{lc_end}
\end{figure*}

%\listofobjects

\end{document}